%% file: authors.tex
\documentclass[
  journal=pasa,
  manuscript=research-paper, 
  year=2022,
  volume=xx,
]{cup-journal}

\usepackage{microtype,siunitx,booktabs,graphicx,color,rotating,amsmath,amssymb,tabularx, xcolor, subcaption}
\title{Low-Frequency Turnover Star Forming Galaxies I: Radio Continuum Observations and Global Properties}

\author{J. A. Grundy}
\affiliation{International Centre for Radio Astronomy Research, Curtin University, Bentley, WA6102, Australia}
\alsoaffiliation{ATNF, CSIRO Space and Astronomy, PO Box 1130, Bentley, WA6102, Australia}
\email[J. A. Grundy]{joe.grundy@postgrad.curtin.edu.au}

\author{N. Seymour}
\affiliation{International Centre for Radio Astronomy Research, Curtin University, Bentley, WA6102, Australia}

\author{O. I. Wong}
\affiliation{ATNF, CSIRO Space and Astronomy, PO Box 1130, Bentley, WA6102, Australia}
\alsoaffiliation{International Centre for Radio Astronomy Research, University of Western Australia, 35 Stirling Hwy, Crawley, WA6009, Australia}

\author{K. Lee-Waddell}
\affiliation{International Centre for Radio Astronomy Research, University of Western Australia, 35 Stirling Hwy, Crawley, WA6009, Australia}
\alsoaffiliation{ATNF, CSIRO Space and Astronomy, PO Box 1130, Bentley, WA6102, Australia}
\alsoaffiliation{International Centre for Radio Astronomy Research, Curtin University, Bentley, WA6102, Australia}

\author{T. J. Galvin}
\affiliation{ATNF, CSIRO Space and Astronomy, PO Box 1130, Bentley, WA6102, Australia}

\author{M. Cluver}
\affiliation{Centre for Astrophysics and Supercomputing, Swinburne University of Technology, Hawthorn, VIC3122, Australia}

\doi{xx} 

\received {dd Mmm YYYY}
\revised  {dd Mmm YYYY}
\accepted {dd Mmm YYYY}
\published{dd Mmm YYYY}


\begin{document}

\input{lft}


\bibliography{lft}

\appendix

\input{appendix}

\end{document}

%% file: lft.tex
\begin{abstract}
There is growing evidence that the broad-band radio spectral energy distributions (SEDs) of star-forming galaxies (SFGs) contain a wealth of complex physics. In this paper we aim to determine the physical emission and loss processes causing radio SED curvature and steepening to see what observed global astrophysical properties, if any, are correlated with radio SED complexity. To do this we have acquired radio continuum data between 70\,MHz and 17\,GHz for a sample of 19 southern local ($z < 0.04$) SFGs. Of this sample 11 are selected to contain low-frequency (< 300\,MHz) turnovers (LFTOs) in their SEDs and eight are control galaxies with similar global properties.  We model the radio SEDs for our sample using a Bayesian framework whereby radio emission (synchrotron and free-free) and absorption or loss processes are included modularly. We find that without the inclusion of higher frequency data (> 17\,GHz) single synchrotron power-law based models are always preferred for our sample; however, additional processes including free-free absorption (FFA) and synchrotron losses are often required to accurately model radio SED complexity in SFGs. The fitted synchrotron spectral indices range from -0.45 to -1.07 and are strongly anticorrelated with stellar mass suggesting that synchrotron losses are the dominant mechanism acting to steepen the spectral index in larger/more massive nearby SFGs. We find that LFTOs in the radio SED are independent from the inclination of SFGs, however higher inclination galaxies tend to have steeper fitted spectral indices indicating losses to diffusion of cosmic ray electrons into the galactic halo. Four of five of the merging systems in our SFG sample have elevated specific star formation rates and flatter fitted spectral indices with unconstrained LFTOs. Lastly, we find no significant separation in global properties between SFGs with or without modelled LFTOs. Overall these results suggest that LFTOs are likely caused by a combination of FFA and ionisation losses in individual recent starburst regions with specific orientations and interstellar medium properties that, when averaged over the entire galaxy, do not correlate with global astrophysical properties.
\end{abstract}

\section{Introduction}

Galaxy spectral energy distributions (SEDs) contain crucial information about the astrophysical processes occurring within them, with varying contributions from gas, dust, stars, and, active galactic nuclei (AGN) at different wavelengths \citep{Walcher2011, Conroy2013}. Different emission mechanisms dominate at distinct wavelength regimes and vary over time allowing us to measure certain astrophysical quantities and estimate their histories \citep[e.g.][]{Thorne2023}. For example, stellar masses can be measured in the optical and/or near-infrared regime where stellar emission is dominant \citep{Taylor2011}. Star formation rates (SFRs) can be measured instantaneously at ultraviolet (UV) wavelengths due to the emission being dominated by short-lived high mass OB stars (HMS; $M_{\odot} \geq 8\,M_{\odot}$) \citep{Kennicutt2012}. However optical and UV photons are absorbed by intervening dust which is heated, reradiating this energy at infrared (IR) wavelengths \citep{draine2003}. Thus corrections, or a combination of IR and UV emission, are often required to provide accurate estimates of the SFR \citep{Bell2005, Kennicutt2009, Davies2016, Delvecchio2021}. Radio continuum emission however is impervious to the effects of dust attenuation due to its long wavelength and can be used as a direct probe into the star-formation activity of ``normal'' SFGs \citep[galaxies without AGN; ][]{Condon1992}. 

The shape of a galaxy's radio SED can provide further insight into the astrophysical processes occurring within but has been often simplified and underutilised due to lack of observations and poor spectral sampling (often including only two photometric points over four orders of magnitude in radio frequency). The spectral index ($\alpha$, where flux density, $S_{\nu} \propto \nu^{\alpha}$) of a galaxy's radio emission can be used to explore radiation laws and examine the interplay between the heating and cooling mechanisms of the interstellar medium (ISM), magnetic fields and relativistic cosmic rays \citep{Murphy2008, Lacki2010, roth2023, Thorne2023}. In galaxies that contain an AGN, the accretion activity and timescales, as well as how AGN jets impact the ISM of the host galaxy, can also be explored through the spectral index and variability over time \citep{Fabian2012, Slob2022, Ross2023}. More widely studied is the direct proportionality between the radio luminosity and SFR in normal SFGs \citep{Condon1992, Bell2003, Davies2017, Heesen2022a}.

Radio continuum emission from normal SFGs is produced as a result of two different emission processes of which both are dependent on HMS formation. Thermal free-free emission is produced in the HII envelopes around HMS's that are ionised by UV flux and provides a direct, near-instantaneous measure of SFR. Thermal emission therefore strongly depends on the number of ionising UV photons with only a weak dependence on electron temperature \citep[see Eq. 2; ][]{Condon1992}. Non-thermal synchrotron emission is produced by relativistic cosmic ray electrons (CREs), accelerated by the shocks produced after Type II and Type Ib supernova, interacting with the large scale magnetic fields within a galaxy. Synchrotron emission typically dominates below 30\,GHz and is delayed by at least $\approx$10\,Myr \citep{Condon1992}. The CREs that produce the synchrotron emission lose energy as they propagate away from star-forming regions with lifetimes of $\sim$50-300\,Myr and scale lengths up to $\approx$7\,kpc dependent on their injection energy and magnetic field strength \citep{Condon1992, Murphy2008, Heesen2023, heesen2024}. Synchrotron emission has a characteristic power-law emission spectrum ($\alpha \approx$ -0.8) whilst free-free emission has almost a flat spectral index ($\alpha \approx$ -0.1). These differences in emission spectrum can be used to separate their contribution to a galaxy's radio SED.

Current radio continuum SFR indicators, mostly related to the 1.4\,GHz luminosity, are typically calibrated against far-infrared (FIR) or total-infrared (TIR) measurements using the FIR/TIR to radio correlation \citep[FRC/TRC; ][]{Condon1992, Yun2001, Bell2003, Molnar2021, Heesen2022a}. The FRC/TRC has been shown to have a tight linear relationship over many orders of magnitude in FIR/TIR and radio luminosities for SFGs \citep{Condon1992, Yun2001, Bell2003, Molnar2021}. This strong correlation arises as a result of the common origin of HMS formation for both radio and IR emission and is often interpreted using calorimetric models \citep{Volk1989, Lacki2010}. These calorimetric models assume that galaxies are optically thick to UV emission such that all UV emission is absorbed and re-emitted as FIR radiation \citep{holwerda2005} and that CREs release all their energy as synchrotron emission before they leave the galaxy. Current research however, suggests a ``conspiracy'' which maintains the FRC/TRC in non-calorimetric models. For example lower mass galaxies with UV continuum leakage are also small enough such that their CRE diffusive escape time is less than than the synchrotron cooling time \citep{Niklas1997, Murphy2009, Lacki2010, Basu2015, Heesen2022a}. Recent findings by \cite{Cook2024} have also shown the SFR history is an important factor when considering the FRC/TRC for calibration of the radio SFR. Synchrotron emission relies on core-collapse supernova from HMS to accelerate the CREs such that the emission is delayed compared to IR emission which is partially responsible for the non-linearity at low masses and scatter in the relationship. Current research suggests the FRC/TRC for SFGs decreases with redshift \citep{Sargent2010, magnelli2015, Delhaize2017, Delvecchio2021, Molnar2021} however this is primarily thought to be a selection effect of sampling bias towards higher mass/luminosity galaxies at higher redshift.

Radio continuum emission can also impacted by a range of loss and absorption processes that leads to the increased complexity in their radio SEDs. At low frequencies free-free absorption (FFA) and ionisation losses are thought to contribute to low-frequency turnovers (LFTOs) recently observed in the radio SEDs of some ultra-luminous/luminous IR galaxies \citep[ULIRGS/LIRGS][]{Clemens2010, Galvin2018, Dey2022, Dey2024}. FFA is dependent on the electron density along our line of sight. As such, high density or heavily obscured starburst regions are expected to absorb the radio emission toward lower frequencies. There is little to no evidence of a relationship between FFA and galactic inclination \citep{hummel1991, chyzy2018}, which therefore suggests that the dense ISM within individual star-forming regions is the primary absorber rather than the lower density ISM within the galactic disk. Low energy CREs can also lose energy due to the ionisation of atomic and molecular hydrogen and the energy loss is directly proportional to the number density of neutral atoms and molecules \citep{Longair2011}. Ionisation losses act to flatten the low frequency spectral index \cite[$\Delta\alpha \leq$ 0.5;][]{Basu2015} if the ionisation loss timescale is less than the synchrotron loss timescale \citep[$\approx$100\,Myr; ]{Murphy2009, Heesen2023} which can occur in regions of a relatively low total gas mass surface density \citep[$\Sigma_{gas}$ = 2.5 - 5 M$_\odot\,$pc$^{-2}$; ]{Basu2015}.

At higher frequencies synchrotron cooling and inverse-Compton (IC) losses become more dominant with their combined effect gradually steepening the synchrotron spectral index by $\Delta\alpha$ = -0.5 \citep[see Eq. 12;][]{Condon1992}. IC losses occur as the synchrotron emitting CREs are scattered by photons and is dependent on the photon energy density, both from IR ISM emission and the cosmic microwave background (CMB), CRE energy and redshift \citep[in the case of CMB photons; ][]{Lacki2010, roth2023}. Synchrotron cooling losses however are primarily dependent on a galaxy's magnetic field and size such that the synchrotron electrons lose energy as they propagate away from their injection site before escaping into the galactic halo \citep{Murphy2008, Lacki2010, heesen2024, roth2024}. Therefore higher frequency spectral steepening in nearby SFGs is expected to be more prevalent in both larger galaxies and extremely compact starbursts with high gas surface densities. Bremsstrahlung, adiabatic and diffusion losses also play a factor in reducing the overall radio emission observed but they do not impact the curvature of synchrotron spectral index as they are generally subdominant and their CRE loss timescales are almost energy independent \citep{roth2023}.

Somewhat surprisingly, the combined effect of these emission, loss and absorption processes result in an observed SFG spectral index of $\alpha \approx$ -0.7 at the typically surveyed 1.4\,GHz radio frequency. As all-sky radio continuum surveys such as GLEAM \citep{Hurley2017} and the LOw-Frequency ARray (LOFAR) Two-metre Sky Survey (LoTSS) \citep{Shimwell2017} have begun to fill out the low frequency regime (50 - 300\,MHz) and targeted observations cover individual sources or regions above 1.4\,GHz, the formerly hidden complexity of the radio SED is being revealed. Recent work has begun to uncover and explain this radio SED complexity with the inclusion of FFA components \citep{Clemens2010, Galvin2018, Dey2022, Dey2024} however the samples are limited to unresolved LIRGS/ULIRGS and do not comprehensively explore other loss processes and their possible relationship with galactic astrophysical properties. With our improved spectral sampling and a sample of more nearby, resolvable lower SFR SFGs, one of the primary goals of this paper is to explore how the shapes of radio SEDs and their parameters are related to a galaxy's global astrophysical properties. This will allow us to infer the dominant physical processes, including emission processes and cooling in the ISM, for a diverse sample of SFGs.

In this study we select a sample of twenty nearby (z < 0.04) SFGs including twelve with LFTOs and eight controls to investigate whether their global astrophysical properties differ. We perform detailed radio SED modelling by constructing a series of increasingly complex modular, radio emission models including loss and absorption processes. We then compare the model results to a number of global properties to investigate how they are connected to the radio SED features observed and the emission and loss processes occurring within SFGs. In sections \ref{sec:sample} and \ref{sec:data} we describe the sample selection and data acquisition. Section \ref{sec:SED} details the SED model construction, fitting and selection. Sections \ref{sec:results} and \ref{sec:discussion} present the results and discussion of these results. Lastly section \ref{sec:conclusion} presents our conclusions. Throughout this paper we assume a Hubble constant of 70\,km s$^{-1}$ Mpc$^{-1}$ (\textit{h} = 0.70), and matter and cosmological constant density parameters of $\Omega_{M}$ = 0.3 and $\Omega_{\Lambda}$ = 0.7.

\section{Sample and Ancillary Data}
\label{sec:sample}
\subsection{Sample Selection}

We initially select SFGs from the GLEAM-6dFGS catalogue \citep{franzen2021} of $z\leq 0.1$ sources selected at 200\,MHz (1590 total sources). Galaxies are categorised as SF based on their optical spectrum containing H$\alpha$ and H$\beta$ emission lines typical of HII regions (427 SFGs or 26.9\%). The candidate LFTO galaxies are selected based on the measured spectral index between 76 - 227\,MHz ($\alpha_{\rm L} \geq -0.2$, 15 sources) which are then visually inspected and confirmed to exhibit flattening/peaks in their radio SED. Visual inspection is necessary due to the large uncertainties in flux density for some GLEAM photometry in a few sources. GLEAM sources with any negative measured sub-band flux densities were not fitted with spectral indices but are also considered upon visual inspection (54 sources with no measured $\alpha_{\rm L}$, 27 of which have $\alpha^{\rm GLEAM}_{\rm low} \geq$ -0.2). The candidate LFTO galaxies were then further selected with declination < $-14^{\circ}$ to minimise elongation of the synthesised beam during Australian Telescope Compact Array \citep[ATCA;][]{frater1992, wilson2011} observations and to maintain comparable beamshapes to {\it Widefield Infrared Survey Explorer} \citep[WISE;][]{Wright2010} observations. Using the high frequency spectral index ($\alpha_{\rm H}$) between 227 - 843/1400\,MHz (depending on whether Sydney University Molonglo Sky Survey \citep[SUMSS; ][]{mauch2003} or NRAO VLA Sky Survey \citep[NVSS; ][]{Condon1998} observations are available) we estimated 9.5\,GHz flux density and performed a flux cut of $S_\nu=$8\,mJy to select brighter SFGs allowing them to be observed by ATCA and ensure sensitivity within the requested observing time. Further LFTO sample restriction was then performed by limiting $z \leq 0.04$. Lastly visual inspection of the RACS 888\,MHz and NVSS/SUMSS emission for the remaining LFTO candidates was performed to remove sources which were confused in the large ($\sim 2'$) GLEAM beam which left a final sample of 11 LFTO SFGs.

A control sample of eight sources is selected first by limiting the low-frequency spectral index to steeper values ($\alpha_{\rm L} \leq -0.5$). The distribution of their redshifts, $K$-band absolute magnitude and estimated 9.5\,GHz flux density are then limited in the same fashion as to the LFTO sample and their distributions matched. Control galaxies are then chosen to be within 10$^{\circ}$ of LFTO galaxies to allow the use of the same phase calibrator for multiple sources and reduce overheads during ATCA observations. We then perform the same visual inspection to remove GLEAM confused sources. The complete SFG sample including common parameters and a flag for whether they are LFTO or control galaxies are given in Table \ref{tab:commonprops}.

\begin{table*}[hbt!]
\centering
\begin{threeparttable}
\caption{SFG Sample Properties.}\label{tab:commonprops}
\begin{tabular}{c|l|c|c|c|c|c|c|c|c|c}
	\toprule
	GLEAM ID & Common ID & Sample & R.A & Dec. & z & D$_L$ & K$_{Mag}$ & Log$_{10}$(M$_{*}$) & Inc. & Morphology\\
     &  &  & ($^{\circ}$) & ($^{\circ}$) &  & (Mpc) & (M) & (M$_{\odot}$) & ($^{\circ}$) & \\
	\midrule
	GLEAM J002238-240737 & ESO473-G018 & Con & 5.66210 & -24.12705 & 0.033 & 141.0 & -25.0 & $10.8\pm0.1$ & 70 & Sbc \\
	GLEAM J003652-333315 & ESO350-IG038 & LFTO & 9.21863 & -33.55467 & 0.021 & 85.9 & -23.0 & $10.1\pm0.1$ & 55$^{c}$ & I-Merger \\
	GLEAM J011408-323907 & IC 1657 & Con & 18.52921 & -32.65090 & 0.012 & 51.8 & -24.1 & $10.5\pm0.1$ & 71 & SBbc \\
    GLEAM J012121-340345 & NGC 0491 & Con & 20.33502 & -34.06330 & 0.013 & 52.5 & -24.3 & $10.6\pm0.1$ & 44 & SBb \\
	GLEAM J034056-223353 & NGC 1415 & LFTO & 55.23695 & -22.56463 & 0.005 & 24.6 & -23.6 & $10.4\pm0.1$ & 63 & S0-a \\
	GLEAM J035545-422210 & NGC 1487 & LFTO & 58.93893 & -42.36540 & 0.003 & 10.9 & -20.5 & $9.2\pm0.1$ & 48$^c$ & I-Merger \\
	GLEAM J040226-180247 & ESO549-G049 & Con & 60.60695 & -18.04759 & 0.026 & 113.7 & -25.0 & $10.9\pm0.1$ & 35 & Sbc \\
	GLEAM J041509-282854 & NGC 1540 & Con & 63.79102 & -28.48390 & 0.019 & 79.9 & -22.1 & $10.1\pm0.1$ & 46$^{b,c}$ & N/A-Merger \\
	GLEAM J042905-372842 & ESO303-G021 & Con & 67.27464 & -37.47952 & 0.029 & 127.9 & -24.6 & $10.7\pm0.1$ & 37 & SBab \\
	GLEAM J072121-690005 & NGC 2397 & LFTO & 110.33320 & -69.00146 & 0.005 & 22.7 & -23.0 & $10.3\pm0.1$ & 53 & SBb \\
	GLEAM J074515-712426 & NGC 2466 & LFTO & 116.31823 & -71.41043 & 0.018 & 78.8 & -24.3 & $10.8\pm0.1$ & 37 & Sc \\
	GLEAM J090634-754935 & ESO036-G019 & Con & 136.64885 & -75.82571 & 0.015 & 80.5 & -24.5 & $10.8\pm0.1$ & 68 & SBbc \\
	GLEAM J120737-145835 & MCG-02-31-019 & LFTO & 181.90985 & -14.96961 & 0.018 & 83.2 & -24.6 & $10.7\pm0.1$ & 56 & Sb \\
	GLEAM J142112-461800 & IC 4402 & LFTO & 215.30450 & -46.29787 & 0.006 & 26.7 & -23.7 & $10.5\pm0.1$ & 74 & Sb \\
	GLEAM J150540-422654 & IC 4527 & LFTO & 226.42091 & -42.44951 & 0.017 & 63.1 & -24.6 & $10.6\pm0.1$ & 69 & Sbc \\
	GLEAM J184747-602054 & AM 1843-602 & LFTO & 281.93881 & -60.34822 & 0.036 & 156.7 & -24.9 & $10.9\pm0.1$ & 44$^c$ & I-Merger \\
	GLEAM J203047-472824 & NGC 6918 & LFTO & 307.69632 & -47.47370 & 0.017 & 23.6 & -24.2 & $9.4\pm0.1$ & 56 & S0-a \\
	GLEAM J205209-484639 & NGC 6970 & Con & 313.03973 & -48.77777 & 0.017 & 73.8 & -24.8 & $10.7\pm0.1$ & 33 & SBa \\
	GLEAM J213610-383236 & ESO343-IG013 & LFTO & 324.04446 & -38.54343 & 0.019 & 81.2 & -24.0 & $10.5\pm0.1$ & 53$^{a, c}$ & Sbc-Merger \\
	\bottomrule
\end{tabular}
\begin{tablenotes}[para]
    \item[]Note: Column (1): GLEAM source ID. Column (2): Common ID from NED. Column (3): SFG sample membership. Column (4-5): Right ascension and declination (J2000) of the source from the WXSC \citep{Jarrett2019}. Column (6): Redshift from GLEAM-6dFGS as in \citet{franzen2021}. Column (7): Luminosity distance from the WXSC \citep{Jarrett2019}. Column (8): \textit{K}-band absolute magnitude from \citet{franzen2021}. Column (9): Stellar mass from the WXSC calculated using the light and colours method \citep{jarrett2023}. Column (10): Estimated source inclination calculated using \textit{K}-band light axis ratios from the Two Micron All-Sky Survey \citep{jarrett2000}. Column (11): Lyon-Meudon Extragalactic Database \citep[LEDA; ][]{Paturel2003} morphology.\\
    \item[$^{a}$]\textit{K}-band light axis ratios were unavailable so light axis ratios from the Morphological catalogue of galaxies \citep{voro1974} were used.\\
    \item[$^{b}$]\textit{K}-band light axis ratios were unavailable so light axis ratios from the ESO-Uppsala galaxies catalogue \citep{lauberts1989} were used.\\
    \item[$^{c}$]Merging systems; inclination estimates are unreliable.
\end{tablenotes}
\end{threeparttable}
\end{table*}

\subsection{Radio}

We make use of the radio data included in the GLEAM-6dFGS SFG catalogue \citep{franzen2021} which includes 20 flux measurements in the GLEAM frequency bands (72-232\,MHz) with channel widths of 7.68\,MHz. A declination dependent SUMSS \citep[843\,MHz;]{mauch2003} or NVSS \citep[1.4\,GHz;]{Condon1998} measurement is also included. The GLEAM-6dFGS SFG catalogue \citep{franzen2021} also includes a measurement of the GLEAM spectral index as measured in \cite{Hurley2017} by linear least squares fitting a single power law spectrum to sources with no missing or negative flux density values. We measure the GLEAM spectral index for all sources after removal of the negative flux values using the same method as \cite{Hurley2017}. This results in three LFTO sources with measured GLEAM spectral indices $\alpha^{\rm GLEAM}_{\rm low} \leq$ -0.2 primarily due to large GLEAM flux density uncertainties at low frequencies. The new $\alpha^{\rm GLEAM}_{\rm low}$ values for the remaining sources agree with the values in \cite{Hurley2017} and new values are marked in Table \ref{tab:radiomeasured2}. The GLEAM to SUMSS/NVSS spectral index is also updated using RACS-mid flux densities to improve consistency and labelled $\alpha^{\rm RACS}_{\rm mid}$.

We create new 31-MHz bandwidth flux densities by combining the flux densities in five sets of four adjacent sub-bands. These are determined by using a weighted average:
\begin{equation}
    S_{\nu} = \frac{\Sigma_{i}~S_{\nu,i}/\sigma_{i}^{2}}{\Sigma_{i}~1/\sigma_{i}^{2}},
	\label{eq:wavg} 
\end{equation}
 where S$_{\nu,i}$ are the flux measurements of the individual 7.68\,MHz sub-bands with uncertainties $\sigma_{i}$. This is done to increase model fitting accuracy and convergence times as a number of individual sub-bands contain significant outliers including negative flux densities. It also avoids the need for the SED fitting to deal with the correlated noise in each set of four adjacent sub-bands which were deconvolved at the same time. 

Flux densities and images from the Rapid Australian SKA Pathfinder (ASKAP) Continuum Survey \citep[RACS;][]{McConnell2016}, a survey performed by the ASKAP currently covering dec. $< +30^{\circ}$ at 888\,MHz \citep{Hale2021} and 1.37\,GHz \citep{Duchesne2024}, are also utilised. Lastly we collated any supplementary radio data in the NASA Extragalactic Database (NED\footnote{https://ned.ipac.caltech.edu}). These flux values are presented in Table A.\ref{tab:radiomeasured2}.

\subsection{Infrared}

FIR flux density measurements at 60 and 100\,$\mu$m from {\it Infrared Astronomy Satellite} \citep[\textit{IRAS}][]{neg1984}  are collated from the IRAS Faint Source Catalogue \citep{moshir1990}. 

Although the GLEAM-6dFGS catalogue also includes {\it WISE} source catalogue measurements \citep{Wright2010} at 3.4, 4.6, 12 and 23\ $\mu$m ({\it WISE} band W1, W2, W3 and W4), due to the morphological complexity of many of the sources in our sample we choose to instead use values from the WISE eXtended Source Catalogue \citep[WXSC][]{Jarrett2019}. The WXSC more accurately encapsulates the emission from extended sources and removes contamination from foreground or background sources. For blended or merging systems the WXSC flux density and Vega magnitude values are measured for the entire system such that they are comparable to the radio continuum observations which have a larger beam. These WXSC flux density values are used to derive the mid-IR based global parameters including the mid-IR SFR \citep[SFR$_{mircor}$; ][]{Cluver2024}, stellar mass \citep{jarrett2023}, and, specific SFR (sSFR$_{mircor}$). SFRs are from \cite{Cluver2024} that updates the calibration in \cite{Cluver2017} which related {\it WISE} W3 and W4 to L$_{TIR}$ (total infrared luminosity). SFRs are determined using an invariance-weighting of SFR$_{12}$ and SFR$_{23}$ ({\it WISE} band W3 and band W4 based SFR) and include a correction to account for star formation in low mass galaxies with relatively low dust content (which makes use of the relationship between dust density and UV to IR emission). The WXSC also provides the luminosity distance D$_L$ for each source. The IR flux density measurements shown in Table A.\ref{tab:IRmeasured} and global derived parameters in Table. \ref{tab:IRderived}.

\subsection{Ancillary Data}

We make use of optical background images to present the stellar distribution of the galaxies in our sample. These optical images are collated from the DataCentral Data Aggregation Service \citep{miszalski2022} \footnote{https://datacentral.org.au/das} and are primarily \textit{g}-band images. Images from the Panoramic Survey Telescope and Rapid Response System (PanSTARRS) data release 2 \citep[PS1-DR2; ][]{flewelling2020}, the Dark Energy Survey \citep[DES; ][]{abbott2021}, the DESI Legacy Survey \citep{dey2019a}, SkyMapper Southern Survey \citep{onken2019} and VISTA Hemisphere Survey \citep[][where we used $K$-band where \textit{g}-band images are not available]{Mcmahon2013} are presented in Figure A.\ref{fig:seds}. Estimated source inclinations were calculated using \textit{K}-band light axis ratios from the Two Micron All-Sky Survey \citep{jarrett2000}.

\section{ATCA Observations and Data Processing}
\label{sec:data}

\subsection{ATCA Observation Details}

\begin{table*}[hbt!]
\centering
\begin{threeparttable}
\caption{ATCA Observation Details.}\label{tab:obs}
\begin{tabular}{c|c|c|c|c|c|c|c|c}
	\toprule
	UT date & ATCA & Antennas & Scan Time & Average Total Integration Time$^a$ & IF1 & IF2 & Bandwidth & LAS\\
	(yyyy-mm-dd) & Configuration &  & (min) & (min) & (GHz) & (GHz) & (GHz) & ('')\\
	\midrule
	2022-04-20 & 1.5A & 6 & 6 & 36 & 5.5 & 9.5 & 2.048 & 52\\
	2022-05-07 & 750D & 6 & 6 & 36 & 5.5 & 9.5 & 2.048 & 259\\
    2022-07-27 & H168 & 5 & 11.5 & 46 & 17 & 18.8$^b$ & 2.048 & 73$^b$\\
    2023-03-08 & 750C & 5 & 6 & 18 & 5.5 & 9.5 & 2.048 & 173\\
    2023-03-11 & 750C & 5 & 6 & 18 & 5.5 & 9.5 & 2.048 & 173\\

	\bottomrule
\end{tabular}
\begin{tablenotes}[para]
    \item[]Note: Column (1): universal time date. Column (2): ATCA configuration. Column(3): Number of antennas used to observe. Column (4): Integration time per {\it u-v} scan on source. Column (5): Average total integration time on source. Column (6-7): Central frequencies of ATCA observations. Column (8): Bandwidths at each central frequency. Column (9): The largest angular scale probed at IF2\\
    \item[$^{a}$]Due to scheduling and source rise and set times some sources may have more or fewer scans resulting in total integration times varying.\\
    \item[$^{b}$]Due to radio frequency interference the 18.8\,GHz observations were discarded.
\end{tablenotes}
\end{threeparttable}
\end{table*}
  
ATCA observations were performed across three 24 hour periods and two 14 hour periods (project ID C3483) using four different ATCA configurations and are summarised in Table~\ref{tab:obs}. Different configurations are chosen to obtain adequate {\it (u,v)} coverage and total integration time for accurate flux measurements for SED modelling as well as high ($\sim$6'') resolution observations at 5.5 and 9.5\,GHz. 

\subsection{ATCA Data Processing}

ATCA data were calibrated and imaged using {\tt MIRIAD} \citep{Sault1995} software package. The {\tt PGFLAG} and {\tt BLFLAG} flagging routines within {\tt MIRIAD} were used for automated and manual flagging respectively in conjunction with the traditional {\tt MIRIAD} calibration tasks to perform the initial data reduction. A frequency dependent calibration solution was determined using the {\tt NFBIN} option given the wide bandwidth of the Compact Array Broadband Backend (CABB) system \citep{wilson2011}. Flux and bandpass calibration were performed on PKS 1934-638 whilst gain calibration was transferred from the phase calibrators. The calibration solutions for each individual day of observations were applied to each source. The calibrated observations different configurations at 5.5 and 9.5\,GHz were then merged to provide the best \textit{u-v} sampling and highest sensitivity.

\subsubsection{Total Flux Density Measurements}

To measure the total flux density across all frequency bands, each source was imaged using their complete bandwidth ($\Delta\nu = 2.048$\,GHz, minus the 100 edge channels automatically flagged by {\tt ATLOD}). Natural weighting (Briggs robust parameter value of 2) was used to provide the maximum signal-to-noise for total flux measurements. {\tt MFCLEAN} \citep{Sault1994} was used to deconvolve the multi-frequency synthesised dirty map. Then {\tt RESTOR} and {\tt LINMOS} were used to restore the full bandwidth images and perform the primary beam correction whilst accounting for the spectral index of the clean components. 

We used an iterative procedure to perform sub-band splitting to allow for improved spectral sampling and modelling. Each CABB band was initially imaged as described above. The peak flux density was extracted and then, provided the signal-to-noise ratio (SNR) was above 8 or 10, the CABB dataset would be split into 2 or 3 sub-bands respectively and reprocessed. To safely use {\tt MFCLEAN}, we ensured that each sub-band maintained a fractional bandwidth larger than 10$\%$. This resulted in a maximum of three sub-bands at 5.5\,GHz, two at 9.5\,GHz and no sub-band splitting at $>10$\,GHz. Sub-bands were split using the {\tt LINE} option in {\tt INVERT} with their new central frequencies recorded individually for each source depending on their unflagged channels. Total fluxes for each image were then measured using {\tt PROFOUND} task within the {\tt ProFound} software package \citep{Robotham2018}. 

\begin{figure*}[hbt!]
    \centering
        \begin{subfigure}[b]{0.43\textwidth}
        \centering
        \includegraphics[width=\textwidth]{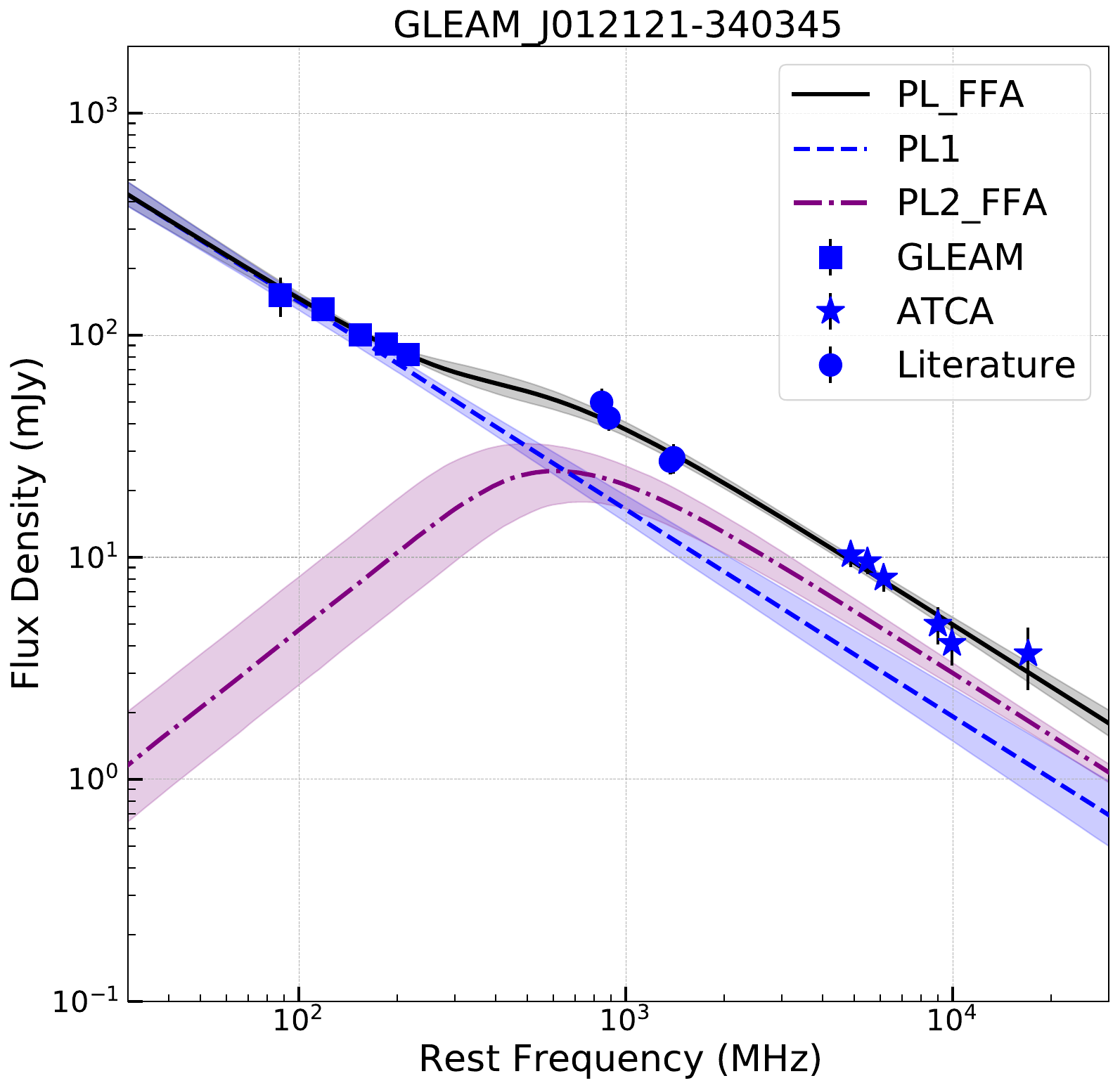}
    \end{subfigure}
    \begin{subfigure}[b]{0.47\textwidth}
        \centering
        \includegraphics[width=\textwidth]{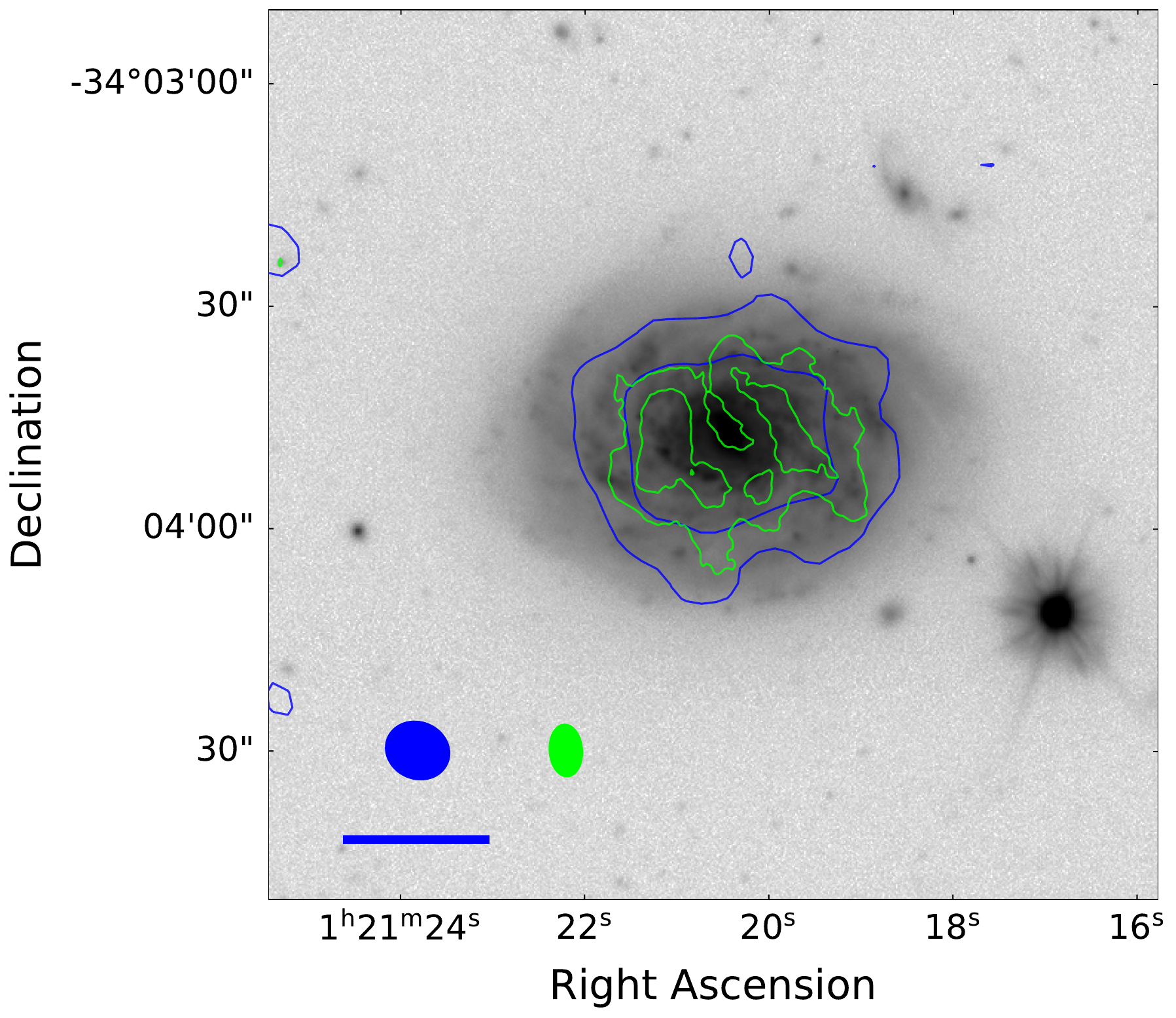}
    \end{subfigure}
    \caption{Left: The preferred model SED of GLEAM J012121-340345 with observed data points. The overlaid black line indicates the full model whilst the dotted blue line indicates the first PL component and purple dashed line indicates the second PL component which is free-free absorbed. The highlighted regions represent the 1-$\sigma$ uncertainties sampled by {\tt EMCEE}. Right: The DES \textit{g}-band optical image of GLEAM J012121-340345 showing the stellar extent and morphology overlaid with contours from RACS-mid at 1.37\,GHz in blue and ATCA 9.5\,GHz in green. Radio contours for both frequencies start at the 4$\sigma$ level and increase by factors of $\sqrt{3}$. The FWHM beams for RACS-mid and ATCA are given by the blue and green ellipses respectively. The scale bar at the bottom left denotes 5\,kpc.}
    \label{fig:j0121}
\end{figure*}

\begin{figure*}[hbt!]
    \centering
        \begin{subfigure}[b]{0.43\textwidth}
        \centering
        \includegraphics[width=\textwidth]{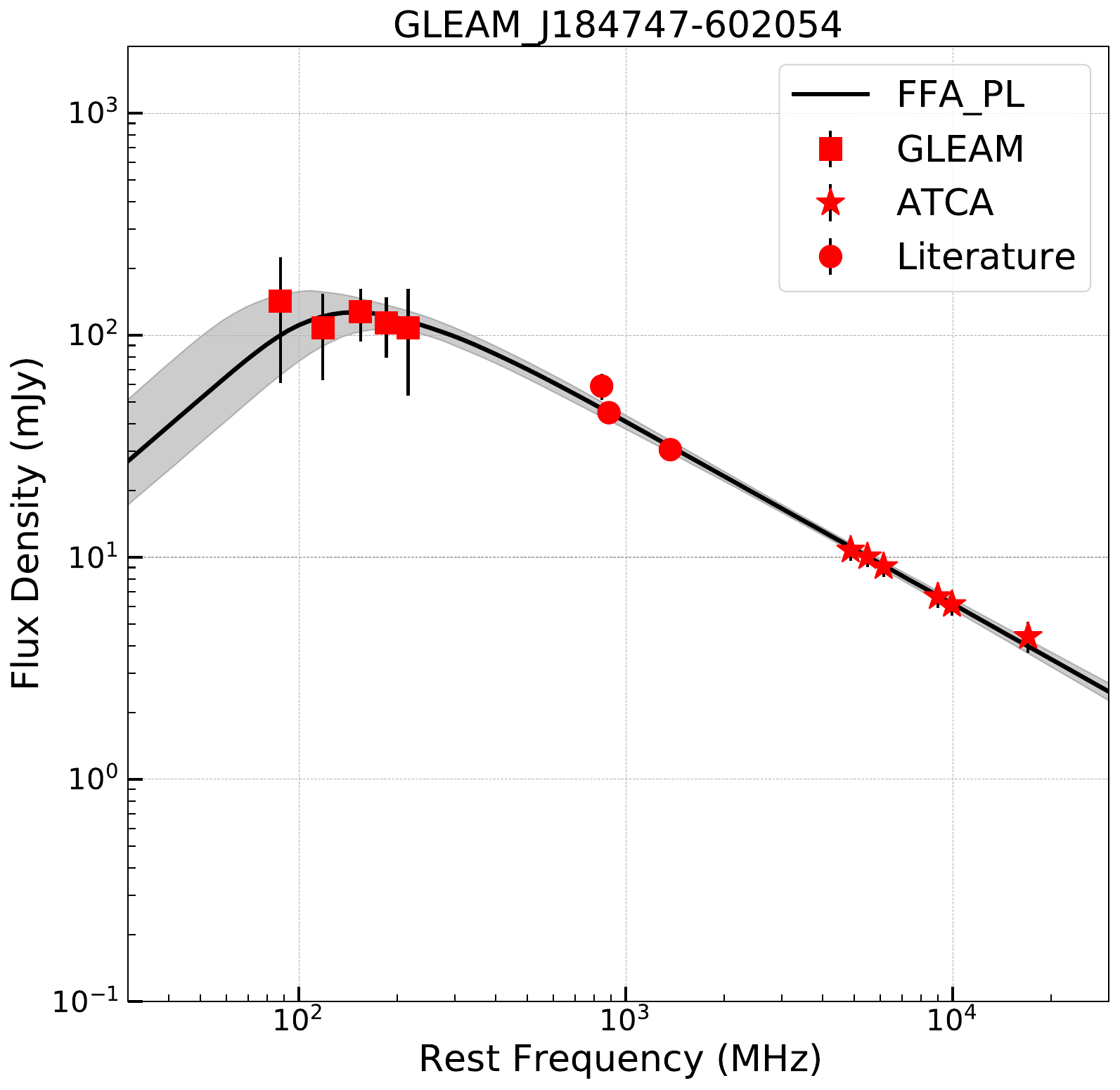}
    \end{subfigure}
    \begin{subfigure}[b]{0.47\textwidth}
        \centering
        \includegraphics[width=\textwidth]{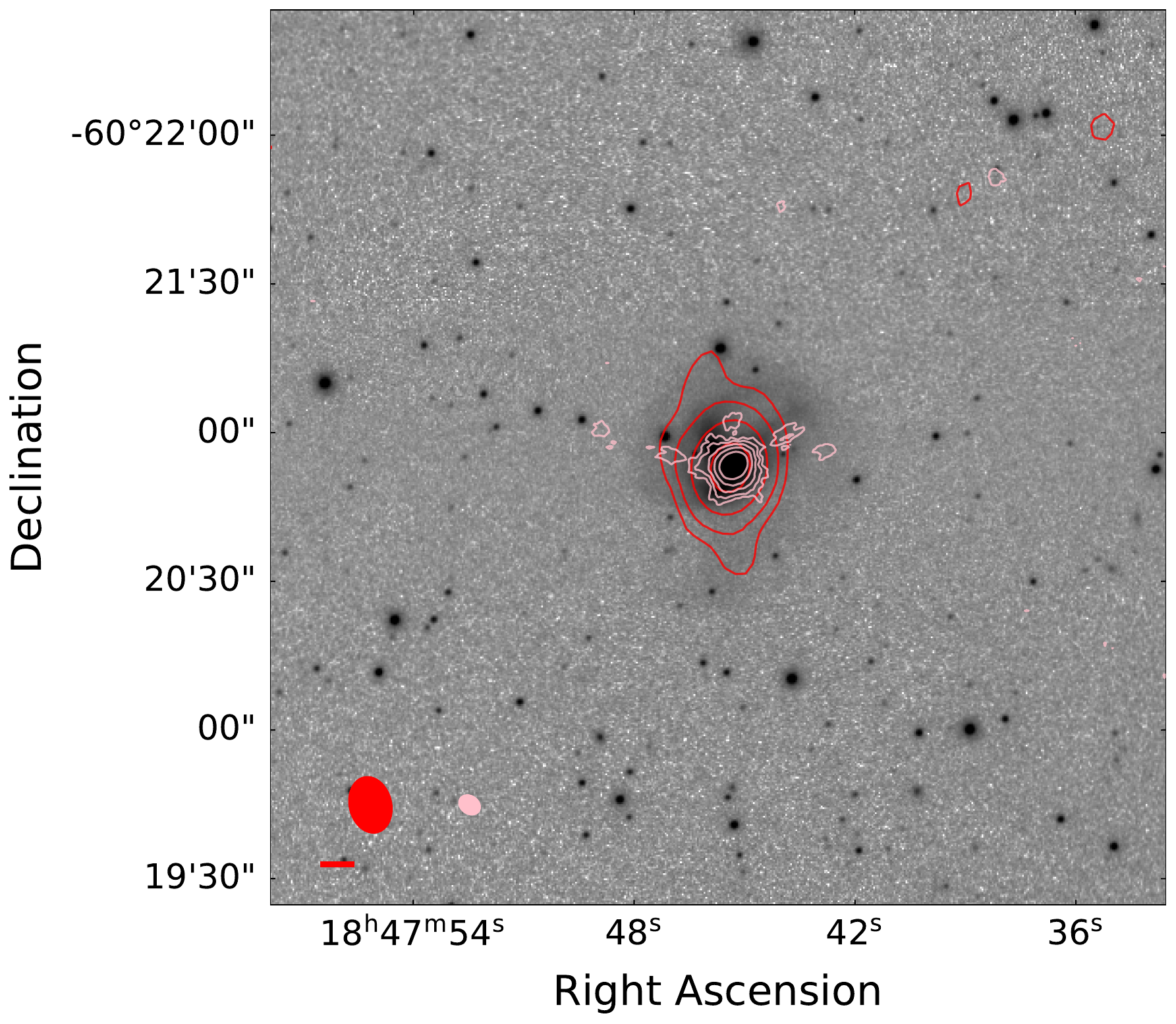}
    \end{subfigure}
    \caption{Left: The preferred model SED of GLEAM J184747-602054 with observed data points. The overlaid black line indicates the full {\tt FFA\_PL} model. The highlighted region represents the 1-$\sigma$ uncertainties sampled by {\tt EMCEE}. Right: The \textit{g}-band optical image of GLEAM J184747-602054 showing the stellar extent and morphology overlaid with contours from RACS-mid at 1.37\,GHz in red and ATCA 9.5\,GHz in pink. Radio contours for both frequencies start at the 4$\sigma$ level and increase by factors of $\sqrt{3}$. The FWHM beams for RACS-mid and ATCA are given by the red and pink ellipses respectively. The scale bar at the bottom left denotes 5\,kpc.}
    \label{fig:j1847}
\end{figure*}

High resolution imaging was also separately performed at 5.5 and 9.5\,GHz using the entire CABB bandwidth. A Briggs robust parameter of 0.5 was chosen to achieve a balance between sensitivity to diffuse emission and angular resolution. The radio contours for the high resolution images are presented in Figures. \ref{fig:j0121}, \ref{fig:j1847}, and A.\ref{fig:seds}.

In order to account for flux scale errors between GLEAM and the other radio data, we applied a conservative 10\% error in quadrature to all non-GLEAM flux density measurements. This 10\% error was chosen such that it encompasses the $\approx8\%$ flux scale error measured for the GLEAM survey \citep{Hurley2017} and accounts for possible source variability, although SFGs generally do not show significant variability at radio frequencies on timescales of decades at the sensitivities observed in our sample \citep{Mooley2016}. Lastly we do not attempt to match the \textit{u-v} coverage or resolution of multi-frequency images used when modelling the radio SEDs due to the large range in resolutions between GLEAM observations ($\sim2'$) and other radio data ($\sim10-45''$). ATCA observations are performed in a number of configurations to ensure we have sufficient short baseline \textit{u-v} coverage for extended flux measurements. Measured ATCA flux densities are given in Table A.\ref{tab:radiomeasured1} with other radio fluxes presented in Table A.\ref{tab:radiomeasured2}.

\section{SED Modelling}
\label{sec:SED}

\subsection{Modular Radio Continuum Models}
We choose to fit a series of increasingly complex models to our sources loosely following \citet{Galvin2018} with all modelling being performed in the frequency rest-frame\footnote{This has negligible impact due to the our sample having $z \leq 0.04$} with a reference frequency of $\nu_{0}$ = 1.4\,GHz. We deviate from the method of \citet{Galvin2018} such that radio continuum models are constructed using a modular approach whereby base single and two component power-law models are modified by prefix models at GLEAM frequencies ($\nu \leq$ 300\,MHz) and suffix models non-GLEAM frequencies ($\nu \geq$ 300\,MHz). The prefix and suffix models encapsulate the physical loss and absorption processes occurring within SFGs including FFA, IC losses and synchrotron losses.

We also make no attempt in this work to include a FIR dust heating component in our current SED modelling. FIR emission increases model complexity and does not help to constrain our radio SED model parameters, as, at our current highest radio frequency measurements (17\,GHz) warm dust contributes $\ll1\%$ to the total flux.

\subsubsection{Base Model: Power Law ({\tt PL})}
Firstly we fit a simple single component power law to radio flux density measurements with the form of:
\begin{equation}
    S_{\nu} = A\bigg(\frac{\nu}{\nu_{0}}\bigg)^{\alpha},
	\label{eq:PL}
\end{equation}
where A is a normalisation component and the spectral index, $\alpha$, is the gradient in logarithmic space with both being treated as free parameters. 

\subsubsection{Base Model: Synchrotron and Free-free Emission ({\tt SFG})}
The radio continuum emission is usually modelled as the sum of two distinct power laws with one representing the flat spectrum thermal free-free emission and the second representing the steep spectrum non-thermal synchrotron emission. This model takes the form:
\begin{equation}
    S_{\nu} = A\bigg(\frac{\nu}{\nu_{0}}\bigg)^{\alpha} + B\bigg(\frac{\nu}{\nu_{0}}\bigg)^{-0.1},
	\label{eq:SFG}
\end{equation}
where the free parameters A and B are the synchrotron and free-free normalisation components respectively. The free parameter $\alpha$ is the synchrotron spectral index which depends on the cosmic ray injection history and is known to vary \citep{Niklas1997}. The spectral index of the free-free emission is well approximated by $-0.1$ over the range of interest \citep{Condon1992}.

\subsubsection{Prefix: Free-free Absorption ({\tt FFA\_})}
Synchrotron emission can be attenuated by FFA processes when in a coextensive environment with free-free emission producing spectral curvature at primarily low frequencies. The attenuation is influenced by the density, flux density and spatial distribution of the ionised free-free emission in comparison to the synchrotron emission. 
The free-free optical depth can be approximated as $\tau_1$ = $(\nu/\nu_{t,1})^{-2.1}$ with $\nu_{t,1}$ being the turn-over frequency where the optical depth reaches unity (see \S~\ref{em}). This model modifies the SFG base model resulting in the full model {\tt FFA\_SFG}:
\begin{equation}
    S_{\nu} = (1-e^{-\tau_1})\bigg(B + A\bigg(\frac{\nu}{\nu_{t,1}}\bigg)^{0.1+\alpha}\bigg)\bigg(\frac{\nu}{\nu_{t,1}}\bigg)^{2},
	\label{eq:FFASFG}
\end{equation}
following \citet{Condon1992} and \citet{Clemens2010}. $\nu_{t,1}$ is limited to GLEAM frequencies (i.e. $\leq$ 300\,MHz). A, B and $\alpha$ are the synchrotron and free-free normalisation components and synchrotron spectral index respectively which are fitted simultaneously with $\nu_{t,1}$. Model degeneracy is minimised by replacing the $\nu_{0}$ term with the turnover frequency parameter for each component following \citet{Galvin2018}. Setting $B = 0$ gives the full model {\tt FFA\_PL}.

\subsubsection{Suffix: Free-free Absorption ({\tt \_FFA})}
The models only including the {\tt FFA\_} prefix assume a single volume of thermal free-free plasma mixed with relativistic electrons which produce synchrotron emission. This model was derived from observations by \citet{Condon1990} of the irregular clumpy galaxy Markarian 325, however \citet{Clemens2010}, \citet{Galvin2018} and \citet{Dey2022, Dey2024} present a set of LIRGS/ULIRGS which show higher frequency ``kinks'' in their radio SEDs which could also be attributed to FFA. Their interpretations suggest that when multiple SF regions of different geometric orientations or composition are integrated over a large synthesised beam the observed radio continuum can be complex. Thus following \citet{Galvin2018} we introduce a model which includes a single relativistic electron population which produces synchrotron emission that is inhomogenously mixed with two distinct star forming regions of two distinct optical depths of which only $\tau_2$ becomes optically thick within the observed frequency range. This situation is described by the full model {\tt SFG\_FFA}:
\begin{multline}
\label{eq:C2 1SAN}
    S_{\nu} = A\bigg(\frac{\nu}{\nu_{0}}\bigg)^{\alpha} + B\bigg(\frac{\nu}{\nu_{0}}\bigg)^{-0.1}\\
    + (1-e^{-\tau_{2}})\bigg(D+C\bigg(\frac{\nu}{\nu_{t,2}}\bigg)^{0.1+\alpha}\bigg)\bigg(\frac{\nu}{\nu_{t,2}}\bigg)^{2},
\end{multline}
where the second component of the {\tt SFG\_FFA} model represents the second distinct star forming region which becomes optically thick below a certain frequency. $\nu_0$ is the reference frequency of 1.4\,GHz and $\tau_2$ describes the optical depth of the second component parameterised by the turnover frequency $\nu_{t,2}$ which is limited to non-GLEAM frequencies ($\geq$ 300\,MHz). A and C are the normalisation parameters for the synchrotron emission and B and D govern the free-free emission component. $\alpha$ is the spectral index of a single synchrotron emission population. Setting B = 0 and D = 0 gives the full model {\tt PL\_FFA}.

We can replace the first line in this model with Eq.~\ref{eq:FFASFG} to account for the situation where there is also a LFTO caused by FFA at low frequencies resulting in the {\tt FFA\_SFG\_FFA} and (when B = 0 and D = 0) {\tt FFA\_PL\_FFA} full models. 

We then allow for two distinct electron populations by relaxing the single spectral index constraint. This is physically motivated by galaxy merger situations which can trigger new bursts of SF such that the electron distribution can be made up of a populations of newly injected and older non-thermal relativistic electrons. This is done by allowing each component in the {\tt FFA\_SFG\_FFA} full model to have different synchrotron spectral index values $\alpha$ and $\alpha_{2}$ which are free parameters in the full model {\tt FFA\_SFG\_FFA2}.

\subsubsection{Suffix: Synchrotron and Inverse-Compton Losses ({\tt \_SIC})}
The spectral steepening observed at higher frequencies may not be related to FFA at different optical depths and instead be due to synchrotron or IC losses. Synchrotron losses take place as CREs age and lose energy as they propagate in a galaxy's large scale magnetic field whilst IC losses are generally dependent on FIR or CMB photon scattering. Synchrotron and IC losses under constant electron injection from massive SF both act to gradually steepen the synchrotron spectral index by $\Delta\alpha$ = -0.5 around a ``break'' frequency $\nu_b$. This results in the full model {\tt SFG\_SIC} having the form:
\begin{equation}
    S_{\nu} = A\bigg(\frac{\nu}{\nu_{0}}\bigg)^{\alpha}\bigg(\frac{1}{1+(\nu/\nu_b)^{\Delta\alpha}}\bigg) + B\bigg(\frac{\nu}{\nu_{0}}\bigg)^{-0.1},
	\label{eq:SFGSIC}
\end{equation}
where the free parameters A and B are the synchrotron and free-free normalisation components respectively. The free parameter $\alpha$ is the low-frequency synchrotron spectral index which gradually steepens by $\Delta\alpha$ = -0.5 around the ``break'' frequency $\nu_b$ which is limited to non-GLEAM frequencies ($\geq$ 300\,MHz). Setting B = 0 gives the full model {\tt PL\_SIC}.

We also model the situation where there is a LFTO caused by FFA and gradual spectral steepening caused by synchrotron and IC losses in the {\tt FFA\_SFG\_SIC} full model which takes the form:
\begin{equation}
    S_{\nu} = (1-e^{-\tau_1})\bigg(B + A\bigg(\frac{\nu}{\nu_{t,1}}\bigg)^{0.1+\alpha}\bigg(\frac{1}{1+(\nu/\nu_b)^{\Delta\alpha}}\bigg)\bigg)\bigg(\frac{\nu}{\nu_{t,1}}\bigg)^{2},
	\label{eq:CIC}
\end{equation}
where the free parameters are the same as in Eq.~\ref{eq:SFGSIC} with the addition that $\nu_{t,1}$ is the turnover frequency where the optical depth ($\tau_1$) reaches unity and is limited to GLEAM frequencies ($\leq$ 300\,MHz). Setting B = 0 gives the full model {\tt FFA\_PL\_SIC}.

We do not make any attempt to model both the Suffix models simultaneously (i.e. two synchrotron components, one with a spectral break and both with FFA) as we do not have the spectral sampling frequency or range to be able to draw accurate conclusions. It is likely however, that both FFA and loss processes are occurring in most SFGs and that the two separate Suffix models will improve the fitting of radio SEDs which display an obvious ``kink'' and gradual spectral steepening at higher frequencies.

\begingroup
\renewcommand{\arraystretch}{2}
\begin{table*}[hbt!]
\centering
\begin{threeparttable}
\caption{Modular Radio Continuum Models}\label{tab:models}
\begin{tabular}{c|c|c|c|c|c|c}
	\toprule
	Prefix & Base & Suffix & Full & Radio & Free & G+18\\
    Model & Model & Model & Model & Spectrum & Parameters & Eq.\\
	\midrule
	 & {\tt PL} &  & {\tt PL} & \(\displaystyle S_{\nu} = A\bigg(\frac{\nu}{\nu_{0}}\bigg)^{\alpha} \) & A, $\alpha$ & PL\\
	 & {\tt SFG} &  & {\tt SFG} & \(\displaystyle S_{\nu} = A\bigg(\frac{\nu}{\nu_{0}}\bigg)^{\alpha} + B\bigg(\frac{\nu}{\nu_{0}}\bigg)^{-0.1} \) & A, B, $\alpha$ & SFG\_NC\\
    {\tt FFA\_} & {\tt PL} &  & {\tt FFA\_PL} & \(\displaystyle S_{\nu} = (1-e^{-\tau_1})A\bigg(\frac{\nu}{\nu_{t,1}}\bigg)^{0.1+\alpha}\bigg(\frac{\nu}{\nu_{t,1}}\bigg)^{2} \) & A, $\alpha$, $\nu_{t,1}$ & \\
    {\tt FFA\_} & {\tt SFG} &  & {\tt FFA\_SFG} & \(\displaystyle S_{\nu} = (1-e^{-\tau_1})\bigg(B + A\bigg(\frac{\nu}{\nu_{t,1}}\bigg)^{0.1+\alpha}\bigg)\bigg(\frac{\nu}{\nu_{t,1}}\bigg)^{2} \) & A, B, $\alpha$, $\nu_{t,1}$ & C\\
     & {\tt PL} & {\tt \_FFA} & {\tt PL\_FFA} & \(\displaystyle S_{\nu} = A\bigg(\frac{\nu}{\nu_{0}}\bigg)^{\alpha} \) & A, $\alpha$ & \\
     & & & & \(\displaystyle +~(1-e^{-\tau_{2}})C\bigg(\frac{\nu}{\nu_{t,2}}\bigg)^{0.1+\alpha}\bigg(\frac{\nu}{\nu_{t,2}}\bigg)^{2} \) & C, $\nu_{t,2}$ & \\
     & {\tt PL} & {\tt \_SIC} & {\tt PL\_SIC} & \(\displaystyle S_{\nu} = A\bigg(\frac{\nu}{\nu_{0}}\bigg)^{\alpha}\bigg(\frac{1}{1+(\nu/\nu_b)^{\Delta\alpha}}\bigg) \) & A, $\alpha$, $\nu_b$ & \\
     & {\tt SFG} & {\tt \_FFA} & {\tt SFG\_FFA} & \(\displaystyle S_{\nu} = A\bigg(\frac{\nu}{\nu_{0}}\bigg)^{\alpha} + B\bigg(\frac{\nu}{\nu_{0}}\bigg)^{-0.1} \) & A, B, $\alpha$ & C2\_1SAN\\
     & & & & \(\displaystyle +~(1-e^{-\tau_{2}})\bigg(D+C\bigg(\frac{\nu}{\nu_{t,2}}\bigg)^{0.1+\alpha}\bigg)\bigg(\frac{\nu}{\nu_{t,2}}\bigg)^{2} \) & C, D, $\nu_{t,2}$ & \\
     & {\tt SFG} & {\tt \_SIC} & {\tt SFG\_SIC} & \(\displaystyle A\bigg(\frac{\nu}{\nu_{0}}\bigg)^{\alpha}\bigg(\frac{1}{1+(\nu/\nu_b)^{\Delta\alpha}}\bigg) + B\bigg(\frac{\nu}{\nu_{0}}\bigg)^{-0.1} \) & A, B, $\alpha$, $\nu_b$ & \\
    {\tt FFA\_} & {\tt PL} & {\tt \_FFA} & {\tt FFA\_PL\_FFA} & \(\displaystyle S_{\nu} = (1-e^{-\tau_1})A\bigg(\frac{\nu}{\nu_{t,1}}\bigg)^{0.1+\alpha}\bigg(\frac{\nu}{\nu_{t,1}}\bigg)^{2} \) & A, $\alpha$, $\nu_{t,1}$ & \\
    & & & & \(\displaystyle +~(1-e^{-\tau_{2}})C\bigg(\frac{\nu}{\nu_{t,2}}\bigg)^{0.1+\alpha}\bigg(\frac{\nu}{\nu_{t,2}}\bigg)^{2} \) & C, $\nu_{t,2}$ & \\
    {\tt FFA\_} & {\tt PL} & {\tt \_SIC} & {\tt FFA\_PL\_SIC} & \(\displaystyle S_{\nu} = (1-e^{-\tau_1})\bigg(A\bigg(\frac{\nu}{\nu_{t,1}}\bigg)^{0.1+\alpha}\bigg(\frac{1}{1+(\nu/\nu_b)^{\Delta\alpha}}\bigg)\bigg)\bigg(\frac{\nu}{\nu_{t,1}}\bigg)^{2} \) & A, $\alpha$, $\nu_{t,1}$, $\nu_b$ & \\
    {\tt FFA\_} & {\tt SFG} & {\tt \_FFA} & {\tt FFA\_SFG\_FFA} &  \(\displaystyle S_{\nu} = (1-e^{-\tau_1})\bigg(B + A\bigg(\frac{\nu}{\nu_{t,1}}\bigg)^{0.1+\alpha}\bigg)\bigg(\frac{\nu}{\nu_{t,1}}\bigg)^{2} \) & A, B, $\alpha$, $\nu_{t,1}$ & C2\_1SA\\
     & & & & \(\displaystyle +~(1-e^{-\tau_{2}})\bigg(D+C\bigg(\frac{\nu}{\nu_{t,2}}\bigg)^{0.1+\alpha}\bigg)\bigg(\frac{\nu}{\nu_{t,2}}\bigg)^{2} \) & C, D, $\nu_{t,2}$ & \\
    {\tt FFA\_} & {\tt SFG} & {\tt \_FFA2} & {\tt FFA\_SFG\_FFA2} & \(\displaystyle S_{\nu} = (1-e^{-\tau_1})\bigg(B + A\bigg(\frac{\nu}{\nu_{t,1}}\bigg)^{0.1+\alpha_1}\bigg)\bigg(\frac{\nu}{\nu_{t,1}}\bigg)^{2} \) & A, B, $\alpha_1$, $\nu_{t,1}$ & C2\\
     & & & & \(\displaystyle +~(1-e^{-\tau_{2}})\bigg(D+C\bigg(\frac{\nu}{\nu_{t,2}}\bigg)^{0.1+\alpha_2}\bigg)\bigg(\frac{\nu}{\nu_{t,2}}\bigg)^{2} \) & C, D, $\alpha_2$ $\nu_{t,2}$ & \\
    {\tt FFA\_} & {\tt SFG} & {\tt \_SIC} & {\tt FFA\_SFG\_SIC} & \(\displaystyle S_{\nu} = (1-e^{-\tau_1})\bigg(B + A\bigg(\frac{\nu}{\nu_{t,1}}\bigg)^{0.1+\alpha}\bigg(\frac{1}{1+(\nu/\nu_b)^{\Delta\alpha}}\bigg)\bigg)\bigg(\frac{\nu}{\nu_{t,1}}\bigg)^{2} \) & A, B, $\alpha$, $\nu_{t,1}$, $\nu_b$ & \\
	\bottomrule
\end{tabular}
\begin{tablenotes}[para]Note: Column (1-3): The prefix, base and suffix model names. Column (4): The full model label. Column (5): The function representing the modelled radio spectrum. Column (6): Free parameters. Column (7): \citet{Galvin2018} equivalent model.
\end{tablenotes}
\end{threeparttable}
\end{table*}
\endgroup

\subsection{Fitting and Selection}
\subsubsection{Model Fitting}
Briefly we use the ``affine invariant'' Markov chain Monte Carlo ensemble sampler \citep{Goodman2010} implemented as the {\tt EMCEE PYTHON} package \citep{Foreman2013} to constrain each of the radio continuum models for each source in our sample. The log likelihood function that {\tt EMCEE} attempts to minimise relies on the assumption that measurements are independent with normally distributed errors.

\subsubsection{Model Priors}
We choose physically motivated uninformative (uniform) priors to constrain our models within a Bayesian framework. Throughout our model fitting we ensure that the normalisation parameters A, B, C, and D remain positive (A/B/C/D > 0). The spectral index parameters $\alpha$ and $\alpha_{2}$ remain in the range of -1.8 $\leq \alpha \leq$ -0.2. The LFTO ($\nu_{t,1}$) and spectral ``kink'' ($\nu_{t,2}$) frequencies are limited to between 10\,MHz to 300\,MHz and 300\,MHz to 17\,GHz respectively. The spectral ``break'' frequency ($\nu_{b}$) is also limited to between 300\,MHz and 17\,GHz. These priors are founded on the sound assumptions that flux densities are positive emission processes and we can only constrain turnovers within the frequency range that we have data, however we note that some SEDs may begin to flatten before the optical depth reaches unity. The spectral index parameters are limited to allow for the range of values found for synchrotron dominated emission in literature \citep{Condon1990, Clemens2010, Galvin2018, An2021, Dey2022}.
 
\subsubsection{Model Selection}
In order to objectively test whether the introduction of additional model complexity is justified by an improved fit and is not just a symptom of overfitting we make use of an estimate of the evidence value $Z$. The evidence value is defined as the ratio between the integral of the posterior volume over the prior volume \citep{Skilling2004} and is computationally difficult to compute but can be reliably estimated using recent algorithms. {\tt DYNESTY} \citep{Speagle2020} uses a dynamic nested sampling method \citep{Higson2019} to obtain an estimate of the evidence value. Given the evidence values of competing models, one is able to determine whether a model is preferred over another for a given set of data. The natural logarithm of the Bayes odds ratio between evidence values $Z_{1}$ and $Z_{2}$ for models $M_{1}$ and $M_{2}$ is described by
\begin{equation}
    {\rm ln}(\Delta Z) = {\rm ln}(Z_1) - {\rm ln}(Z_2).
	\label{eq:evidence}
\end{equation}
A value of ${\rm ln}(\Delta Z)$ < -5 provides very strong evidence for $M_{1}$ over $M_{2}$ whilst -5 < ${\rm ln}(\Delta Z)$ < -3 and -3 < ${\rm ln}(\Delta Z)$ < -1.1 provide strong and positive evidence respectively. When ${\rm ln}(\Delta Z)$ > -1.1 the models are indistinguishable from each other. This scale was established by \citet{Kass1995} and is considered the standard for model selection.

The prior parameter space searched by {\tt DYNESTY} is limited to a uniform distribution within the uncertainties given by the 1st and 99th percentiles of the samples posterior distribution as determined by {\tt EMCEE}. This limitation of priors is necessary as {\tt DYNESTY} requires bounded priors and different types of models are explored. Because the evidence value is entirely dependent on the "size" of the prior volume \citep{Skilling2004} setting arbitrarily large priors on normalisation components would heavily bias the evidence values against models with extra normalisation parameters. Within {\tt DYNESTY} we also select "Single ellipsoid bounds" and "random walk" samplers (rather than uniform samplers) to improve processing time and maintain consistency with the posteriors estimated by {\tt EMCEE}. The results of the Bayes odds ratio tests for all models as defined relative to the lowest value are summarised in Table. \ref{tab:modelsel} with the most preferred model and any other competing models highlighted, i.e. $ln(\Delta Z)$ > -1.1.

\begin{table*}[hbt!]
\centering
\begin{threeparttable}
\caption{Bayes Odds Ratio Table.}
\label{tab:modelsel}
\begin{tabular}{c|c|c|c|c|c|c|c|c|c|c|c|c|c|c}
	\toprule
     &  &  &  & {\tt FFA\_} & {\tt FFA\_} &  &  &  &  & {\tt FFA\_} & {\tt FFA\_} & {\tt FFA\_} & {\tt FFA\_} & {\tt FFA\_}\\
	Source & Sample & {\tt PL} & {\tt SFG} & {\tt PL} & {\tt SFG} & {\tt PL} & {\tt PL} & {\tt SFG} & {\tt SFG} & {\tt PL} & {\tt PL} & {\tt SFG} & {\tt SFG} & {\tt SFG}\\
     &  &  &  &  &  & {\tt \_FFA} & {\tt \_SIC} & {\tt \_FFA} & {\tt \_SIC} & {\tt \_FFA} & {\tt \_SIC} & {\tt \_FFA} & {\tt \_FFA2} & {\tt \_SIC}\\
	\midrule
	GLEAM J002238-240737 & \cellcolor[HTML]{7fa3fa} Con & \cellcolor[HTML]{e2e2e2} -0.44 & -2.47 & -2.98 & -4.63 & -1.22 & \cellcolor[HTML]{9cff99} 0 & -6.83 & -1.39 & -4.42 & -1.89 & -6.42 & -5.74 & -3.55 \\
	GLEAM J003652-333315 & \cellcolor[HTML]{f17ffa} LFTO & \cellcolor[HTML]{9cff99} 0 & -2.92 & -1.48 & -3.73 & -4.27 & \cellcolor[HTML]{e2e2e2} -1.06 & -4.91 & -2.11 & -4.57 & -2.54 & -8.34 & -8.87 & -3.82 \\
	GLEAM J011408-323907 & \cellcolor[HTML]{7fa3fa} Con & -3.68 & -5.79 & -6.91 & -8.47 & \cellcolor[HTML]{9cff99} 0 & -2.58 & -5.76 & -4.01 & -4.93 & -4.64 & -7.56 & -4.88 & -6.23 \\
	GLEAM J012121-340345 & \cellcolor[HTML]{7fa3fa} Con & -3.94 & -6.06 & -6.68 & -8.34 & \cellcolor[HTML]{9cff99} 0 & -3.22 & -6.67 & -4.64 & -5.06 & -4.98 & -8.51 & -7.80 & -6.51 \\
	GLEAM J034056-223353 & \cellcolor[HTML]{fe6767} LFTO$^*$ & \cellcolor[HTML]{e2e2e2} -0.27 & -2.46 & \cellcolor[HTML]{9cff99} 0 & -2.62 & -4.32 & \cellcolor[HTML]{e2e2e2} -0.38 & -7.09 & -1.53 & -4.11 & -1.38 & -5.05 & -5.38 & -2.65 \\
	GLEAM J035545-422210 & \cellcolor[HTML]{f17ffa} LFTO & \cellcolor[HTML]{9cff99} 0 & -2.16 & -2.46 & -4.09 & -4.53 & \cellcolor[HTML]{e2e2e2} -1.10 & -8.19 & -2.18 & -5.45 & -3.02 & -7.51 & -7.78 & -4.45 \\
	GLEAM J040226-180247 & \cellcolor[HTML]{7fa3fa} Con & \cellcolor[HTML]{9cff99} 0 & -2.04 & -3.42 & -4.97 & -1.91 & \cellcolor[HTML]{e2e2e2} -0.18 & -3.95 & -1.42 & -3.49 & -2.21 & -5.96 & -6.49 & -3.51 \\
	GLEAM J041509-282854 & \cellcolor[HTML]{7fa3fa} Con & \cellcolor[HTML]{9cff99} 0 & -2.48 & -2.70 & -4.36 & -1.37 & \cellcolor[HTML]{e2e2e2} -0.15 & -2.42 & -1.34 & -3.94 & -1.93 & -7.46 & -7.71 & -3.44 \\
	GLEAM J042905-372842 & \cellcolor[HTML]{7fa3fa} Con & -4.25 & -6.24 & -1.92 & -4.54 & \cellcolor[HTML]{9cff99} 0 & -3.30 & -1.78 & -4.59 & -5.13 & -3.66 & -7.09 & -7.04 & -5.62 \\
	GLEAM J072121-690005 & \cellcolor[HTML]{fe6767} LFTO$^*$ & -2.38 & -4.28 & \cellcolor[HTML]{9cff99} 0 & -1.32 & \cellcolor[HTML]{e2e2e2} -0.27 & -1.25 & -2.47 & -2.53 & -3.64 & \cellcolor[HTML]{e2e2e2} -0.16 & -5.43 & -5.94 & -2.93 \\
	GLEAM J074515-712426 & \cellcolor[HTML]{fe6767} LFTO & -2.84 & -4.52 & \cellcolor[HTML]{9cff99} 0 & -1.20 & -5.28 & -2.21 & -5.01 & -3.32 & -4.33 & -2.71 & -6.21 & -6.32 & -4.11 \\
	GLEAM J090634-754935 & \cellcolor[HTML]{7fa3fa} Con & -7.88 & -9.52 & -2.68 & -6.41 & \cellcolor[HTML]{9cff99} 0 & -3.55 & -6.46 & -4.97 & -2.01 & -2.26 & -4.54 & -3.18 & -5.52 \\
	GLEAM J120737-145835 & \cellcolor[HTML]{fe6767} LFTO & -2.07 & -3.90 & \cellcolor[HTML]{9cff99} 0 & -1.14 & -4.96 & -1.71 & -4.90 & -2.81 & -4.68 & -2.53 & -6.48 & -6.33 & -4.34 \\
	GLEAM J142112-461800 & \cellcolor[HTML]{f17ffa} LFTO$^*$ & \cellcolor[HTML]{e2e2e2} -0.71 & -2.39 & \cellcolor[HTML]{e2e2e2} -0.93 & -2.24 & -3.18 & \cellcolor[HTML]{9cff99} 0 & -3.56 & -1.20 & -3.67 & -1.46 & -6.12 & -5.91 & -3.12 \\
	GLEAM J150540-422654 & \cellcolor[HTML]{f17ffa} LFTO$^*$ & \cellcolor[HTML]{9cff99} 0 & -1.63 & \cellcolor[HTML]{e2e2e2} -1.10 & -2.43 & -3.95 & \cellcolor[HTML]{e2e2e2} -0.26 & -3.90 & -1.37 & -4.67 & -2.40 & -7.37 & -7.17 & -3.82 \\
	GLEAM J184747-602054 & \cellcolor[HTML]{fe6767} LFTO$^*$ & \cellcolor[HTML]{e2e2e2} -0.83 & -2.48 & \cellcolor[HTML]{9cff99} 0 & \cellcolor[HTML]{e2e2e2} -0.17 & -1.16 & \cellcolor[HTML]{e2e2e2} -1.08 & -2.25 & -1.98 & -4.37 & -2.53 & -6.88 & -6.59 & -3.30 \\
	GLEAM J203047-472824 & \cellcolor[HTML]{fe6767} LFTO$^*$ & \cellcolor[HTML]{e2e2e2} -0.91 & -2.67 & \cellcolor[HTML]{9cff99} 0 & -1.54 & -3.07 & \cellcolor[HTML]{e2e2e2} -0.71 & -4.06 & -1.70 & -4.66 & -2.04 & -5.78 & -6.21 & -3.47 \\
	GLEAM J205209-484639 & \cellcolor[HTML]{7fa3fa} Con & \cellcolor[HTML]{9cff99} 0 & -1.90 & -2.10 & -3.43 & -3.45 & \cellcolor[HTML]{e2e2e2} -0.24 & -3.94 & -1.24 & -5.36 & -2.15 & -7.47 & -7.00 & -3.36 \\
	GLEAM J213610-383236 & \cellcolor[HTML]{f17ffa} LFTO & \cellcolor[HTML]{9cff99} 0 & -2.37 & -1.90 & -3.44 & \cellcolor[HTML]{e2e2e2} -1.10 & \cellcolor[HTML]{e2e2e2} -0.17 & -2.43 & -1.32 & -3.61 & -1.90 & -7.47 & -7.20 & -3.42 \\
	\bottomrule
\end{tabular}
\begin{tablenotes}[para]An overview of the natural log of the Bayes odds ratio from the {\tt DYNESTY} fitting of each model to every source. For each source, the values presented below are the evidence values for each model divided by the most preferred model. The natural log of the ratio is presented, such that the most preferred models have values in this table equal to ln(1)$=0$ (highlighted green). Models that are indistinguishable from the most preferred model correspond to $<$ln(3)$=1.1$ (highlighted grey). Less preferred models therefore have more negative numbers. Initial SFG sample membership is denoted by LFTO or control. Control galaxies with favoured models that do not include an LFTO have their sample membership highlighted in blue. LFTO galaxies that with {\tt FFA\_} prefix models are shown in red. LFTO galaxies which do not have their most favoured model with the {\tt FFA\_} prefix (i.e. the favoured model does not include a LFTO) are shown in pink. An * indicates a competing model of the opposite class.
\end{tablenotes}
\end{threeparttable}
\end{table*}

\section{Results}
\label{sec:results}
\subsection{Model Results}

\begin{table*}[hbt!]
\centering
\begin{threeparttable}
\caption{Preferred Model Parameter Table.}\label{tab:modelvals}
\begin{tabular}{c|c|c|c|c|c|c|c}
	\toprule
	Source & Model & A & $\alpha$ & $\nu_{t,1}$ & C & $\nu_{t,2}$ & $\nu_b$ \\
     &  & mJy &  & GHz & mJy & GHz & GHz \\
	\midrule
    GLEAM J002238-240737 & {\tt PL\_SIC}$^a$ & $77.0_{-20}^{+15}$ & $-1.14_{-0.03}^{+0.05} (-0.89)$ & & & & $8.5_{-5.3}^{+5.7}$ \\
	GLEAM J003652-333315 & {\tt PL} & $25.5_{-0.8}^{+0.8}$ & $-0.45_{-0.02}^{+0.02}$ & & & & \\
	GLEAM J011408-323907 & {\tt PL\_FFA} & $15.1_{-2.6}^{+2.7}$ & $-0.94_{-0.09}^{+0.08}$ & & $36.8_{-8.4}^{+8.2}$ & $0.98_{-0.31}^{+0.32}$ & \\
	GLEAM J012121-340345 & {\tt PL\_FFA} & $12.1_{-1.8}^{+2.2}$ & $-0.93_{-0.06}^{+0.06}$ & & $40.4_{-11}^{+13}$ & $0.60_{-0.14}^{+0.21}$ & \\
	GLEAM J034056-223353 & {\tt FFA\_PL} & $140_{-8.4}^{+14}$ & $-0.71_{-0.04}^{+0.04}$ & $0.13_{-0.04}^{+0.03}$ & & & \\
	GLEAM J035545-422210 & {\tt PL} & $21.2_{-0.8}^{+0.9}$ & $-0.63_{-0.02}^{+0.02}$ & & & & \\
	GLEAM J040226-180247 & {\tt PL} & $21.1_{-0.7}^{+0.7}$ & $-0.72_{-0.02}^{+0.02}$ & & & & \\
	GLEAM J041509-282854 & {\tt PL} & $35.1_{-1.2}^{+1.2}$ & $-0.56_{-0.02}^{+0.02}$ & & & & \\
	GLEAM J042905-372842 & {\tt PL\_FFA} & $9.7_{-1.9}^{+2.1}$ & $-0.88_{-0.07}^{+0.06}$ & & $37.6_{-10}^{+15}$ & $0.57_{-0.17}^{+0.27}$ & \\
	GLEAM J072121-690005 & {\tt FFA\_PL} & $242_{-18}^{+21}$ & $-0.80_{-0.05}^{+0.05}$ & $0.17_{-0.04}^{+0.03}$ & & & \\
	GLEAM J074515-712426 & {\tt FFA\_PL} & $326_{-31}^{+41}$ & $-0.93_{-0.04}^{+0.04}$ & $0.15_{-0.03}^{+0.03}$ & & & \\
	GLEAM J090634-754935 & {\tt PL\_FFA} & $15.7_{-2.9}^{+3.1}$ & $-1.07_{-0.10}^{+0.08}$ & &  $72.0_{-15}^{+30}$ & $0.83_{-0.04}^{+0.04}$ & \\
	GLEAM J120737-145835 & {\tt FFA\_PL} & $165_{-19}^{+24}$ & $-0.82_{-0.05}^{+0.05}$ & $0.16_{-0.04}^{+0.04}$ & & & \\
	GLEAM J142112-461800 & {\tt PL\_SIC}$^a$ & $146_{-38}^{+27}$ & $-1.01_{-0.04}^{+0.06} (-0.76)$ & & & & $9.1_{-5.5}^{+5.4}$ \\
	GLEAM J150540-422654 & {\tt PL} & $23.5_{-1.3}^{+1.3}$ & $-0.81_{-0.04}^{+0.04}$ & & & & \\
	GLEAM J184747-602054 & {\tt FFA\_PL} & $202_{-29}^{+40}$ & $-0.83_{-0.04}^{+0.04}$ & $0.15_{-0.04}^{+0.04}$ & & & \\
	GLEAM J203047-472824 & {\tt FFA\_PL} & $198_{-23}^{+32}$ & $-0.73_{-0.04}^{+0.04}$ & $0.14_{-0.04}^{+0.04}$ & & & \\
	GLEAM J205209-484639 & {\tt PL} & $45.9_{-1.6}^{+1.6}$ & $-0.76_{-0.02}^{+0.02}$ & & & & \\
	GLEAM J213610-383236 & {\tt PL} & $22.4_{-0.7}^{+0.7}$ & $-0.55_{-0.02}^{+0.02}$ & & & & \\
	\bottomrule
\end{tabular}
\begin{tablenotes}[para]The most preferred models, as judged by their evidence value with their constrained parameter values and 1$\sigma$ uncertainties. We use the 50th percentile of the samples posterior distribution as the nominal value, and use the 16th and 84th percentiles to provide the 1$\sigma$ uncertainties. Parameters not included in the models are left blank.\\
    \item[$^a$]By construction the final values of $\alpha$ for this model are $\alpha + \Delta\alpha$ (i.e the steeper value at higher frequencies). In brackets we include the value of $\alpha$ at the break frequency $\nu_b$.
\end{tablenotes}
\end{threeparttable}
\end{table*}

We find that all eight galaxies selected to be controls are favoured to be modelled with no LFTO (highlighted blue in Table. \ref{tab:modelsel}) and no control galaxies have models with the prefix {\tt FFA\_} as being indistinguishable suggesting that their selection criteria were appropriate. Of these eight control galaxies three are best modelled by simple {\tt PL} models, four by {\tt PL\_FFA} models and one by a {\tt PL\_SIC} model. There is however competition between primarily the {\tt PL} and {\tt PL\_SIC} models with many being indistinguishable from each other. The {\tt PL\_SIC} models often provide a better fit to the data by accounting for both flattening and steepening of the SED, however the addition of an extra free parameter which is difficult to constrain ($\sigma(\nu_{b}) \sim5$\,GHz) results in {\tt PL} models being more favoured as selected purely by their log-likelihood values. It is possible that models which include synchrotron or IC losses are more physically motivated at the expense of an additional parameter. We discuss the implications of this in the discussion. {\tt PL\_FFA} models do not have any competing models in the four sources they are preferred for. Indeed they do all visually have distinctly offset power law emission of a similar spectral index suggesting either distinct components as suggested in \cite{Dey2024} or perhaps GLEAM flux scale discrepancies.

Six of the 11 galaxies in the LFTO sample have models with LFTOs ({\tt FFA\_} prefix, highlighted in red in Table. \ref{tab:modelsel}) of which only four do not have competing simpler models with no curvature. Despite their initial selection as sources with LFTOs five sources have no {\tt FFA\_} prefix in their most favoured models (highlighted in pink). It is often the case that the LFTO galaxies are better fit by {\tt FFA\_} prefix models however this is for sources with large GLEAM flux density uncertainties or less spectral curvature. Simpler models compete well due to the large impact of adding free parameters. We find that GLEAM\_J072121-690005 is the only source to have a competing full model with both a prefix and suffix components, this source has a distinct LFTO and steepening spectral index as shown in Figure. A.\ref{fig:seds}.

We find our different models are more similar (i.e. the Bayes odds ratios are smaller) than those in \cite{Galvin2018} and \cite{Dey2022, Dey2024}. This is primarily a result of model construction such that our prior bounds are smaller with the overall frequency space being a factor of 2.5 times smaller which is then further constrained by the prefix and suffix model conditions only allowing $\nu_{t,1}$ to take values between 10 - 300\,MHz and 300\,MHz $\leq$ $\nu_{t,2}$ or $\nu_{b}$ $\leq$ 17\,GHz. In the case of \cite{Galvin2018} we have fewer free parameters due to their inclusion of FIR dust and GLEAM covariance models. We are also lacking data above 17\,GHz which helps constrain and distinguish models, especially those with thermal emission components \citep{Galvin2018, Dey2022, Dey2024}.

All favoured models do not include thermal emission, this is likely due to the lack of high frequency data needed to constrain the thermal emission component. GLEAM\_J184747-602054 is the only source that has a competitive {\tt SFG} based model suggesting that it may be the only source in our sample that has a significant thermal fraction at 17\,GHz. We do not observe spectral flattening in our other sources towards higher frequencies and generally find the modelled spectral indices for our SFG sample agree with the commonly accepted range of $-0.8 \geq \alpha \leq -0.7$ \citep{Condon1992}. These results are summarised in Table. \ref{tab:modelsel} and Figures \ref{fig:j0121}, \ref{fig:j1847} and A.\ref{fig:seds}.

\subsubsection{Emission Measure}
\label{em}

The turnover frequency due to FFA is dependent on where the optical depth reaches unity which occurs at the turnover frequency ($\nu_{t,1}$ or $\nu_{t,2}$) measured during SED modelling. The emission measure (EM) is then calculated assuming that the emitting HII regions form a cylinder orientated along of line of sight with constant temperature and density \citep{Condon1992}. The free–free opacity is then well approximated by:
\begin{equation}
    \tau_{\nu} = 3.28\times10^{-7} \bigg(\frac{T_e}{10^4 \rm~K}\bigg)\bigg(\frac{\nu}{\rm GHz}\bigg)^{-2.1}\bigg(\frac{\rm EM}{\rm pc~cm^{-6}}\bigg),
	\label{eq:EM}
\end{equation}
where $T_e$ is the electron temperature of the HII emitting region, typically taken as $10^4$\,K, and EM is the emission measure at a depth $s$, defined as:
\begin{equation}
    \frac{\rm EM}{\rm pc~cm^{-6}} = \int_{los} \bigg(\frac{n_e}{\rm cm^{-3}}\bigg)^2d\bigg(\frac{s}{\rm pc}\bigg).
	\label{eq:EMdens}
\end{equation}
 Where the EM is the integral of the electron density, $n_e$, along the line of sight, $los$, of a HII region of depth $s$. Using the turnovers constrained by our modelling, we have estimated the EMs of our sources, outlined in Table. \ref{tab:EM}, using Equation. \ref{eq:EM}. We label the corresponding EM of $\nu_{t,1}$ and $\nu_{t,2}$ for all models as EM$_1$ and EM$_2$, respectively. We find that the EM values obtained for our sources are consistent with the more luminous LIRG and ULIRG samples of \citet{Clemens2010, Galvin2016, Dey2022, Dey2024} suggesting that the depth and orientation of the HII regions within these galaxies are similar despite their lower starburst activity.

\begin{table}[hbt!]
\centering
\begin{threeparttable}
\caption{Emission Measures}\label{tab:EM}
\begin{tabular}{c|c|c}
	\toprule
	Source & EM$_1$ & EM$_2$ \\
      & $10^6$ cm$^{-6}$ pc & $10^6$ cm$^{-6}$ pc \\
	\midrule
	GLEAM J011408-323907 & & $2.9_{-1.6}^{+2.3}$ \\
	GLEAM J012121-340345 & & $1.0_{-0.4}^{+0.9}$ \\
	GLEAM J034056-223353 & $0.04_{-0.02}^{+0.02}$ &  \\
	GLEAM J042905-372842 & & $0.9_{-0.5}^{+1.2}$ \\
	GLEAM J072121-690005 & $0.07_{-0.03}^{+0.03}$ &  \\
	GLEAM J074515-712426 & $0.06_{-0.02}^{+0.03}$ & \\
	GLEAM J090634-754935 & & $2.1_{-1.5}^{+2.3}$ \\
	GLEAM J120737-145835 & $0.06_{-0.03}^{+0.04}$ &  \\
	GLEAM J184747-602054 & $0.06_{-0.03}^{+0.04}$ &  \\
	GLEAM J203047-472824 & $0.05_{-0.03}^{+0.03}$ &  \\
    \bottomrule
\end{tabular}
\begin{tablenotes}[para]An overview of the emission measures (EM) derived for each source from the model most supported by the evidence. Objects without an emission measure constrained are not listed.
\end{tablenotes}
\end{threeparttable}
\end{table}

\subsection{Global Radio Properties}

\begin{table*}[hbt!]
\centering
\begin{threeparttable}
\caption{Derived radio continuum properties.}\label{tab:radio}
\begin{tabular}{c|c|c|c|c|c|c}
	\toprule
	GLEAM ID & $\alpha^{\rm GLEAM}_{\rm low}$ & $\alpha^{\rm RACS}_{\rm mid}$ & $\alpha^{\rm ATCA}_{\rm high}$ & Log$_{10}(L_{\rm 1.4GHz}$) & SFR$_{\rm 1.4GHz}$ & Log$_{10}(\Sigma$SFR$_{\rm 1.4GHz}$) \\
     &  &  &  & (W m$^2$ Hz$^{-1}$) & (M$_{\odot}$ yr$^{-1}$) & (M$_{\odot}$ yr$^{-1}$ kpc$^{-2}$) \\
	\midrule
	GLEAM J002238-240737 & $-0.64\pm0.08$ & $-0.72\pm0.05$ & $-1.10\pm0.27$ & $22.8\pm0.3$ & $17.91\pm1.72$ & $-1.24\pm0.04$ \\
	GLEAM J003652-333315 & $-0.30\pm0.09^a$ & $-0.43\pm0.06$ & $-0.46\pm0.18$ & $22.4\pm0.3$ & $8.91\pm0.80$ & $-1.24\pm0.04$ \\
	GLEAM J011408-323907 & $-0.84\pm0.09$ & $-0.58\pm0.05$ & $-1.07\pm0.19$ & $22.0\pm0.4$ & $4.03\pm0.54$ & $-1.86\pm0.06$ \\
	GLEAM J012121-340345 & $-0.71\pm0.09$ & $-0.60\pm0.05$ & $-0.82\pm0.29$ & $21.9\pm0.3$ & $3.62\pm0.40$ & $-1.92\pm0.05$ \\
	GLEAM J034056-223353 & $-0.09\pm0.08$ & $-0.59\pm0.05$ & $-0.82\pm0.18$ & $21.3\pm0.4$ & $1.10\pm0.14$ & $-2.18\pm0.05$ \\
	GLEAM J035545-422210 & $-0.20\pm0.19$ & $-0.68\pm0.06$ & $-0.85\pm0.19$ & $20.5\pm0.4$ & $0.23\pm0.03$ & $-1.27\pm0.06$ \\
	GLEAM J040226-180247 & $-1.09\pm0.18$ & $-0.53\pm0.06$ & $-0.97\pm0.21$ & $22.7\pm0.2$ & $14.10\pm1.24$ & $-1.69\pm0.04$ \\
	GLEAM J041509-282854 & $-0.63\pm0.09$ & $-0.43\pm0.05$ & $-0.82\pm0.21$ & $22.5\pm0.3$ & $10.20\pm1.09$ & $-1.63\pm0.05$ \\
	GLEAM J042905-372842 & $-0.52\pm0.21$ & $-0.49\pm0.06$ & $-0.75\pm0.19$ & $22.7\pm0.2$ & $15.99\pm1.41$ & $-1.85\pm0.04$ \\
	GLEAM J072121-690005 & $-0.18\pm0.18$ & $-0.58\pm0.07$ & $-1.03\pm0.21$ & $21.4\pm0.3$ & $1.33\pm0.15$ & $-2.01\pm0.05$ \\
	GLEAM J074515-712426 & $-0.06\pm0.18$ & $-0.81\pm0.07$ & $-1.10\pm0.21$ & $22.5\pm0.3$ & $10.11\pm1.24$ & $-1.92\pm0.05$ \\
	GLEAM J090634-754935 & $-0.59\pm0.15$ & $-0.62\pm0.06$ & $-1.29\pm0.14$ & $22.6\pm0.3$ & $12.30\pm1.45$ & $-1.71\pm0.05$ \\
	GLEAM J120737-145835 & $0.01\pm0.31$ & $-0.70\pm0.07$ & $-0.88\pm0.21$ & $22.4\pm0.3$ & $8.35\pm0.83$ & $-1.85\pm0.04$ \\
	GLEAM J142112-461800 & $-0.29\pm0.33^a$ & $-0.60\pm0.10$ & $-0.93\pm0.14$ & $21.5\pm0.3$ & $1.48\pm0.15$ & $-2.35\pm0.04$ \\
	GLEAM J150540-422654 & $-0.22\pm0.46^a$ & $-0.78\pm0.13$ & $-0.87\pm0.27$ & $22.1\pm0.3$ & $4.94\pm0.56$ & $-1.93\pm0.05$ \\
	GLEAM J184747-602054 & $-0.05\pm0.29$ & $-0.70\pm0.11$ & $-0.75\pm0.17$ & $23.0\pm0.2$ & $25.30\pm2.17$ & $-1.52\pm0.04$ \\
	GLEAM J203047-472824 & $-0.14\pm0.34^a$ & $-0.64\pm0.08$ & $-0.79\pm0.14$ & $21.4\pm0.2$ & $1.35\pm0.12$ & $-1.43\pm0.04$ \\
	GLEAM J205209-484639 & $-0.58\pm0.12$ & $-0.68\pm0.06$ & $-0.87\pm0.17$ & $22.5\pm0.3$ & $11.33\pm1.20$ & $-1.70\pm0.05$ \\
	GLEAM J213610-383236 & $-0.70\pm0.15^a$ & $-0.43\pm0.06$ & $-0.63\pm0.17$ & $22.3\pm0.3$ & $7.21\pm0.68$ & $-2.08\pm0.04$ \\
	\bottomrule
\end{tabular}
\begin{tablenotes}[para]
    \item[]Note: Column (1): GLEAM source ID. Column (2): GLEAM (72 - 231\,MHz) spectral index. Column (3): GLEAM-RACS-mid (200\,MHz to 1.37\,GHz) spectral index. Column (4): ATCA (5.5 - 17\,GHz) spectral index. Column (5): RACS-mid luminosity. Column (6): Total radio SFR from RACS-mid luminosity and Equation. \ref{eq:Molnar}. Column (7): Total radio star formation rate surface density calculated based on radio SFR and $K$-band light axis ratios (see Equation. \ref{eq:SFRD}).
    \item[$^{a}$]Recalculated values of $\alpha^{\rm GLEAM}_{\rm low}$ due to a GLEAM sub-band containing negative flux.
\end{tablenotes}
\end{threeparttable}
\end{table*}

\subsubsection{Radio Star-Formation Rates}

RACS-mid flux densities at 1.37\,GHz from the 25$''$ catalogue \citep{Duchesne2024} are used to estimate the 1.4\,GHz radio-SFR as they are available for all of our SFGs at a higher resolution than NVSS and its flux scale is consistent with that of NVSS within <1\%. We use the relationship from \cite{Molnar2021}:
\begin{equation}
    \log\bigg(\frac{\rm SFR_{\rm 1.4\,GHz}}{\rm M_{\odot}\,yr^{-1}}\bigg)= (0.823\pm0.009)\log\bigg(\frac{L_{1.4}}{\rm W\,Hz^{-1}}\bigg) - (17.5\pm0.2)
	\label{eq:Molnar}
\end{equation}
where $L_{1.4}$ is the 1.4\,GHz radio luminosity in ${\rm W~Hz^{-1}}$ and is calibrated against the $q_{\rm TIR}$ derived SFR to estimate the radio SFR. The 1.4\,GHz luminosity is calculated using:
\begin{equation}
    \bigg(\frac{\rm L_{\rm 1.4\,GHz}}{\rm W\,Hz^{-1}}\bigg)=9.52\times10^{15}\bigg(\frac{4\pi}{(1+z)^{1+\alpha}}\bigg)\,\bigg(\frac{\rm D_{L}}{\rm Mpc}\bigg)^{2}\bigg(\frac{S_{\rm 1.4\,GHz}}{\rm mJy}\bigg)
	\label{eq:luminosity}
\end{equation}
where $z$ is the redshift, $\alpha$ is the modelled spectral index, D$_{\rm L}$ is the luminosity distance from the WXSC in Mpc and $S_{\rm 1.4\,GHz}$ is the RACS-mid integrated flux density in mJy. 

The radio SFR surface density is given by:
\begin{equation}
    \frac{\Sigma \rm SFR_{\rm 1.4\,GHz}}{\rm M_{\odot}\,yr^{-1}\,kpc^{-2}}= \bigg(\frac{\rm SFR_{\rm 1.4\,GHz}}{\rm M_{\odot}\,yr^{-1}}\bigg)\bigg(\frac{\pi k^2 A B}{\rm kpc^2}\bigg)^{-1}
	\label{eq:SFRD}
\end{equation}
where k is the conversion factor from angular to physical scale in kpc arcsec$^{-1}$ as calculated based on D$_L$ and A and B are the $K$-band light major and minor axis radii in arcseconds. All these values are presented in Table~\ref{tab:radio}.

\subsubsection{Spectral Indices}

\begin{figure*}[hbt!]
	\includegraphics[width=\textwidth]{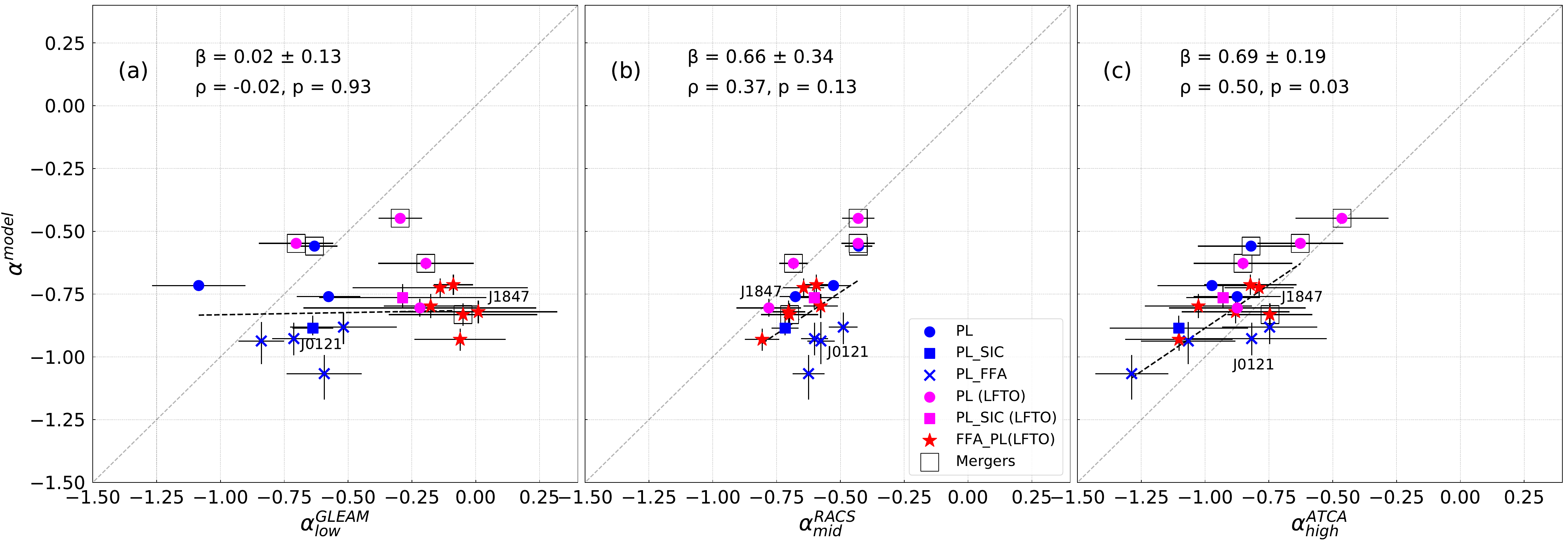}
    \centering
    \caption{Comparisons between the modelled spectral index and GLEAM, GLEAM to RACS-mid and ATCA spectral indices in panels (a), (b), and (c) respectively. The slope of the weighted linear fit and its 1$\sigma$ uncertainty and the Spearman's rank correlation test $\rho$ and p-values are given inside each panel. {\tt PL\_SIC} models have had their $\alpha^{model}$ values increased by 0.25 to be comparable due to their model construction. We do not include the outlier GLEAM J003652-333315 in our statistical analysis as it is Haro-11 the Lyman-continuum leaker with extreme IR properties.}
    \label{fig:alphacomps}
\end{figure*}

To determine the relationship between modelled spectral index $\alpha$ (hereafter $\alpha^{model}$) and the GLEAM, GLEAM to RACS-mid, and ATCA spectral indices ($\alpha^{\rm GLEAM}_{\rm low}$, $\alpha^{\rm RACS}_{\rm mid}$ and $\alpha^{\rm ATCA}_{\rm high}$ respectively) for our samples we perform a Spearman's rank correlation test. We find that $\alpha^{model}$ is most strongly correlated to $\alpha^{\rm ATCA}_{\rm high}$ and slightly correlated to $\alpha^{\rm RACS}_{\rm mid}$ as based on their p-values which reject the null hypothesis of no correlation. This is likely due to the lower error budget and increased sampling weighting the modelled spectral index towards $\alpha^{\rm ATCA}_{\rm high}$ in comparison to $\alpha^{\rm RACS}_{\rm mid}$. 

Comparing our modelled spectral index (as presented in Table. \ref{tab:modelvals}) to $\alpha^{\rm GLEAM}_{\rm low}$, $\alpha^{\rm RACS}_{\rm mid}$ and $\alpha^{\rm ATCA}_{\rm high}$ at GLEAM, RACS-mid and ATCA frequencies in Figure. \ref{fig:alphacomps} we see positive correlation between the modelled spectral index and $\alpha^{\rm RACS}_{\rm mid}$ and $\alpha^{\rm ATCA}_{\rm high}$. The modelled spectral index tends to be steeper than $\alpha^{\rm RACS}_{\rm mid}$ but flatter than $\alpha^{\rm ATCA}_{\rm high}$. By construction the modelled spectral index will differ based on the most favoured model as they each account for spectral flattening or steepening differently to simple unmodified power-law models. The overall effect of this is that basic {\tt PL} only models have flatter spectral indices as they do not account for low frequency flattening or include a second emission component as in {\tt PL\_FFA} models.

The low frequency spectral index is not related to the modelled spectral index for the total sample due to the construction of the {\tt FFA\_PL} model including a low frequency turnover. For the non-{\tt FFA\_PL} sample $\alpha^{model}$ is not correlated to $\alpha^{\rm GLEAM}_{\rm low}$ and we see that $\alpha^{\rm GLEAM}_{\rm low}$ is much flatter. All but one source exhibits some spectral index flattening between $\alpha^{\rm ATCA}_{\rm high}$ and $\alpha^{\rm GLEAM}_{\rm low}$ indicating that loss and absorption processes play a key role in modifying the synchrotron spectral index. Merging systems are found to have flatter modelled spectral indices than non-merging systems.

\subsubsection{Modelled Spectral Index Correlations}

\begin{figure*}[hbt!]
	\includegraphics[width=\textwidth]{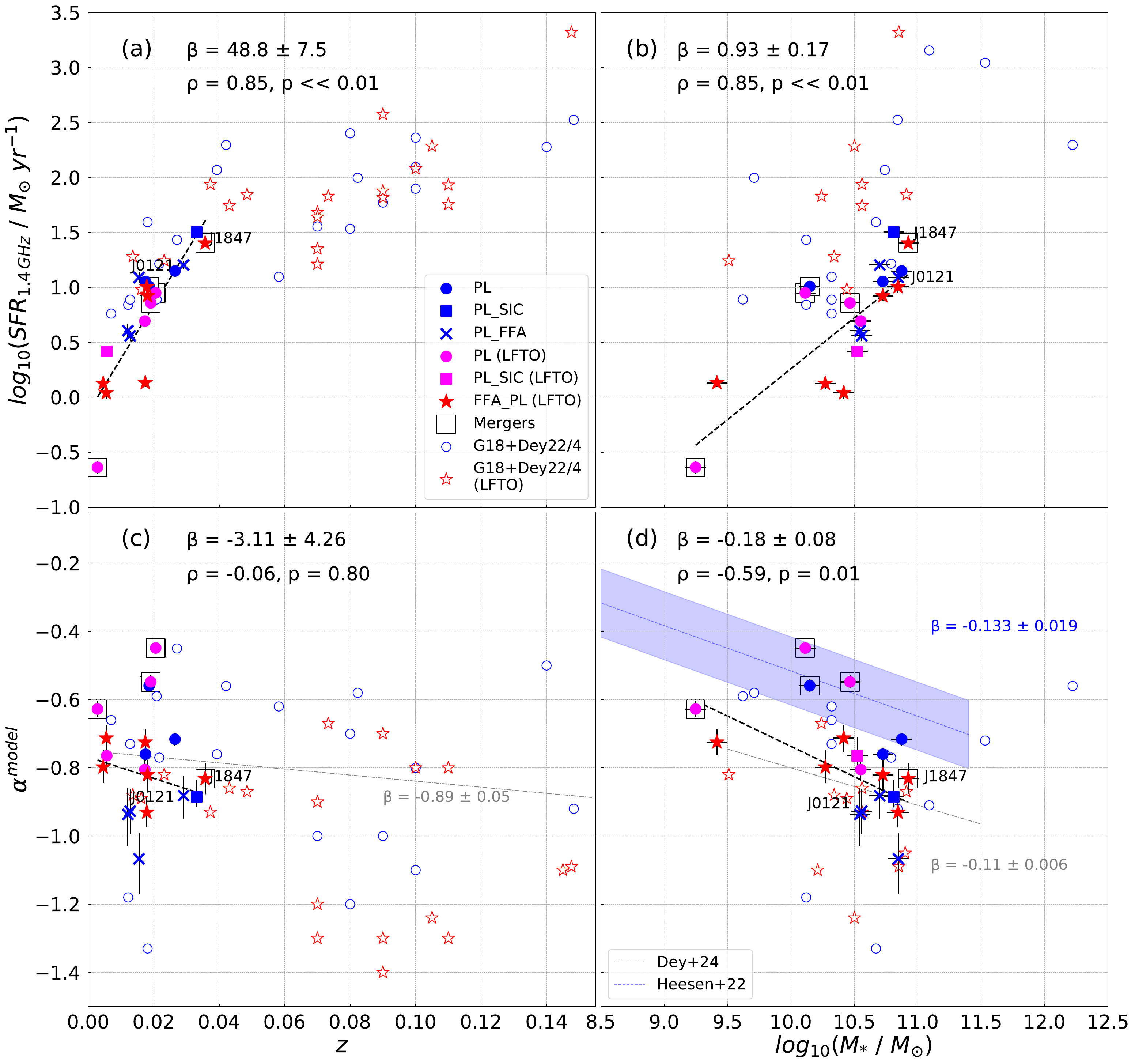}
    \centering
    \caption{Comparisons between the radio SFR versus redshift and stellar mass in panels (a) and (b) and modelled spectral index versus redshift and stellar mass in panels (c) and (d) respectively. The slope of the weighted linear fit and its 1$\sigma$ uncertainty and the Spearman's rank correlation test $\rho$ and p-values are given inside each panel. We do not include the outlier source GLEAM J003652-333315 in our statistical analysis. The open points are compiled from the LIRG/ULIRG samples of \cite{Galvin2018}, \cite{Dey2022} and \cite{Dey2024} and are separated based on their most favoured radio SED model with red stars being sources which include LFTOs in their radio SEDs and blue circles being all other sources. Grey dashed lines and relationships are the fits from \cite{Dey2024} to compare to our SFG sample. The blue dashed line is from \cite{Heesen2022a} and measures $\alpha^{0.15}_{1.4}$ against total galaxy mass (i.e. it probes a flatter part of the radio SED). {\tt PL\_SIC} models have had their $\alpha^{model}$ values increased by 0.25 to be comparable due to their model construction.}
    \label{fig:alphamodzmass}
\end{figure*}

\begin{figure}[hbt!]
	\includegraphics[width=\columnwidth]{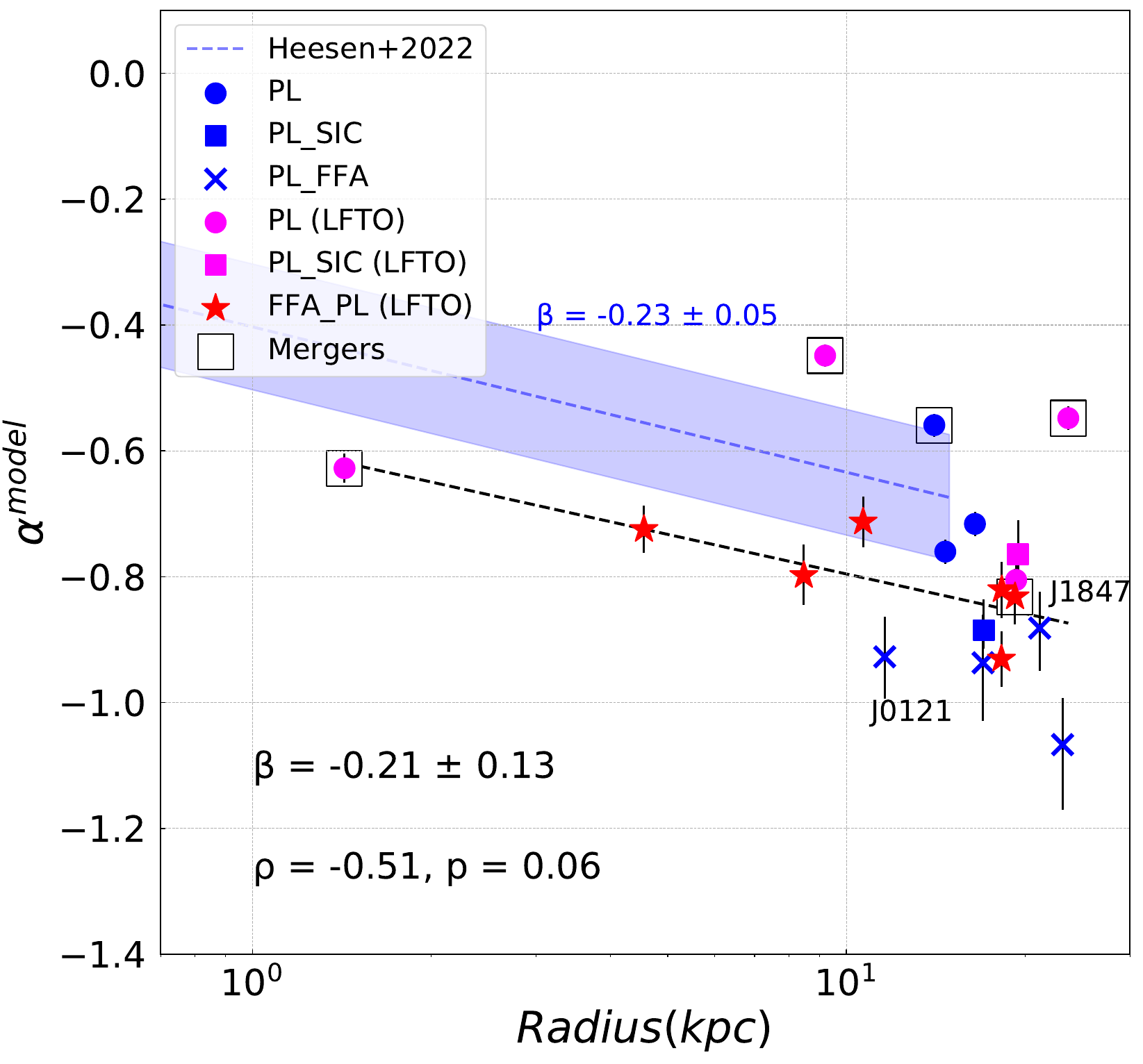}
    \centering
    \caption{The modelled spectral index in comparison to the $K$-band light major axis radius. The slope of the weighted linear fit and its 1$\sigma$ uncertainty and the Spearman's rank correlation test $\rho$ and p-values are given. We do not include the outlier source GLEAM J003652-333315 in our statistical analysis. The blue dashed line is from \cite{Heesen2022a} and measures $\alpha^{0.15}_{1.4}$ against the star formation radius (i.e. it probes a flatter part of the radio SED). {\tt PL\_SIC} models have had their $\alpha^{model}$ values increased by 0.25 to be comparable due to their model construction.}
    \label{fig:alphasize}
\end{figure}

We compare the modelled spectral index with a number of properties to provide insight as to the possible different physical processes occurring within our LFTO and control samples. Our sample lies within the range of values of the spectral index for LIRGS/ULIRGS with higher SFR galaxies being more easily detected at higher redshifts \citep{Galvin2018, Dey2024}. Figure. \ref{fig:alphamodzmass} also shows that there is no significant relationship (as determined by the p-value of the Spearman rank correlation test) between the modelled spectral index and redshift indicating that IC losses due to CMB photons are unlikely to be a dominant mechanism acting to steepen the spectral index as expected due to the low redshifts of our SFG sample. \cite{Galvin2018} and \cite{Dey2024} also find no significant relationship between spectral index and redshift for LIRGS/ULIRGS even out to a redshift of 0.4.

On the other hand we do find a statistically significant relationship between stellar mass and the modelled spectral index whereby more massive galaxies have a higher radio SFR and steeper modelled spectral index. This effect is seen in \citet{An2021}, \citet{Heesen2022a}, \citet{An2024}, and \citet{Dey2024} with negative correlations between stellar mass and spectral index. The steepening of the spectral index with increasing stellar mass is a result of higher mass galaxies generally being larger \citep{Gurkan2018} which causes the CREs to take longer to escape from the galaxy thus synchrotron cooling losses become important. This effect is highlighted by the negative correlation in Figure. \ref{fig:alphasize} between $\alpha^{model}$ and the galactic radius which is also seen in \cite{Heesen2022a} (however their results probe a flatter part of the radio SED hence the vertical offset).

\begin{figure*}[hbt!]
	\includegraphics[width=\textwidth]{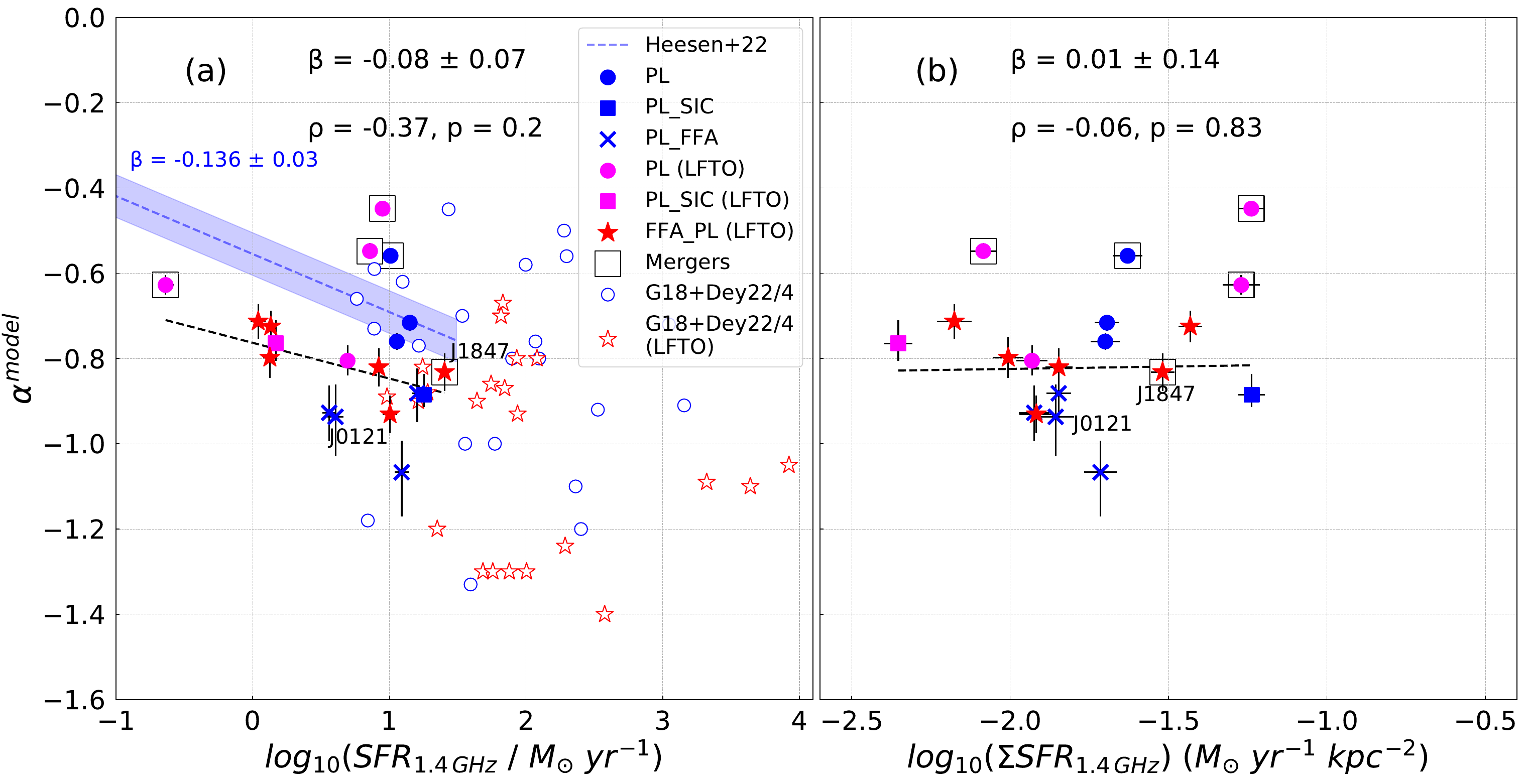}
    \centering
    \caption{Comparisons between the modelled spectral index with radio SFR and radio star formation rate surface density in panels (a) and (b) respectively. The slope of the weighted linear fit and its 1$\sigma$ uncertainty and the Spearman's rank correlation test $\rho$ and p-values are given inside each panel. The blue dashed line is from \cite{Heesen2022a} and measures $\alpha^{0.15}_{1.4}$ against TIR SFR (i.e. it probes a flatter part of the radio SED). {\tt PL\_SIC} models have had their $\alpha^{model}$ values increased by 0.25 to be comparable due to their model construction. We do not include the outlier source GLEAM J003652-333315 in our statistical analysis.}
    \label{fig:SFRD}
\end{figure*}

The modelled spectral index decreases with increasing radio SFR as shown in Figure. \ref{fig:SFRD} however the correlation is not statistically significant similar to the result of \citet{Dey2024} for the IR derived SFR. This modelled spectral index-SFR relationship is secondary with galaxy size and synchrotron cooling primarily driving the correlation between these variables as shown by the Spearman rank correlation coefficient for the $\alpha^{model}$-Radius correlation being -0.51 (p = 0.06) compared to -0.37 (p=0.2) for the $\alpha^{model}$-SFR$_{1.4\,GHz}$. We do not find a statistically significant relationship between radio SFR surface density and modelled spectral index similar to \cite{Heesen2022a}. These results are contrary to the expectation \citep[and results in][]{Tabatabaei2017} that galaxies with higher SFRs and younger more energetic CREs have a flatter spectral index. Overall we do not see any separation in radio derived properties between LFTO and non-LFTO (or {\tt FFA\_PL} and non-{\tt FFA\_PL}) samples or those in \cite{Galvin2018} or \cite{Dey2022, Dey2024}.

\subsection{Global Mid-Infrared Properties}

\begin{figure}[hbt!]
	\includegraphics[width=\columnwidth]{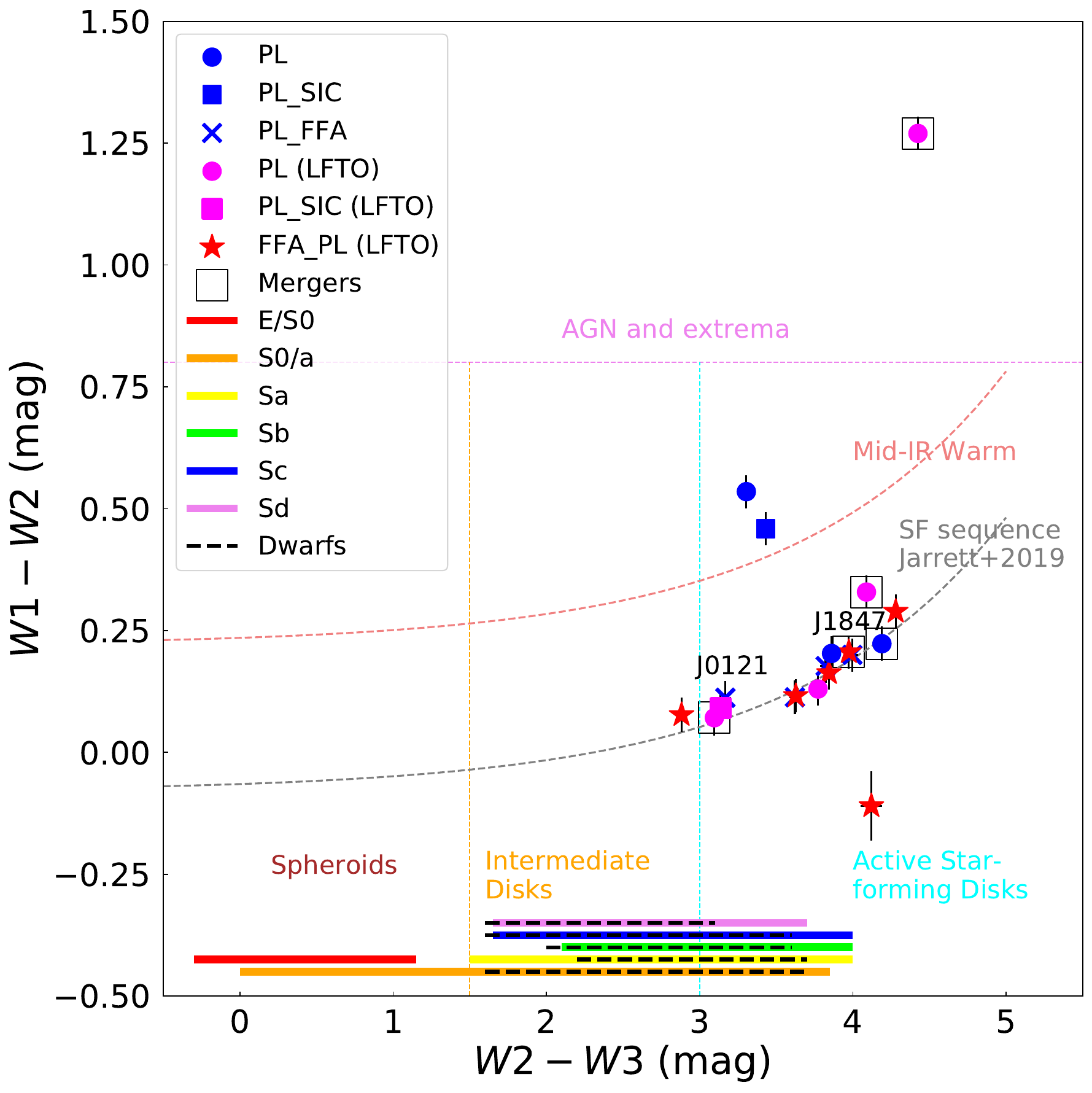}
    \centering
    \caption{WISE colour-colour diagram for our SFG sample. Magnitudes are in the Vega system with calibration described in \citet{Jarrett2011}. Regions roughly delineate source types into the labelled categories with AGN and extrema including luminous dust-obscured starbursts (GLEAM J003652-333315). The grey dotted line indicates the "star formation sequence" identified by the 100 largest galaxies in the WXSC \citep{Jarrett2019}.}
    \label{fig:WISECCD}
\end{figure}

The {\it WISE} colour-colour diagram (see Figure. \ref{fig:WISECCD}) provides a diagnostic for the dominant mid-IR emission mechanism and activity of nearby galaxies. Our sample primarily lies in the region occupied by star-forming disk galaxies at the top end of the SF sequence identified in \citet{Jarrett2019} which is expected based on their morphology and measured SFRs. There is no significant separation between the LFTO/control or {\tt FFA\_PL}/non-{\tt FFA\_PL} samples in this parameter space. Two control galaxies GLEAM J040226-180247 and GLEAM J002238-240737 have warmer mid-IR colours but show no evidence for significant AGN emission in their mid-IR, radio or optical spectra. GLEAM J074515-712426 lies below the SF sequence likely due to contamination by a saturated foreground star which causes a deficit of W1-W2 emission throughout the field. Lastly GLEAM J003652-333315 is found to be an extremely mid-IR bright dust-obscured source also known as Haro 11. Haro 11 has been shown to be a starbursting blue compact galaxy with both dust obscured star-forming regions and Lyman-$\alpha$ leakage \citep{ostlin15, ostlin21}. GLEAM J003652-333315 will therefore have somewhat overestimated \citep[but still comparable to those of][]{ostlin15} mid-IR based parameters including M$_*$, SFR$_{mircor}$ and sSFR$_{mircor}$ due to the dust-obscuration.

\begin{table*}[hbt!]
\centering
\begin{threeparttable}
\caption{SFG Sample Derived IR Properties.}\label{tab:IRderived}
\begin{tabular}{c|c|c|c|c|c|c}
	\toprule
	GLEAM ID & W1-W2 & W2-W3 & SFR$_{mircor}$ & log$_{10}$(sSFR$_{mircor}$) & log$_{10}$(L$_{60\,\mu m}$) & q$_{FIR}$ \\
     & (mag) & (mag) & (M$_{\odot}$ yr$^{-1}$) & (yr$^{-1}$) & (L$_{\odot}$) &  \\
	\midrule
	GLEAM J002238-240737 & $0.46\pm0.03$ & $3.43\pm0.04$ & $10.65\pm3.51$ & $-9.78\pm0.05$ & $10.68\pm0.18$ & $2.01\pm0.09$ \\
	GLEAM J003652-333315 & $1.27\pm0.03$ & $4.43\pm0.04$ & $53.84\pm20.96$ & $-8.38\pm0.02$ & $10.87\pm0.14$ & $2.41\pm0.08$ \\
	GLEAM J011408-323907 & $0.11\pm0.04$ & $3.17\pm0.04$ & $1.95\pm0.48$ & $-10.25\pm0.23$ & $10.06\pm0.16$ & $2.17\pm0.08$ \\
	GLEAM J012121-340345 & $0.11\pm0.04$ & $3.62\pm0.04$ & $3.65\pm0.90$ & $-10.00\pm0.13$ & $10.09\pm0.14$ & $2.30\pm0.08$ \\
	GLEAM J034056-223353 & $0.08\pm0.04$ & $2.88\pm0.04$ & $1.36\pm0.33$ & $-10.28\pm0.33$ & $9.69\pm0.12$ & $2.52\pm0.08$ \\
	GLEAM J035545-422210 & $0.07\pm0.04$ & $3.10\pm0.05$ & $0.23\pm0.06$ & $-9.88\pm1.86$ & $8.78\pm0.09$ & $2.40\pm0.09$ \\
	GLEAM J040226-180247 & $0.54\pm0.03$ & $3.31\pm0.04$ & $14.93\pm5.16$ & $-9.70\pm0.04$ & $10.76\pm0.12$ & $2.14\pm0.09$ \\
	GLEAM J041509-282854 & $0.22\pm0.04$ & $4.19\pm0.04$ & $6.31\pm1.58$ & $-9.35\pm0.08$ & $10.52\pm0.09$ & $2.01\pm0.08$ \\
	GLEAM J042905-372842 & $0.20\pm0.03$ & $4.00\pm0.04$ & $15.28\pm4.88$ & $-9.52\pm0.04$ & $10.79\pm0.18$ & $2.11\pm0.09$ \\
	GLEAM J072121-690005 & $0.12\pm0.03$ & $3.63\pm0.03$ & $2.15\pm0.53$ & $-9.94\pm0.21$ & $9.77\pm0.09$ & $2.43\pm0.07$ \\
	GLEAM J074515-712426 & $-0.11\pm0.07$ & $4.12\pm0.07$ & $6.81\pm1.89$ & $-10.01\pm0.07$ & $10.50\pm0.41$ & $2.17\pm0.13$ \\
	GLEAM J090634-754935 & $0.18\pm0.03$ & $3.82\pm0.04$ & $9.42\pm3.03$ & $-9.87\pm0.05$ & $10.54\pm0.14$ & $2.09\pm0.09$ \\
	GLEAM J120737-145835 & $0.16\pm0.03$ & $3.84\pm0.04$ & $7.90\pm2.30$ & $-9.83\pm0.06$ & $10.56\pm0.14$ & $2.29\pm0.09$ \\
	GLEAM J142112-461800 & $0.09\pm0.03$ & $3.14\pm0.04$ & $1.88\pm0.46$ & $-10.25\pm0.24$ & $9.96\pm0.14$ & $2.48\pm0.09$ \\
	GLEAM J150540-422654 & $0.13\pm0.03$ & $3.77\pm0.04$ & $4.47\pm1.11$ & $-9.90\pm0.10$ & $10.14\pm0.16$ & $2.16\pm0.11$ \\
	GLEAM J184747-602054 & $0.21\pm0.03$ & $3.97\pm0.04$ & $28.53\pm11.00$ & $-9.47\pm0.02$ & $11.05\pm0.21$ & $2.15\pm0.09$ \\
	GLEAM J203047-472824 & $0.29\pm0.04$ & $4.28\pm0.04$ & $1.27\pm0.31$ & $-9.31\pm0.35$ & $9.91\pm0.14$ & $2.54\pm0.07$ \\
	GLEAM J205209-484639 & $0.20\pm0.04$ & $3.86\pm0.04$ & $9.10\pm2.75$ & $-9.77\pm0.06$ & $10.58\pm0.14$ & $2.09\pm0.08$ \\
	GLEAM J213610-383236 & $0.33\pm0.04$ & $4.09\pm0.04$ & $14.33\pm4.33$ & $-9.31\pm0.04$ & $10.78\pm0.16$ & $2.50\pm0.08$ \\
	\bottomrule
\end{tabular}
\begin{tablenotes}[para]
    \item[]Note: Column (1): GLEAM source ID. Column (2): \textit{WISE} band W1 subtracted from \textit{WISE} band W2 magnitude. Column (3): \textit{WISE} band W2 subtracted from \textit{WISE} band W3 magnitude. Column (4): Mid-IR + UV corrected SFR from \cite{Cluver2024}. Column (5): Mid-IR + UV corrected specific SFR from \cite{Cluver2024}. Column (6): Bolometric 60 $\mu$m luminosity from \citet{moshir1990}. Column (7): q$_{FIR}$ as calculated following \citet{Yun2001} using RACS-mid 1.4\,GHz radio flux densities.
    \item[$^{a}$]This source is undetected in \textit{IRAS} source catalogues.
\end{tablenotes}
\end{threeparttable}
\end{table*}

\begin{figure}[hbt!]
	\includegraphics[width=\columnwidth]{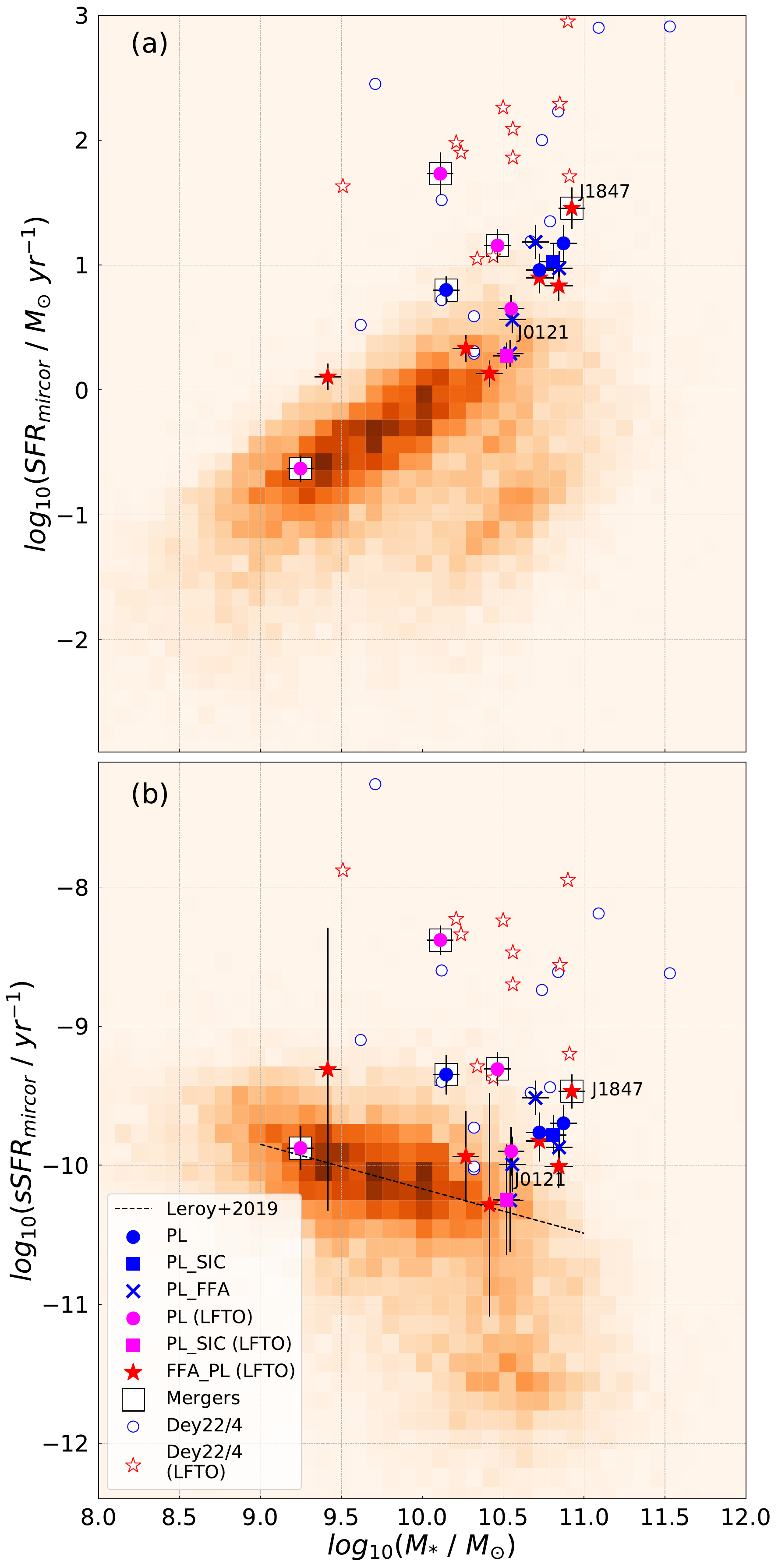}
    \centering
    \caption{(a): WISE mid-IR+UV corrected SFR \citep{Cluver2024} versus stellar mass for our SFG sample. (b): WISE mid-IR+UV corrected specific SFR \citep{Cluver2024} versus stellar mass for our SFG sample. The orange background sample and black dashed SFG main sequence best fit is from the WISE and GALEX Atlas of Local Galaxies \citep{Leroy2019}. The open points are compiled from the LIRG/ULIRG samples of \cite{Dey2022} and \cite{Dey2024} and are separated based on their most favoured radio SED model with red stars being sources which include LFTOs in their radio SEDs and blue circles being all other sources.}
    \label{fig:WISESFR}
\end{figure}

Our sample lies above the SFR-M$_{*}$ main sequence in the WISE and GALEX Atlas of Local Galaxies \citep{Leroy2019}. Our sample also has high sSFRs (see Figure. \ref{fig:WISESFR}) primarily due to their selection criteria requiring that their fluxes are measurable in C and X band ATCA radio observations. The minor separation in M$_*$ between the LFTO and control samples is observable with the LFTO sample probing masses up to an order of magnitude lower than the control sample. GLEAM J003652-333315 has SFR$_{mircor}$ and sSFR$_{mircor}$ more comparable to the more distant LIRGs/ULIRGs in \cite{Dey2022, Dey2024} due to its dusty obscured starbursts. Overall we see no distinct separation between galaxies which contain LFTOs and those which do not in both our sample and those in \cite{Dey2022, Dey2024}  (selected based on their most favoured model containing a low-$\nu$ FFA component).

\subsection{Inclination and Mergers}
\label{sec:morph}
\begin{figure}[hbt!]
	\includegraphics[scale=0.27]{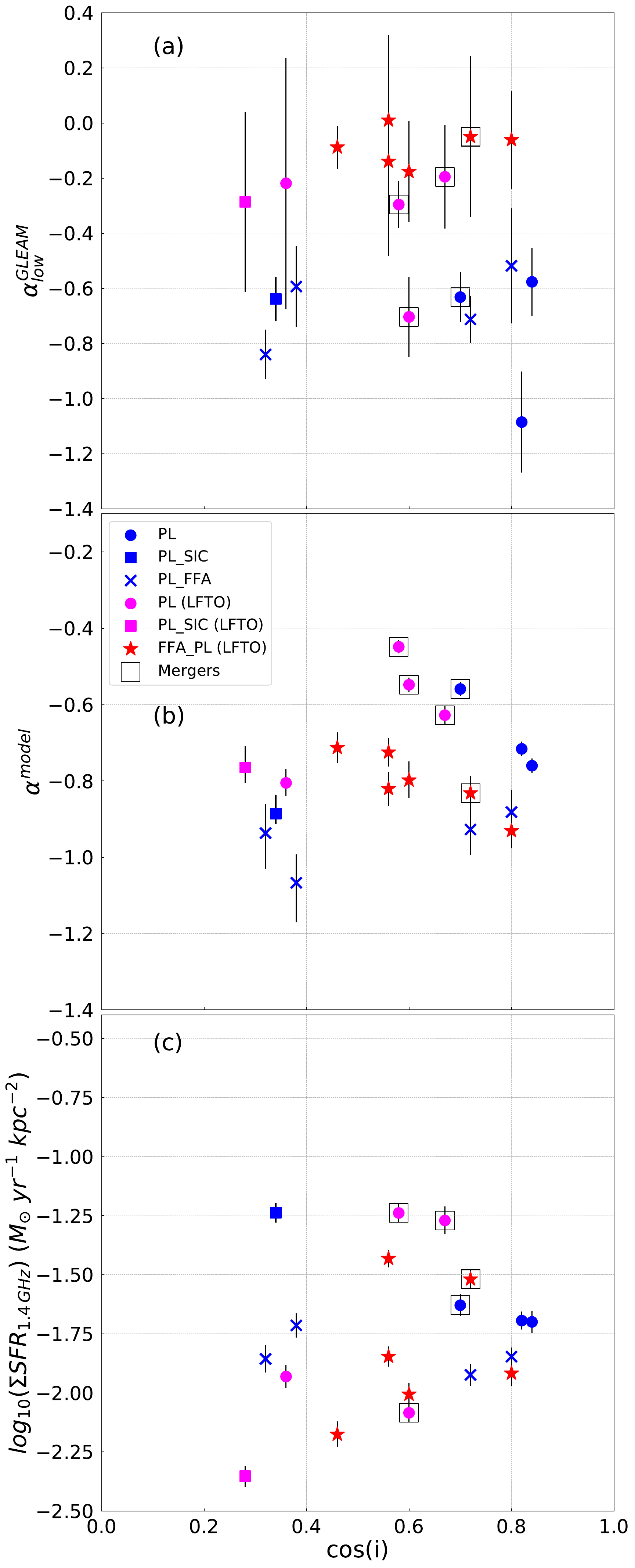}
    \centering
    \caption{Comparisons between the inclination and GLEAM spectral index, modelled spectral index and star formation rate surface density in panels (a), (b), and (c) respectively. Edge-on sources have cos(i) $\sim$ 0 whilst face-on sources have cos(i) $\sim$ 1. {\tt PL\_SIC} models have had their $\alpha^{model}$ values increased by 0.25 to be comparable due to their model construction.}
    \label{fig:incplots}
\end{figure}

We investigate whether there are any relationships between the morphology or inclination and the radio SED features observed. Inclinations for these galaxies are estimated using the ratio between major and minor \textit{K}-band light axes where axes are compiled from NED with sources given in Table.~\ref{tab:commonprops}. Inclinations are compared to the GLEAM and modelled spectral index with merging systems flagged in Figure.~\ref{fig:incplots}. Overall we do not see any significant correlations between inclination and spectral index (either GLEAM or modelled) or whether edge on SFGs are more likely to contain LFTOs agreeing with the findings of \cite{hummel1991}, \cite{chyzy2018} and \cite{Heesen2022a}. The two galaxies best modelled by {\tt PL\_SIC} models (square points) including synchrotron and IC losses are both some of the highest inclination sources in this SFG sample. We also examine whether inclination is correlated to the global star formation rate surface density and find no significant relationship. There is no separation between our samples based on inclination.

\subsection{Global Radio and Infrared Properties}
\label{sec:res2}

\begin{figure}[hbt!]
	\includegraphics[width=\columnwidth]{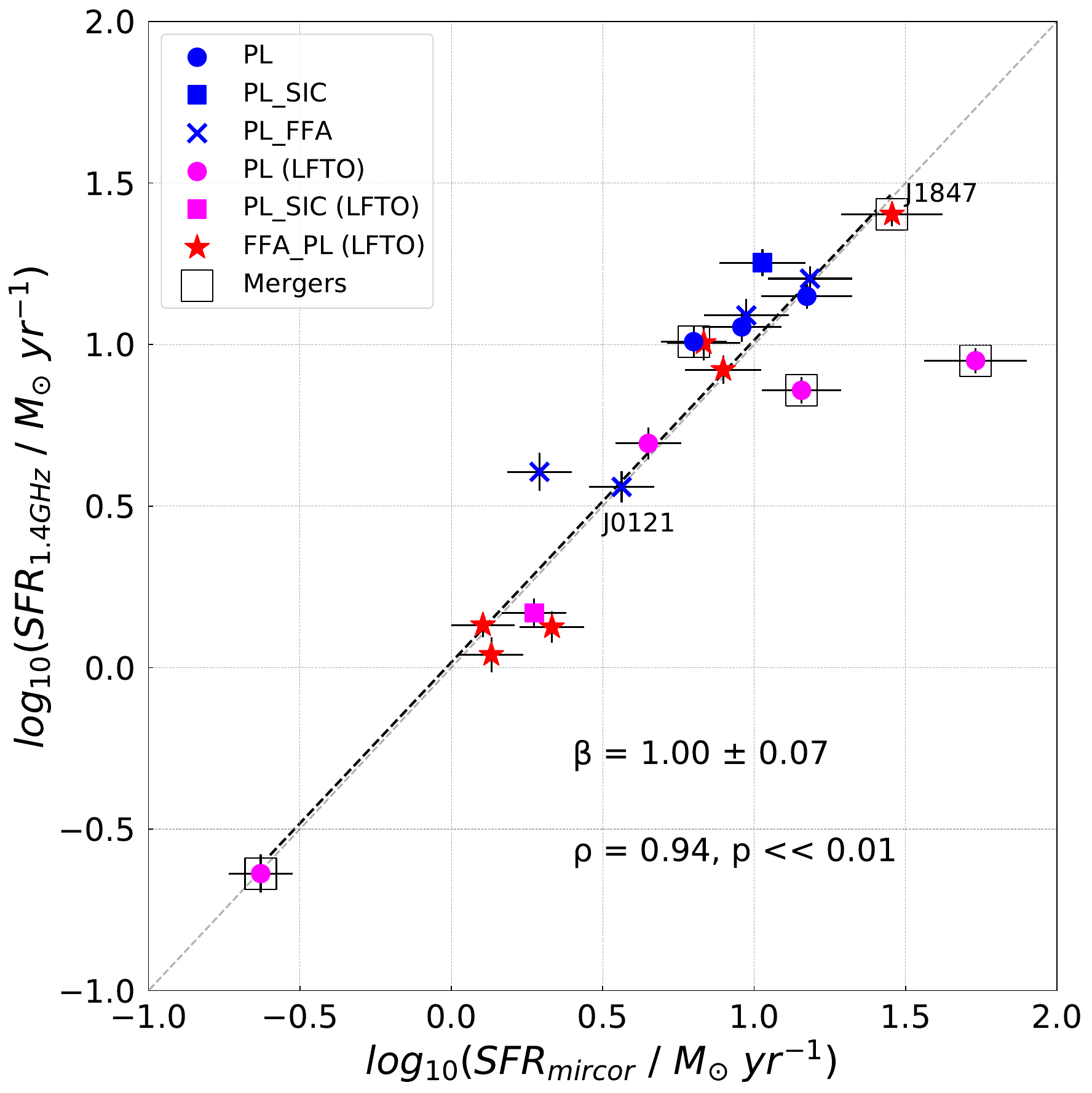}
    \centering
    \caption{The 1.4\,GHz radio-SFR compared to the mid-IR+FUV corrected SFR. The slope of the weighted linear fit with its 1$\sigma$ uncertainty and the Spearman's rank correlation test $\rho$ and p-values are presented.}
    \label{fig:sfrcomp}
\end{figure}

We see that our sample shows a statistically significant linear relationship between radio-SFR and SFR$_{mircor}$ in Figure. \ref{fig:sfrcomp} providing further evidence for the relationship between the IR and radio emission mechanisms. The SFR$_{mircor}$-excess outlier source GLEAM J003652-333315 is the dust obscured blue compact galaxy Haro 11 \citep{ostlin15}. The radio-SFR for this source is potentially underestimated as its compact size allows for the escape of CREs into the galactic halo on timescales shorter than the synchrotron loss timescale. 

The $q_{FIR}$ parameter, which is the logarithmic ratio between the far infrared flux and 1.4\,GHz flux density of an object, is a parameterisation of the FRC, where $q_{FIR}$ is defined as:
\begin{equation}
    q_{FIR} = \log\bigg(\frac{\rm FIR}{\rm 3.75\times10^{12}~W\,m^{-2}}\bigg) - \log\bigg(\frac{S_{1.4}}{\rm W\,m^{-2}\,Hz^{-1}}\bigg).
	\label{eq:qratio}
\end{equation}
Where S$_{1.4}$ is the flux density at $\nu$ = 1.4\,GHz, and FIR is defined as:
\begin{equation}
    \rm FIR = 1.26\times10^{-14} (2.58S_{60} + S_{100})\rm \,W\,m^{-2},
	\label{eq:FIR}
\end{equation}
where S$_{60}$ and S$_{100}$ are the 60 and 100\,$\mu$m band flux
densities from IRAS in Jy \citep{Helou1985}. The mean $q_{FIR}$ value between 60\,µm and 1.4\,GHz is typically taken as 2.34 for SFGs \citep{Yun2001}. $q_{FIR}$ is known to decrease with redshift \citep{magnelli2015, Delhaize2017}\footnote{These works use more sensitive Herschel IR observations and in the case of \cite{Delhaize2017} higher frequency 3\,GHz radio observations} and luminosity \citep{Molnar2021}. Deviations from the typical $q_{FIR}$ value can be a critical diagnostic of the physical processes driving the radio and IR emission. Radio excess objects ($q_{FIR}$ < 1.6) are usually associated with AGN emission whilst IR excess objects ($q_{FIR}$ > 3) may be dust obscured AGN or compact starbursts. With much of the scatter being influenced by the variation in dust temperature, extinction, and the different timescales associated with different SFR indicators.

\begin{figure*}[hbt!]
	\includegraphics[width=\textwidth]{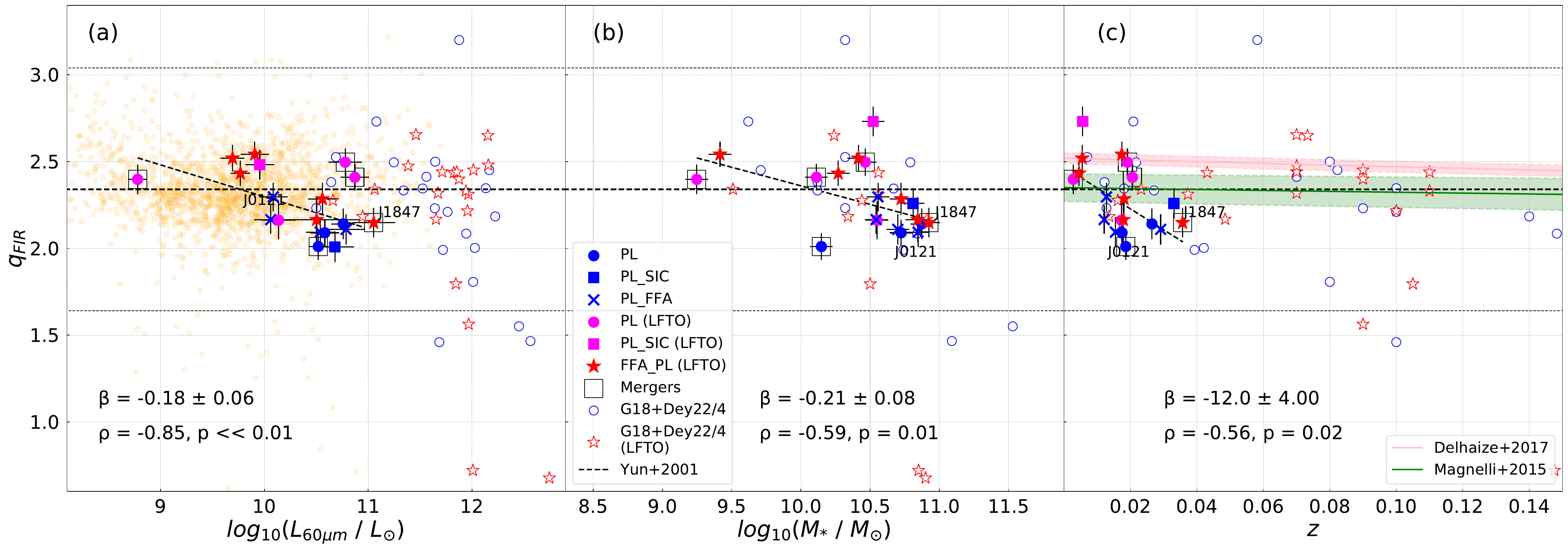}
    \centering
    \caption{$q_{FIR}$ compared to the IRAS 60\,$\mu$m luminosity with the relationship from \citet{Yun2001} shown in panel (a). Panels (b) and (c) show the comparison between $q_{FIR}$ and the stellar mass and redshift respectively. The slope of the weighted linear fit with its 1$\sigma$ uncertainty and the Spearman's rank correlation test $\rho$ and p-values are given inside each panel. {\tt PL\_SIC} models have had their $\alpha^{model}$ values increased by 0.25 to be comparable due to their model construction. We do not include the outlier source GLEAM J003652-333315 in our statistical analysis.}
    \label{fig:qplots}
\end{figure*}

Our SFG sample lies within the large scatter of previous $q_{FIR}$ observations shown in Figure. \ref{fig:qplots} \citep{Yun2001, magnelli2015, Delhaize2017, Galvin2018, Dey2024} with no FIR or radio excess sources (except J232600-815311 which is undetected in the FIR). We extend the samples of SFGs with spectral curvature down 2 orders of magnitude in L$_{60\,\mu m}$ to non-LIRG sources. When comparing $q_{FIR}$ we find the clearest separation between the control (blue) and LFTO (pink and red) galaxy samples with the LFTO sample having a significantly higher mean $q_{FIR}$ value than the control galaxies (see Table. \ref{tab:ttest}). This indicates either an excess in IR emission in LFTO galaxies or a radio deficit at 1.4\,GHz compared to the control sample. This separation is not seen in the sample of LIRGS/ULIRGS from \cite{Galvin2018} and \cite{Dey2024}. There is a significant negative correlation between $q_{FIR}$ and FIR luminosity or stellar mass which agrees with the findings in \cite{Molnar2021}. We also see a significant decrease in $q_{FIR}$ with redshift even over the small redshift range observed however due to our short lookback times this is likely a result of scatter and is not seen in previous research \citep{magnelli2015, Delhaize2017, Galvin2018, Dey2024}. These results are summarised in Figure. \ref{fig:qplots}.

\begin{figure*}[hbt!]
	\includegraphics[width=\textwidth]{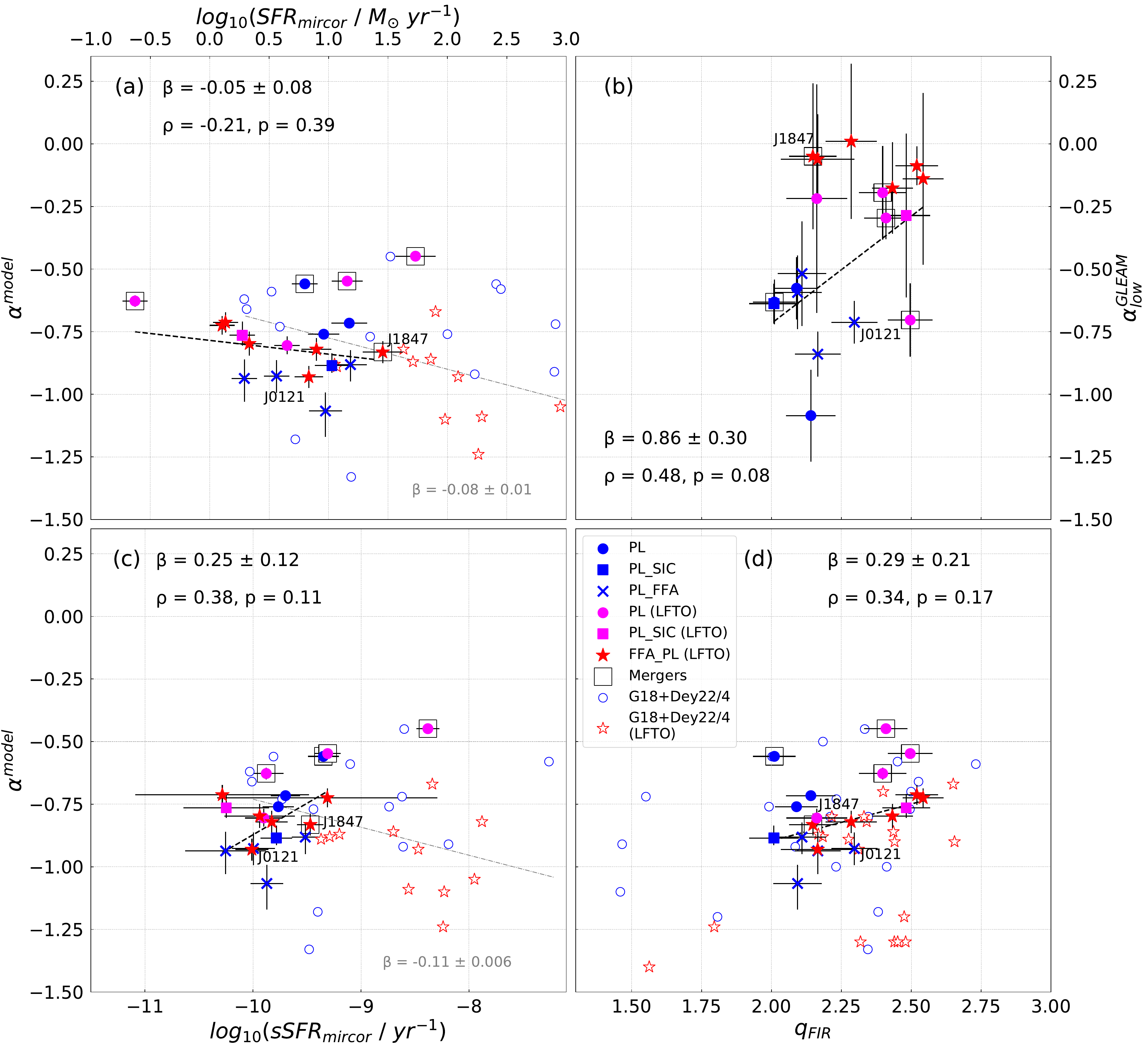}
    \centering
    \caption{Comparisons between the modelled spectral index and the IR SFR, sSFR and $q_{FIR}$ in panels (a), (c), and (d) respectively. Panel (b) compares the GLEAM spectral index to $q_{FIR}$. The slope of the weighted linear fit and its 1$\sigma$ uncertainty and the Spearman's rank correlation test $\rho$ and p-values are given inside each panel. {\tt PL\_SIC} models have had their $\alpha^{model}$ values increased by 0.25 to be comparable due to their model construction. We do not include the outlier source GLEAM J003652-333315 in our statistical analysis.}
    \label{fig:amodIR}
\end{figure*}

Lastly we compare the modelled spectral index to a number of IR derived physical properties in Figure. \ref{fig:amodIR}. In panels (a) and (c) we see that there is no significant correlation between the modelled spectral index and the mid-IR + UV derived SFR or sSFR. We observe no significant separation between the LFTO and control or {\tt FFA\_PL} and non-{\tt FFA\_PL} samples but can clearly see the flatter spectral indices in merging systems as well as their elevated sSFRs. This suggests mergers can trigger new starburst activity which raises the sSFR and injects new CREs with a flatter spectral index \citep{Murphy2009, Murphy2013, Donevski2015}. In panels (b) and (d) we examine the relationship between the GLEAM and modelled spectral indices and $q_{FIR}$ and find that while there is a positive correlation it is not statistically significant above the 5\% level. Panel (b) shows the most significant separation between LFTO and control sources whereby LFTO sources occupy the top right region of the plot having flatter GLEAM spectral indices and higher $q_{FIR}$ values. This provides evidence that the spectral flattening at low frequencies is more strongly tied to the ISM properties of the galaxy than the high frequency spectral index ($\alpha^{model}$).

We present P-values for Welch's T-tests between the LFTO and control, {\tt FFA\_PL} and non-{FFA\_PL}, and, merger and non-merger samples for a number of parameters in Table. \ref{tab:ttest}. We verify that, by selection, the LFTO and control samples have $\alpha_{\rm L}$ mean values that are different at a 1\% significance level. This separation also is observed in the {\tt FFA\_PL} and non-{\tt FFA\_PL} samples (see Figure. \ref{fig:alphacomps}) as SFGs with the flattest $\alpha_{\rm L}$ have most preferred models which include the FFA. We find a 5\% significant separation between LFTO and control galaxies in M$_*$ and SFR$_{1.4GHz}$ with control galaxies having higher values and low variance. This is likely a selection and small sample effect as we see that LIRGs/ULIRGs with high stellar mass and SFRs often contain spectral curvature \citep{Galvin2018, Dey2022, Dey2024}. We find a separation between LFTO and control galaxies in $q_{FIR}$ which does not carry through to the {\tt FFA\_PL} and non-{\tt FFA\_PL} samples. This suggests that the a clearly modelled LFTO doesn't necessarily indicate an elevated $q_{FIR}$ value but SFGs with spectral flattening generally have a higher $q_{FIR}$ as shown in Figure. \ref{fig:amodIR}. We find a separation between merging and non-merging SFGs in $\alpha_{A}$ and $\alpha^{model}$ (at P < 0.05) with merging systems having flatter spectral indices as seen in Figure. \ref{fig:alphacomps}. It has been shown that mergers can trigger star formation flattening the spectral index \citep{Donevski2015} and increasing the sSFR \citep{Murugeshan2021} which shows a 5\% significant separation in sSFR$_{mircor}$.

\begin{table}[hbt!]
\centering
\begin{threeparttable}
\caption{T-Test p-values between samples}\label{tab:ttest}
\begin{tabular}{c|c|c|c}
	\toprule
	Parameter & LFTO & {\tt FFA\_PL} & merger\\
    Compared  & Control & non-{\tt FFA\_PL} & non-merger\\
	\midrule
	$\alpha^{\rm GLEAM}_{\rm low}$ & \cellcolor[HTML]{9cff99}<0.01 & \cellcolor[HTML]{9cff99}<0.01 & 0.88\\
    $\alpha^{\rm RACS}_{\rm mid}$ & 0.15 & 0.13 & 0.40\\
    $\alpha^{\rm ATCA}_{\rm high}$ & 0.24 & 0.80 & \cellcolor[HTML]{e2e2e2}0.02\\
	$\alpha^{model}$ & 0.22 & 0.82 & \cellcolor[HTML]{e2e2e2}0.05\\
    M$_*$ & 0.14 & 0.80 & 0.39 \\
    SFR$_{mircor}$ & 0.13 & 0.80 & 0.96\\
    sSFR$_{mircor}$ & 0.80 & 0.96 & \cellcolor[HTML]{e2e2e2}0.05\\
    SFR$_{1.4GHz}$ & \cellcolor[HTML]{e2e2e2}0.03 & 0.62 & 0.91\\
    $\Sigma$SFR$^{radio}_{total}$ & 0.27 & 0.76 & 0.33\\
    Inclination & 0.50 & 0.68 & - \\
    $q_{FIR}$ & \cellcolor[HTML]{9cff99}<0.01 & 0.13 & 0.92\\
	\bottomrule
\end{tabular}
\begin{tablenotes}[para] Welch's T-test p-values with columns indicating the two samples compared after removal of outlier source GLEAM J003652-333315. Samples which reject the null hypothesis that their means are the same at the 1\% and 5\% level are highlighted in green and grey respectively. We do not compare inclinations between merging and non-merging systems as measuring the inclination of merging systems with axes light ratios is unreliable.
\end{tablenotes}
\end{threeparttable}
\end{table}

\section{Discussion}
\label{sec:discussion}

\subsection{SED modelling}

Accurately modelling the physical processes which occur in SFGs and how they manifest in the radio continuum emission observed is a complex problem caused primarily by the lack of accurate radio data across a wide frequency range. We have observed that over the last few decades the preferred radio SED models have gone from simple single power laws to models encompassing thermal and non-thermal emission \citep{pacholczyk1970, Condon1992} with loss or absorption processes occasionally being invoked in sources which are not well modelled by simple power laws \citep{Clemens2010, Galvin2018, Dey2022, Dey2024}. The limitations in this research follow a similar vein such that it is difficult to conclusively confirm whether the features observed in our radio SEDs are the result of these loss and absorption processes or due to lack of high quality data. For example, of the initial sample of 427 SFGs in the GLEAM-6dFGS catalogue, 54 have negative flux values in at least one subband, whilst the remainder may have photometric errors which can introduce a false sense of curvature in the radio SED. Our sample selection aimed to mitigate these issues by selecting brighter sources with distinct LFTOs such that the photometry is more reliable, however it is likely that more accurate observations with GLEAM-X and the SKA will reveal a larger population of SFGs with LFTOs highlighting the importance of including physical loss and absorption mechanisms during radio SED modelling. Table.~\ref{tab:modelsel} shows that synchrotron only, {\tt PL} based models are most preferred for all of our sources (except J1847) despite the consensus that radio emission in SFGs is due to both non-thermal synchrotron and thermal free-free emission \citep{Condon1992}. We attribute this to lack of high frequency radio data and limitations of the modelling framework as we discuss below. 
 
Thermal free-free emission generally does not contribute significantly to radio emission at low frequencies contributing between 1-10\% at 1.4\,GHz \citep{Condon1992} however more recent work has been leaning towards 2-4\% \citep{Galvin2018, Dey2022, Dey2024}. The flat spectral index of thermal emission is also typically thought to become dominant above $\sim$30\,GHz \citep[e.g. in M82;][]{Condon1992} but again with recent work we find that in other starburst galaxies this is likely an overestimation with \cite{Galvin2018, Dey2022, Dey2024} often finding values of TF$_{40GHz}$ < 0.5. There is a large scatter in thermal fractions which is linked to the star formation timescales with thermal emission being tied to instantaneous star formation with timescales of $\sim$10\,Myr \citep{Condon1992} whilst non-thermal emission is typically delayed and has a timescale of > 30\,Myr \citep{Condon1992, Vollmer2020, heesen2024}. Thus starbursts with increasing or decreasing star formation rates will result in different thermal fractions \citep[and radio SFR estimates; ][]{Cook2024} emphasising the importance of knowing the SFR histories of SFGs. We will explore the impact of SFR histories and their connection to thermal fractions, radio SED curvature and the FRC/TRC in a future paper in this series.

At our current highest frequency point at 17\,GHz the flattening we would expect to see at higher frequencies is typically not observed meaning that the thermal fractions are likely low. This, coupled with the fact that the number of free parameters in our models are currently low without the inclusion of GLEAM flux covariance modelling and FIR emission, means that the addition of model complexity by including a thermal component has a huge impact on the selected model. For example the difference between a model with 2 and 3 free parameters is far greater than the difference between SED models with 10 and 11 free parameters. Thus because we do not observe this spectral flattening and including unconstrained thermal emission in our models increases model complexity a relatively large amount we do not favour {\tt SFG} based models despite their more physically realistic nature. The increased spectral sampling afforded by ALMA 40 and 115\,GHz \footnote{Approved cycle 10 ALMA observations (project code: 2023.1.01342.S) for 16 of these sources completed in 2023 and 2024.} and FIR observations compounded with the increased model complexity when including covariance and FIR emission models would likely result in {\tt SFG} based models being favoured in a future paper in this series. This may also cause the measured model spectral indices to be steeper and more comparable to the work of \citet{Galvin2018} especially for sources with a higher thermal fraction indicating a higher contribution of synchrotron or IC based loss processes.

Our modelled synchrotron spectral index values lie within the range of values found in \cite{Galvin2018} and \cite{Dey2022, Dey2024} with them being found to be, on average, steeper than the canonical value assumed for SFGs $\alpha$ = -0.8 \citep{Condon1992} especially when considering models which account for spectral steepening ({\tt PL\_FFA} and {\tt PL\_SIC}). {\tt PL\_SIC} models are competitive with the pure {\tt PL} model in 13 of the 19 SFGs in our sample indicating that synchrotron losses are potentially an important physical loss mechanism. This mechanism should be considered during radio SED modelling above 1.4\,GHz, especially for large, massive galaxies. One outcome of this is that higher frequency flux estimates derived from two point spectral indices at low frequencies will likely be overestimated and the spectral index used for radio K-corrections may have a mass dependence. Interestingly however {\tt PL\_SIC} models which are made to take into spectral steepening at high frequencies are less likely to be completely preferred than {\tt PL\_FFA} models which have an extra free component. This is likely due to the inability to constrain the break frequency when steepening is so gradual. Despite this we speculate along the same lines as \cite{Dey2022, Dey2024} that it is likely to be synchrotron losses causing this steepness rather than IC losses or FFA.

Over half of our sources are best fit by models including FFA whether at low or high frequencies which supports the growing evidence for more complex radio SED model requirements. The inclusion of FFA accounts for both curvature at low frequencies or kinks at higher frequencies which are often evident in well sampled radio SEDs. In Section. \ref{sec:res2} see no distinct separation in global properties (except $q_{FIR}$) for SFGs which contain either low or high frequency FFA suggesting that the regions in which FFA is occurring are being averaged out and require a resolved analysis. It was also found that a number of our SFG galaxy sample that were initially selected as having LFTOs are best modelled without the inclusion of the {\tt FFA\_} prefix model meant to model this spectral curvature. This is attributed to the large errors on the GLEAM flux densities for some sources which results in simpler models being preferred over the increased complexity added by including a relatively unconstrained FFA turnover. 

Ionisation losses are also theorised to be a significant loss process for low energy non-thermal emission in SFGs \citep{Lacki2010, Longair2011, Basu2015, roth2023, roth2024} which under steady continuous injection of electrons act to gradually flatten the low frequency spectral index by $\Delta \alpha$ = +0.5 towards $\alpha$ = -0.1 \citep{Basu2015}. However for single injection events the resulting spectrum is dependent on the ratio of timescales between injection and ionisation loss \citep{Basu2015}. We chose not to include ionisation losses in this work as the shape of this model and its rate of curvature is currently unconstrained however recent work suggests ionisation losses do play a key role in flattening the spectral index of galaxies, particularly those with high SFRs \citep{roth2023,roth2024}. As this is the case we cannot currently disentangle the contributions of ionisation losses and FFA to the radio SEDs of our sources however only absorption processes allow for a decrease in the measured flux as we move to lower frequencies (i.e. $\alpha$ > 0) so it can still be safely concluded that FFA is occurring in some of these SFGs. 

We also do not include a number of other loss processes within our models as they do not significantly impact the observed spectral index these include adiabatic losses, diffusion losses and bremsstrahlung losses. These three loss mechanisms play a role in the observed radio emission of SFGs over five decades in SFR but due to their relative independence with electron energy in the non-thermal radio regime they do not act to modify the spectral index at different frequencies and result in a constant power-law spectral index with $\alpha$ = -0.6 \citep{Lacki2010, roth2023, roth2024}.

Lastly spectral curvature in the radio SED can also be caused by synchrotron self absorption or the Tsytovitch-Razin effect when the refractive index of the medium is less than unity \citep{israel1990}. Both these effects however occur below the observed frequency range of our GLEAM observations when considering the peak flux and size of our sources or would require unreasonably high magnetic field strengths \citep{kellerman1969} so they can be safely ruled out as the cause of LFTOs in our sample following \citep{Dey2024}.

\subsection{Global Properties}

We look to connect the features of the radio SED to the global properties and allow us to predict the physical processes occurring within SFGs for unresolved galaxies in future large scale radio surveys. With the improved frequency sampling and sensitivity afforded by future surveys we will begin to be able to disentangle the dominant cooling and loss mechanisms and infer the ISM properties of SFGs. We find that the parameter most strongly correlated with the modelled spectral index is the stellar mass whereby more massive galaxies have steeper spectral indices, this however is a secondary effect as it is in fact galactic size which causes this correlation with more massive galaxies tending to be larger and more star-forming \citep{Gurkan2018}. Due to the diffusion scale length and lifetime of CREs larger galaxies are able to retain electrons as they undergo synchrotron losses with higher energy CREs losing energy faster than low energy CREs. This causes the spectral steepness we observe particularly above 1.4\,GHz. This effect is more pronounced in our SFG sample than those of \cite{Heesen2022a} and \cite{Dey2024} as shown by our steeper relationship between $\alpha^{model}$ and $M_{*}$ (see Figure. \ref{fig:alphamodzmass}) with SFGs that have preferred models that account for the spectral steepening ({\tt \_SIC} and {\tt \_FFA}) being generally higher mass. However it is important to note that the spectral index measured in \cite{Heesen2022a} is a two point spectral index at a lower frequency so it does not account for spectral curvature and will be intrinsically flatter than our modelled spectral index. Overall this suggests that synchrotron losses are the primary driver for spectral steepness above 1.4\,GHz for larger nearby SFGs.

Typically the impact of IC losses is unable to be disentangled from synchrotron losses as they result in a similar spectral shape and their loss timescales have similar CRE energy dependence. IC losses due to CMB photons however primarily occur at high redshifts in low density galaxies due to increased CMB photon density \citep{Lacki2010, klein2018}. We however probe low redshift galaxies and find no statistically significant correlation to the spectral index such that we can likely rule out the impact of CMB based IC losses for our sample. \citet{Dey2024} do not find a correlation with redshift out to z$\sim$0.4 due to small sample size and that they also do not probe out to sufficient redshifts for the CMB photon density to become significant. \cite{magnelli2015} and \cite{An2024} also do not find evidence for IC losses finding flat constant (two point) spectral indices out to z$\sim$2, however the spectral indices probed in these papers are below 1.4\,GHz where IC losses have long timescales and so the lack of evidence for IC losses are unsurprising. Another possibility is that IC losses are occurring due to high FIR photon energy densities in optically thick starburst galaxies; \cite{Lacki2010} explore this but find these galaxies begin to violate the FRC, whilst \cite{Basu2015} and \cite{roth2024} show that IC losses are more important for low density systems. A potential test that could be performed to see whether IC losses in starbursts play a role in steepening the radio SED is by testing whether we see enhanced X-ray emission in starbursts which has been upscattered, however disentangling this emission from other sources of X-ray emission is currently not feasible. IC losses will however play a role in steepening the observed radio SED at high frequencies so will need to be considered during high redshift radio surveys particularly above rest frame 1.4\,GHz, and,  radio K-corrections which rely on the radio spectral index above 1.4\,GHz will likely need to have a frequency dependence which acts to steepen the spectral index at higher frequencies.

One would expect however that the spectral index would flatten with increasing SFR as there is a constant injection of young electrons, or that lower SFR galaxies would have a higher proportion of aged CREs. The relationship observed between modelled spectral index and 1.4\,GHz radio SFR (which is a delayed tracer) of star formation however presents the opposite effect (see Figure. \ref{fig:SFRD}). This result is consistent with \cite{Heesen2022a}, \cite{Galvin2019}. and \cite{Dey2022, Dey2024} which all rely on more instantaneous IR SFR tracers. However, as suggested before, this trend is likely a secondary effect to galaxy size or mass with the correlation of spectral index being stronger and more significant with both of these parameters. This observed global correlation of steepening spectral index with increasing SFR also breaks down within galaxies as we see the higher SFR or SFR surface density regions are those with the flattest spectral index \citep{Tabatabaei2017, Heesen2022a}. This global averaging also likely explains the lack of correlation between spectral index and total SFR surface density for our sample, similarly to Figure. 7 in \cite{Heesen2022a} which shows that size is the driving factor. We do however find a slightly stronger correlation between sSFR and spectral index with higher sSFR \citep[related to source compactness][]{Elbaz2011} sources having flatter spectral index, consistent with \cite{Murphy2013} and \cite{An2021}. This again is consistent with larger more massive galaxies having a steeper spectral index due to retaining aging CREs for longer timescales such that they can undergo synchrotron losses.

The only parameter with a large separation between the LFTO and control samples is $q_{FIR}$ (see Figure. \ref{fig:qplots} panel (b)) which implies there may be differences in their SFR timescales, emission sources or ISM properties. Curvature in the low frequency spectral index is also related to the ISM properties of galaxies with both FFA and ionisation losses being dependent on ionised and neutral gas densities respectively \citep{Lacki2010}. This separation then suggests that these LFTO sources that have higher $q_{FIR}$ and flattened spectral indices have different ISM or star formation properties than the control sources. We can likely rule out AGN emission as a reason for this separation as we have no evidence for AGN emission in their optical spectra or significant $q_{FIR}$ deficits. The spectral flattening and turnovers in the LFTO sample imply a higher gas density of ionised gas which is related to HMS formation and/or neutral gas which is the fuel for star formation in both molecular clouds and HI reservoirs. Thus we would expect these galaxies to have high instantaneous SFRs (as the HMS lifetime is short) and/or be gas rich. We unfortunately do not have resolved HI or H$\alpha$ observations for most of these sources so we can not currently determine whether they have enhanced gas densities\footnote{although this will be explored in a future paper in this series as our ALMA observations include J=1-0 CO molecular gas spectral line observations}. 

We do have evidence of these galaxies having high instantaneous SFRs, in their high mid-IR SFRs and their elevated $q_{FIR}$ values. Elevated $q_{FIR}$ values are a result of an excess of FIR compared to radio emission which would be the case in young starbursts or galaxies with increasing SFRs \citep{Galvin2019, Cook2024} as the IR emission is produced 10\,Myr before the synchrotron dominated 1.4\,GHz radio emission. These differences in SFR indicator timescales and contamination by AGN emission are partly responsible for scatter in the $q_{FIR}$ relationship \citep{Galvin2019, Molnar2021} and would be somewhat mitigated by considering the short timescale thermal emission at 40\,GHz (however this introduces errors with radio SED decomposition or H$\alpha$ scaling relationships). Overall this paints a picture of recent compact starbursts in regions of high gas density being the cause of the spectral curvature and flattening in the LFTO sample. Thus comparisons between resolved spectral index, FRC and gas density maps as well as measuring SFR histories are vital to understanding complex radio SED features and will be explored in future papers in this series.

\subsection{Inclination and Mergers}

In Section \ref{sec:morph} we find no relationship between inclination and either modelled or GLEAM spectral index for our SFG sample. This suggests that the spectral index is not dependent on the viewing angle of galaxies and that spectral curvature or flattening is not an effect of looking through the galactic disk. \cite{hummel1991} and \cite{Heesen2022a} also find no connection between the low frequency spectral index and inclination. For FFA to occur a sufficient electron density along a line of sight is required, these dense ionised regions typically occur within individual star forming molecular clouds which exist throughout the galactic disk. It is then the superposition of these individual regions within the large synthesised GLEAM beam that causes us to observe this spectral curvature. Interestingly if we remove the merging systems from Figure. \ref{fig:incplots} panel (b) we see that the modelled spectral index (which is most strongly related to $\alpha^{\rm ATCA}_{\rm high}$) is steeper for edge on galaxies. This may be an effect of ATCA sensitivity or CRE diffusion into the halo and would be the case if off-planar radio emission from a radio halo due to CRE transport was preferentially detected at low frequencies \citep{Vollmer2020} due to their longer lifetimes \citep{heesen2024}.

As a number of sources in our SFG sample are merging we investigate the effects of mergers on the radio SED and other global properties. Mergers have been shown to enhance the sSFR of galaxies as turbulence and gas inflows can trigger star formation \cite{Ellison2018}. These triggered starbursts inject young CREs with flatter spectral indices which have not had the time to undergo synchrotron cooling. We find that the merging systems in our sample have systematically higher sSFRs and flatter spectral indices both modelled and at higher frequencies indicating that they may have a younger population of CREs or enhanced thermal fractions. Four of these merging systems are members of the LFTO sample supporting the evidence that merger driven starbursts can result in FFA and cause spectral flattening or curvature at low frequencies.

\section{Summary and Conclusions}
\label{sec:conclusion}
We created a sample of 19 GLEAM selected SFGs of which 11 display LFTOs at GLEAM frequencies and eight do not. These SFGs are observed with ATCA between 5.5 and 17\,GHz. Their radio SEDs are then modelled between 71\,MHz and 17\,GHz using a modular radio continuum models within a Bayesian framework which encompass combinations of thermal and non-thermal (power-law, {\tt PL}) emission processes as well as FFA, synchrotron and IC losses. We find that:
\begin{enumerate}
    \item {\tt PL} based models are preferred for our entire sample of SFGs over models containing thermal free-free emission due to lack of high frequency radio data displaying spectral flattening and the relatively large increase in model complexity being unfavoured.
    \item Of the originally selected 11 LFTO galaxies five favour SED models with no FFA based LFTO which is attributed to large uncertainties in GLEAM flux densities resulting in the inability to accurately constrain the turnover frequency.
    \item Radio SED models with spectral ``kinks'' or gradual steepening are competitive with simple {\tt PL} only models.
    \item The radio emission is most strongly related to the stellar mass (correlated to physical size) of these galaxies such that more massive or larger galaxies generally have higher SFRs and retain synchrotron emitting CREs for longer steepening the spectral index and lowering $q_{FIR}$.
    \item As both GLEAM spectral index and $q_{FIR}$ depend on SFR timescale with short lived HMSs being responsible for increasing the ionised gas density and FIR emission it is likely that LFTOs are transient and dependent on recent starburst activity.
    \item The merger systems in our sample have elevated sSFRs and flatter spectral indices than our non-merging SFGs indicating that they are undergoing recently triggered starburst activity. Four (of five) are members of the LFTO sample with three of these sources having elevated $q_{FIR}$ values suggesting that LFTOs may be caused by recently triggered compact starbursts.
    \item We find no relationship between the inclination and whether a galaxy contains an LFTO suggesting FFA occurs within specific star-forming regions rather than because we are looking through the galactic disk. Galaxies with favoured models that include synchrotron and IC losses tend to be the most inclined suggesting that CRE diffusion into the galactic halo is responsible for the steep spectral index.
\end{enumerate}
Overall we show that for our sample single power-law radio SED models are often appropriate to model radio continuum emission from SFGs especially with limited sampling below 300\,MHz and above 10\,GHz. However with improved sampling, especially at low and high frequencies more complex models including loss and absorption processes are often needed to explain the spectral features observed. We also find that no specific global properties act as a predictor of LFTOs for our sample suggesting that resolved observations are required to measure the physical properties in specific star-forming regions in which FFA may be occurring. The relationship between the radio SED and resolved emission properties and star-formation history will be explored in future papers in this series.

\section*{Acknowledgements}

 We would like to thank the late Tom Jarrett for his help with the WISE components of this work and valuable discussion as well as contributions to the field overall. We also thank the anonymous referee for their valuable comments and suggestions to improve the quality and accuracy of this manuscript.
 
 This scientific work uses data obtained from Inyarrimanha Ilgari Bundara / the Murchison Radio-astronomy Observatory. We acknowledge the Wajarri Yamaji People as the Traditional Owners and native title holders of the Observatory site. CSIRO’s ASKAP radio telescope is part of the Australia Telescope National Facility (https://ror.org/05qajvd42). Operation of ASKAP is funded by the Australian Government with support from the National Collaborative Research Infrastructure Strategy. ASKAP uses the resources of the Pawsey Supercomputing Research Centre. Establishment of ASKAP, Inyarrimanha Ilgari Bundara, the CSIRO Murchison Radio-astronomy Observatory and the Pawsey Supercomputing Research Centre are initiatives of the Australian Government, with support from the Government of Western Australia and the Science and Industry Endowment Fund. The Australia Telescope Compact Array is part of the Australia Telescope National Facility (https://ror.org/05qajvd42) which is funded by the Australian Government for operation as a National Facility managed by CSIRO. This paper includes archived data obtained through the CSIRO ASKAP Science Data Archive, CASDA (http://data.csiro.au). 

This publication makes use of data products from the Wide-field Infrared Survey Explorer, which is a joint project of the University of California, Los Angeles, and the Jet Propulsion Laboratory/California Institute of Technology, funded by the National Aeronautics and Space Administration.
 
 This research has made use of the Astronomical Society of Australia's Student Travel Assistance Scheme which facilitated international conference attendance and valuable discussions. 
 
 This research has made use of the NASA/IPAC Extragalactic Database (NED), which is operated by the Jet Propulsion Laboratory, California Institute of Technology, under contract with the National Aeronautics and Space Administration. This paper includes data that has been provided by AAO Data Central (datacentral.org.au).


\section*{Data Availability}

The derived data produced in this work can be found in the article and supplementary material, or can be shared upon reasonable request to the corresponding author.

%% file: appendix.tex
\section{Radio Flux Densities}

\begin{table*}[hbt!]
\centering
\begin{threeparttable}
\caption{ATCA Radio Fluxes.}\label{tab:radiomeasured2}
\begin{tabular}{c|c|c|c|c|c|c|c|c}
	\toprule
	Source & ATCA & ATCA & ATCA & ATCA & ATCA & ATCA & ATCA & ATCA\\
    Frequency (GHz) & 5.47 & 4.87 & 5.48 & 6.14 & 9.47 & 9.00 & 9.94 & 16.93\\
    Bandwidth (GHz) & 1.97 & 0.66 & 0.66 & 0.66 & 1.97 & 0.98 & 0.98 & 1.97\\
     & (mJy) & (mJy) & (mJy) & (mJy) & (mJy) & (mJy) & (mJy) & (mJy)\\
    \midrule
        GLEAM J002238-240737 & $7.97\pm0.90$ & $8.82\pm1.04$ & $7.56\pm0.82$ & $7.17\pm0.83$ & $4.14\pm0.56$ & $4.40\pm0.61$ & $3.76\pm0.44$ & $2.29\pm0.65$ \\
		GLEAM J003652-333315 & $14.46\pm1.50$ & $15.29\pm1.59$ & $14.25\pm1.45$ & $13.31\pm1.35$ & $10.43\pm1.19$ & $10.57\pm1.22$ & $9.98\pm1.17$ & $8.56\pm1.52$ \\
		GLEAM J011408-323907 & $12.68\pm1.55$ & $14.05\pm1.60$ & $12.08\pm1.38$ & $10.38\pm1.11$ & $6.73\pm1.19$ & $6.60\pm0.82$ & $6.07\pm0.86$ & $3.80\pm0.65$ \\
		GLEAM J012121-340345 & $9.27\pm1.06$ & $10.29\pm1.26$ & $9.57\pm1.17$ & $8.07\pm1.07$ & $4.52\pm1.07$ & $5.00\pm0.94$ & $4.10\pm0.84$ & $3.68\pm1.15$ \\
		GLEAM J034056-223353 & $10.57\pm1.16$ & $10.67\pm1.31$ & $10.38\pm1.12$ & $9.36\pm1.03$ & $6.31\pm0.85$ & $6.26\pm1.02$ & $5.97\pm0.85$ & $4.17\pm0.72$ \\
		GLEAM J035545-422210 & $10.39\pm1.33$ & $10.60\pm1.21$ & $9.99\pm1.28$ & $8.85\pm1.01$ & $5.09\pm0.99$ & $6.07\pm1.30$ & $4.76\pm1.06$ & $3.97\pm0.70$ \\
		GLEAM J040226-180247 & $7.97\pm0.85$ & $7.97\pm0.85$ & $8.01\pm0.91$ & $7.99\pm0.83$ & $4.99\pm0.72$ & $5.68\pm1.13$ & $4.66\pm1.08$ & $2.65\pm0.57$ \\
		GLEAM J041509-282854 & $17.39\pm1.97$ & $18.00\pm1.99$ & $16.84\pm1.93$ & $15.26\pm1.64$ & $11.49\pm1.53$ & $11.68\pm1.65$ & $11.06\pm1.83$ & $6.89\pm1.42$ \\
		GLEAM J042905-372842 & $8.33\pm0.91$ & $10.15\pm1.12$ & $8.40\pm0.85$ & $6.50\pm0.68$ & $5.23\pm0.71$ & $5.17\pm0.64$ & $4.84\pm0.75$ & $3.58\pm0.64$ \\
		GLEAM J072121-690005 & $15.83\pm1.82$ & $16.22\pm2.15$ & $14.79\pm1.90$ & $13.22\pm1.52$ & $10.91\pm1.29$ & $10.32\pm1.19$ & $10.73\pm1.35$ & $4.97\pm1.03$ \\
		GLEAM J074515-712426 & $12.24\pm1.66$ & $12.85\pm1.66$ & $10.98\pm1.30$ & $10.76\pm1.73$ & $7.58\pm1.17$ & $7.62\pm0.97$ & $7.08\pm0.95$ & $3.52\pm0.69$ \\
		GLEAM J090634-754935 & $14.53\pm1.58$ & $15.37\pm1.63$ & $13.88\pm1.53$ & $12.97\pm1.74$ & $7.68\pm1.26$ & $8.22\pm1.55$ & $6.90\pm0.97$ & $3.39\pm0.41$ \\
		GLEAM J120737-145835 & $9.66\pm1.05$ & $10.86\pm1.28$ & $9.09\pm1.02$ & $8.28\pm0.90$ & $5.67\pm0.75$ & $5.58\pm0.59$ & $5.50\pm0.66$ & $3.57\pm0.76$ \\
		GLEAM J142112-461800 & $17.39\pm2.04$ & $18.14\pm2.32$ & $16.75\pm1.93$ & $15.11\pm1.58$ & $10.39\pm1.55$ & $11.14\pm1.62$ & $9.71\pm1.23$ & $6.08\pm0.69$ \\
		GLEAM J150540-422654 & $8.76\pm0.99$ & $9.11\pm1.01$ & $8.80\pm1.07$ & $7.13\pm0.81$ & $5.08\pm0.79$ & $4.48\pm0.82$ & $4.41\pm0.54$ & $3.26\pm0.91$ \\
		GLEAM J184747-602054 & $10.23\pm1.04$ & $10.79\pm1.12$ & $10.07\pm1.04$ & $9.09\pm0.95$ & $6.46\pm0.73$ & $6.65\pm0.75$ & $6.14\pm0.68$ & $4.40\pm0.70$ \\
		GLEAM J203047-472824 & $13.94\pm1.43$ & $14.81\pm1.55$ & $13.67\pm1.40$ & $12.58\pm1.28$ & $9.26\pm1.14$ & $9.30\pm1.05$ & $8.61\pm0.99$ & $5.72\pm0.65$ \\
		GLEAM J205209-484639 & $17.74\pm2.02$ & $18.68\pm1.96$ & $17.17\pm1.81$ & $15.17\pm1.72$ & $9.23\pm1.21$ & $9.39\pm1.78$ & $8.83\pm1.20$ & $6.61\pm1.03$ \\
		GLEAM J213610-383236 & $11.11\pm1.17$ & $12.06\pm1.34$ & $10.96\pm1.14$ & $9.84\pm1.01$ & $7.17\pm0.84$ & $7.35\pm0.91$ & $6.89\pm0.82$ & $5.47\pm0.86$ \\
   \bottomrule
\end{tabular}
\begin{tablenotes}[para]
    \item[]Note: Column (1): GLEAM catalogue designation. Column (2): WXSC designation. Column (3-6): WXSC W1, W2, W3 and W4 band integrated flux density respectively. Column (7): IRAS 60\,$\mu$m integrated flux density. Column (8): IRAS 100\,$\mu$m integrated flux density.
\end{tablenotes}
\end{threeparttable}
\end{table*}

\begin{sidewaystable}[hbt!]
\centering
\begin{threeparttable}
\caption{Archival Radio Fluxes.}\label{tab:radiomeasured1}
\begin{tabular}{c|c|c|c|c|c|c|c|c|c|c|c}
	\toprule
	Source & GLEAM & GLEAM & GLEAM & GLEAM & GLEAM & GLEAM & SUMSS & RACS-low & RACS-mid & NVSS & VLA\\
    Frequency (MHz) & 200.0 & 87.7 & 118.4 & 154.0 & 185.0 & 215.7 & 843 & 888 & 1368 & 1400 & 8460\\
    Bandwidth (MHz) & 61.4 & 30.7 & 30.7 & 30.7 & 30.7 & 30.7 & 3 & 288 & 144 & 42 & 50\\
     & (mJy) & (mJy) & (mJy) & (mJy) & (mJy) & (mJy) & (mJy) & (mJy) & (mJy) & (mJy) & (mJy)\\
    \midrule
		GLEAM J002238-240737 & $94.3\pm1.7$ & $173\pm42$ & $130\pm15$ & $102.6\pm8.1$ & $104.4\pm5.7$ & $88.5\pm4.7$ &   & $33.6\pm4.2$ & $23.8\pm3.1$ & $26.6\pm3.9$ &   \\
		GLEAM J003652-333315 & $65.0\pm4.6$ & $47\pm40$ & $69\pm15$ & $65.7\pm7.8$ & $61.9\pm5.3$ & $58.6\pm5.0$ & $33.4\pm5.4$ & $38.9\pm4.4$ & $28.4\pm3.3$ & $26.8\pm3.6$ & $10.1\pm1.2$ \\
		GLEAM J011408-323907 & $101.6\pm2.5$ & $201\pm60$ & $156\pm19$ & $129.6\pm7.7$ & $104.3\pm5.0$ & $96.6\pm5.0$ & $46.2\pm7.0$ & $48.8\pm6.2$ & $33.6\pm4.1$ & $34.4\pm5.2$ &   \\
		GLEAM J012121-340345 & $86.1\pm1.9$ & $152\pm30$ & $131\pm12$ & $100.3\pm5.9$ & $91.5\pm4.4$ & $81.8\pm4.4$ & $49.9\pm7.4$ & $42.4\pm5.4$ & $27.1\pm3.5$ & $28.1\pm4.3$ &   \\
		GLEAM J034056-223353 & $83.7\pm1.3$ & $99\pm46$ & $90\pm16$ & $84.1\pm8.7$ & $82.3\pm6.7$ & $84.9\pm6.2$ &   & $35.5\pm4.2$ & $26.7\pm3.5$ & $25.3\pm3.7$ &   \\
		GLEAM J035545-422210 & $74.0\pm2.9$ & $85\pm50$ & $90\pm22$ & $74.0\pm13.1$ & $76.3\pm8.3$ & $72.0\pm7.9$ & $31.5\pm4.7$ & $35.8\pm6.1$ & $19.9\pm3.0$ &   &   \\
		GLEAM J040226-180247 & $83.2\pm6.0$ & $250\pm48$ & $117\pm18$ & $96.2\pm10.4$ & $81.2\pm7.9$ & $79.3\pm7.0$ &   & $37.7\pm4.7$ & $30.2\pm4.4$ & $24.8\pm3.4$ &   \\
		GLEAM J041509-282854 & $100.0\pm2.4$ & $177\pm55$ & $133\pm19$ & $117.0\pm12.3$ & $105.5\pm8.8$ & $95.7\pm8.7$ &   & $59.0\pm7.1$ & $43.8\pm5.8$ & $39.1\pm5.5$ &   \\
		GLEAM J042905-372842 & $67.4\pm2.9$ & $84\pm64$ & $94\pm24$ & $83.1\pm13.1$ & $68.8\pm8.6$ & $63.4\pm7.4$ & $39.2\pm6.1$ & $37.3\pm4.7$ & $26.3\pm3.5$ & $24.9\pm3.7$ &   \\
		GLEAM J072121-690005 & $142.2\pm11.9$ & $133\pm77$ & $154\pm35$ & $144.4\pm20.4$ & $140.0\pm22.7$ & $143.0\pm21.4$ & $65.3\pm10.2$ & $66.7\pm8.1$ & $46.9\pm6.3$ &   &   \\
		GLEAM J074515-712426 & $205.1\pm16.6$ & $221\pm100$ & $213\pm47$ & $168.8\pm31.6$ & $212.5\pm31.9$ & $189.8\pm29.1$ & $62.2\pm9.8$ & $63.0\pm7.3$ & $43.5\pm5.6$ &   &   \\
		GLEAM J090634-754935 & $167.5\pm11.2$ & $211\pm80$ & $214\pm43$ & $195.2\pm21.4$ & $171.2\pm18.6$ & $149.6\pm17.2$ & $63.7\pm10.3$ & $69.7\pm8.4$ & $50.4\pm7.1$ &   &   \\
		GLEAM J120737-145835 & $101.9\pm8.1$ & $72\pm58$ & $95\pm38$ & $121.5\pm30.1$ & $97.5\pm19.9$ & $92.1\pm17.6$ &   & $41.6\pm5.1$ & $26.3\pm3.3$ & $27.3\pm3.9$ &   \\
		GLEAM J142112-461800 & $130.1\pm20.5$ & $172\pm119$ & $60\pm87$ & $145.6\pm50.4$ & $137.7\pm40.7$ & $123.2\pm35.9$ & $61.2\pm8.3$ & $68.3\pm10.9$ & $40.9\pm5.6$ &   &   \\
		GLEAM J150540-422654 & $111.0\pm24.6$ & $90\pm145$ & $96\pm87$ & $73.7\pm50.8$ & $111.7\pm40.3$ & $105.6\pm28.4$ & $42.3\pm6.2$ & $37.2\pm5.4$ & $24.8\pm3.8$ &   &   \\
		GLEAM J184747-602054 & $117.1\pm22.4$ & $142\pm81$ & $108\pm45$ & $127.7\pm34.1$ & $113.7\pm34.4$ & $107.8\pm54.2$ & $59.0\pm7.9$ & $44.8\pm5.1$ & $30.5\pm3.5$ &   &   \\
		GLEAM J203047-472824 & $125.2\pm14.7$ & $79\pm78$ & $144\pm38$ & $110.6\pm25.8$ & $109.0\pm23.7$ & $134.1\pm31.6$ & $53.0\pm7.1$ & $49.2\pm5.5$ & $36.3\pm4.0$ &   &   \\
		GLEAM J205209-484639 & $207.4\pm13.2$ & $328\pm79$ & $257\pm41$ & $237.7\pm27.0$ & $222.5\pm23.0$ & $182.4\pm27.4$ & $68.8\pm9.3$ & $79.2\pm9.2$ & $56.5\pm6.9$ &   &   \\
		GLEAM J213610-383236 & $58.9\pm4.4$ & $45\pm64$ & $77\pm17$ & $78.1\pm9.7$ & $66.3\pm6.9$ & $55.0\pm5.9$ & $35.0\pm5.2$ & $36.0\pm4.4$ & $25.7\pm3.1$ & $25.6\pm3.5$ &   \\
    \bottomrule
\end{tabular}
\begin{tablenotes}[para]
    \item[]Note: Column (1): GLEAM catalogue designation. Columns (2-7): GLEAM flux density. Column (8): SUMSS flux density. Columns (9-10): RACS-low and RACS-mid flux densities. Column (11): NVSS flux density. Column (12): VLA flux density.
\end{tablenotes}
\end{threeparttable}
\end{sidewaystable}

\section{IR Flux Densities}

\begin{table*}[hbt!]
\centering
\begin{threeparttable}
\caption{Measured IR Properties.}\label{tab:IRmeasured}
\begin{tabular}{c|l|c|c|c|c|c|c}
	\toprule
	GLEAM ID & WXSC ID & W1  & W2 & W3 & W4 & IRAS 60$\mu$m & IRAS 100$\mu$m\\
      &  & (mJy) & (mJy) & (mJy) & (mJy) & (Jy) & (Jy)\\
    \midrule
	GLEAM J002238-240737 & J002238.94-240737.0 & $22.9\pm0.5$ & $19.3\pm0.4$ & $78.7\pm2.1$ & $168\pm 5$ & $1.5\pm0.1$ & $3.2\pm0.2$ \\
	GLEAM J003652-333315 & J003652.45-333316.9 & $17.8\pm0.4$ & $31.5\pm0.7$ & $320.4\pm7.6$ & $2099\pm50$ & $6.5\pm0.4$ & $5.0\pm0.4$ \\
	GLEAM J011408-323907 & HIPASSJ0114-32 & $61.5\pm1.4$ & $37.5\pm0.9$ & $120.5\pm3.3$ & $240\pm 7$ & $2.8\pm0.2$ & $7.5\pm0.4$ \\
	GLEAM J012121-340345 & NGC 0491 & $65.2\pm1.5$ & $40.0\pm0.9$ & $196.1\pm5.1$ & $286\pm 8$ & $2.8\pm0.2$ & $8.6\pm0.5$ \\
	GLEAM J034056-223353 & NGC 1415 & $190.1\pm4.3$ & $112.9\pm2.6$ & $292.9\pm8.2$ & $462\pm14$ & $5.3\pm0.3$ & $12.7\pm0.5$ \\
	GLEAM J035545-422210 & NGC1487ab & $66.7\pm1.5$ & $39.1\pm1.0$ & $120.4\pm4.9$ & $290\pm11$ & $3.2\pm0.1$ & $6.4\pm0.3$ \\
	GLEAM J040226-180247 & 2MRS04022567-1802513 & $41.3\pm1.0$ & $37.2\pm0.9$ & $134.3\pm3.8$ & $333\pm 9$ & $2.9\pm0.1$ & $5.1\pm0.4$ \\
	GLEAM J041509-282854 & NGC1540 & $14.5\pm0.3$ & $9.9\pm0.2$ & $79.1\pm2.2$ & $475\pm12$ & $3.3\pm0.1$ & $4.8\pm0.2$ \\
	GLEAM J042905-372842 & 2MRS04290593-3728463 & $18.3\pm0.4$ & $12.1\pm0.3$ & $86.3\pm2.5$ & $289\pm 7$ & $2.4\pm0.2$ & $3.9\pm0.2$ \\
	GLEAM J072121-690005 & NGC 2397 & $182.8\pm4.0$ & $110.5\pm2.5$ & $543.3\pm12.3$ & $897\pm21$ & $7.3\pm0.3$ & $18.8\pm0.6$ \\
	GLEAM J074515-712426 & HIPASSJ0745-71 & $44.1\pm1.4$ & $20.6\pm1.2$ & $155.8\pm4.8$ & $278\pm10$ & $3.3\pm0.6$ & $10.4\pm1.8$ \\
	GLEAM J090634-754935 & HIPASSJ0906-75 & $59.9\pm1.4$ & $38.9\pm0.9$ & $227.6\pm5.4$ & $357\pm 9$ & $3.4\pm0.2$ & $9.8\pm0.6$ \\
	GLEAM J120737-145835 & 2MRS12073835-1458105 & $42.0\pm1.0$ & $26.8\pm0.6$ & $160.8\pm4.3$ & $293\pm 8$ & $3.4\pm0.2$ & $6.5\pm0.8$ \\
	GLEAM J142112-461800 & HIPASSJ1421-46 & $213.5\pm4.8$ & $127.6\pm2.9$ & $400.2\pm9.6$ & $970\pm24$ & $8.2\pm0.5$ & $15.9\pm1.0$ \\
	GLEAM J150540-422654 & 2MRS15054101-4226581 & $46.6\pm1.1$ & $28.7\pm0.7$ & $160.9\pm4.0$ & $264\pm 8$ & $2.2\pm0.2$ & $5.0\pm0.6$ \\
	GLEAM J184747-602054 & GLEAM J184747-602054 & $20.2\pm0.5$ & $13.3\pm0.3$ & $89.3\pm2.5$ & $318\pm12$ & $3.0\pm0.3$ & $5.1\pm0.4$ \\
	GLEAM J203047-472824 & NGC 6918 & $35.3\pm0.8$ & $25.3\pm0.6$ & $227.3\pm5.3$ & $1113\pm29$ & $9.3\pm0.6$ & $13.6\pm0.7$ \\
	GLEAM J205209-484639 & J205210-4846.6 & $56.6\pm1.3$ & $37.7\pm0.9$ & $228.1\pm5.7$ & $436\pm11$ & $4.5\pm0.3$ & $9.0\pm0.5$ \\
	GLEAM J213610-383236 & ESO343-IG013ab & $32.6\pm0.8$ & $24.2\pm0.6$ & $181.4\pm4.5$ & $714\pm18$ & $5.9\pm0.4$ & $8.9\pm0.4$ \\
	GLEAMJ232600-815311 & TJ23255537-8152402 & $7.5\pm0.2$ & $4.9\pm0.1$ & $29.6\pm0.8$ & $60\pm 2$ &  &  \\
	\bottomrule
\end{tabular}
\begin{tablenotes}[para]
    \item[]Note: Column (1): GLEAM catalogue designation. Column (2): WXSC designation. Column (3-6): WXSC W1, W2, W3 and W4 band integrated flux density respectively. Column (7): IRAS 60\,$\mu$m integrated flux density. Column (8): IRAS 100\,$\mu$m integrated flux density.
\end{tablenotes}
\end{threeparttable}
\end{table*}

\section{Best fit SEDS and images}
\begin{figure*}[hbt!]
    \centering
    \begin{subfigure}[b]{0.43\textwidth}
        \centering
        \includegraphics[width=\textwidth]{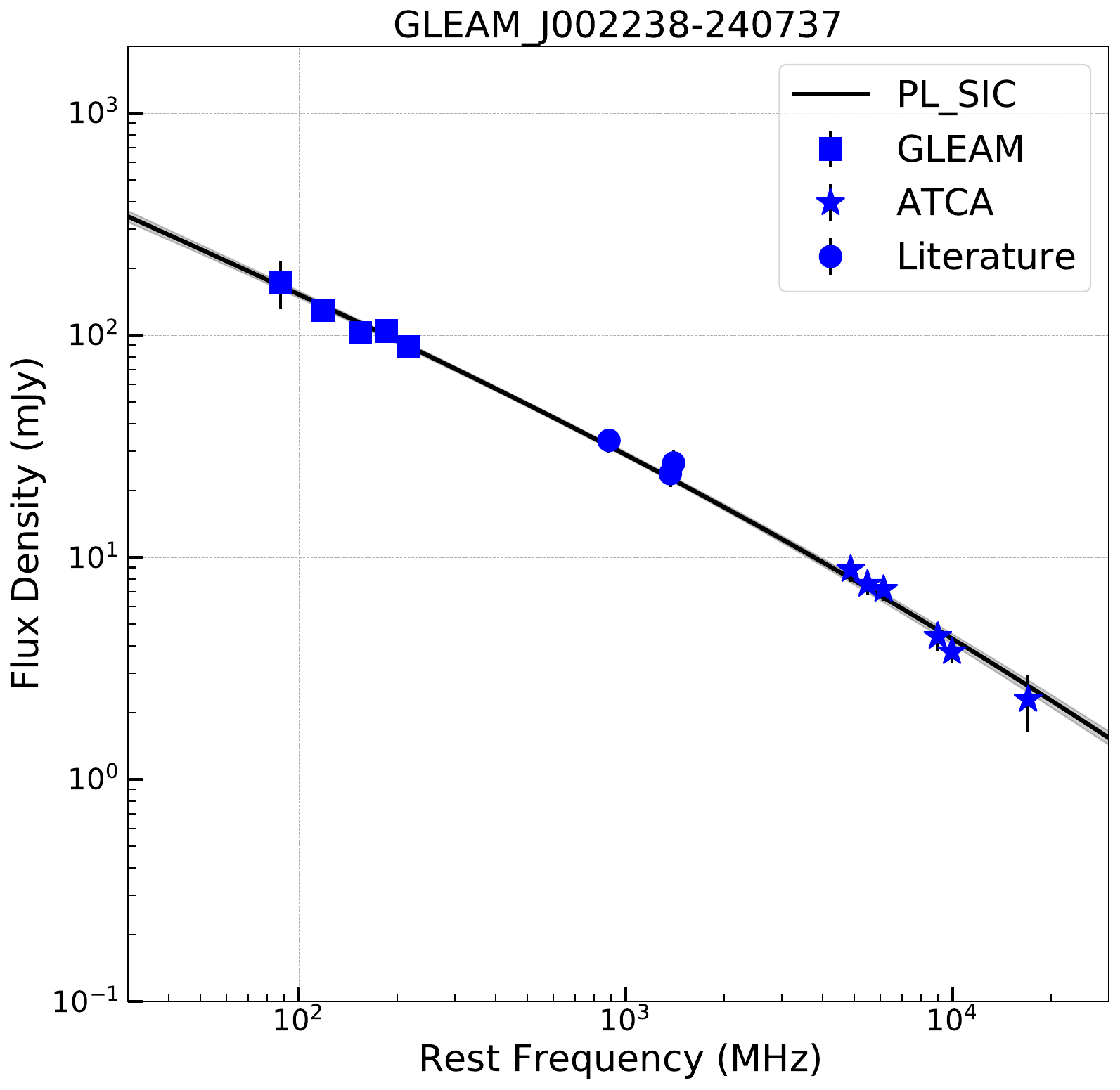}
    \end{subfigure}
    \begin{subfigure}[b]{0.47\textwidth}
        \centering
        \includegraphics[width=\textwidth]{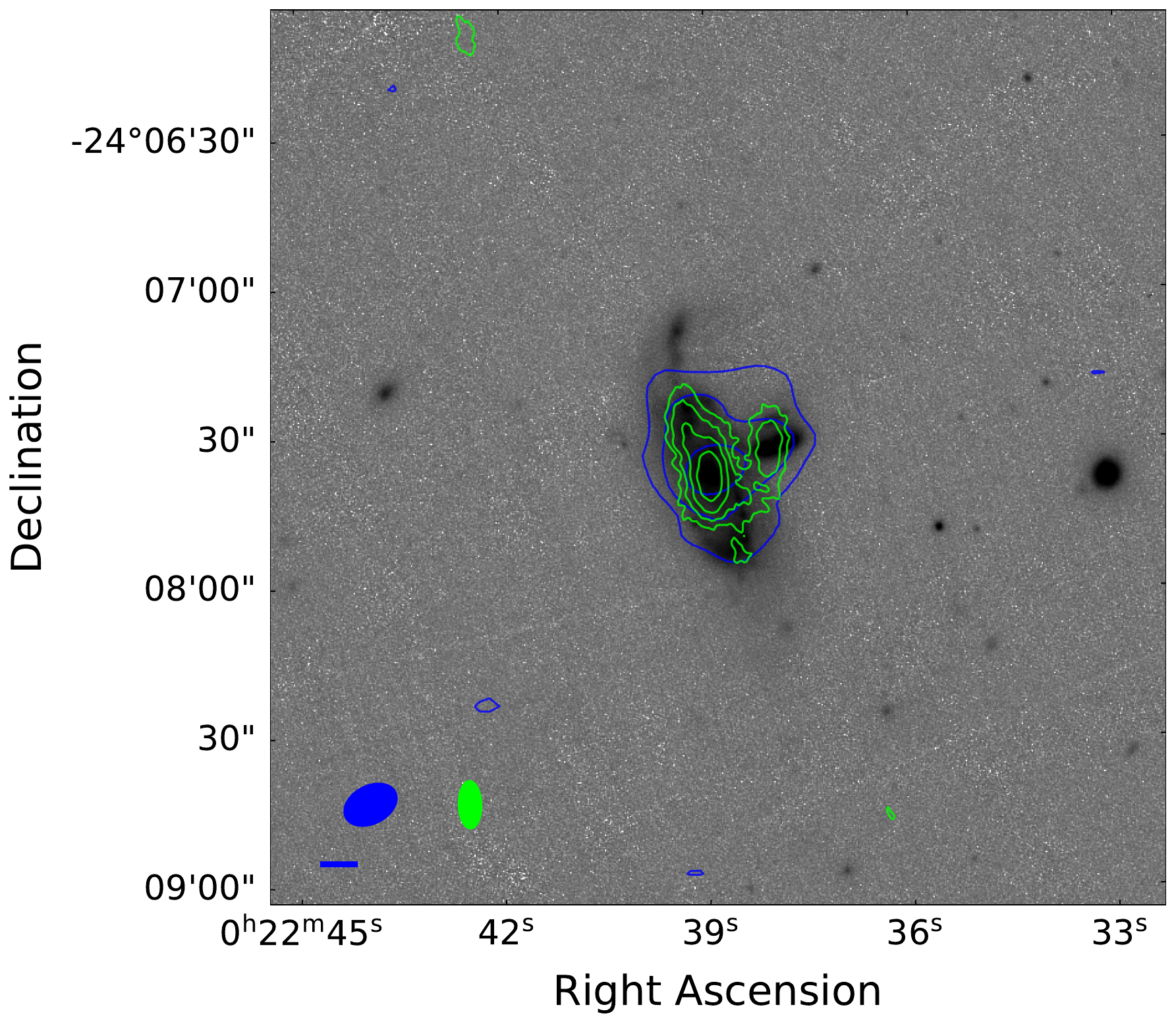}
    \end{subfigure}
    \vspace{0.8em}
    \begin{subfigure}[b]{0.43\textwidth}
        \centering
        \includegraphics[width=\textwidth]{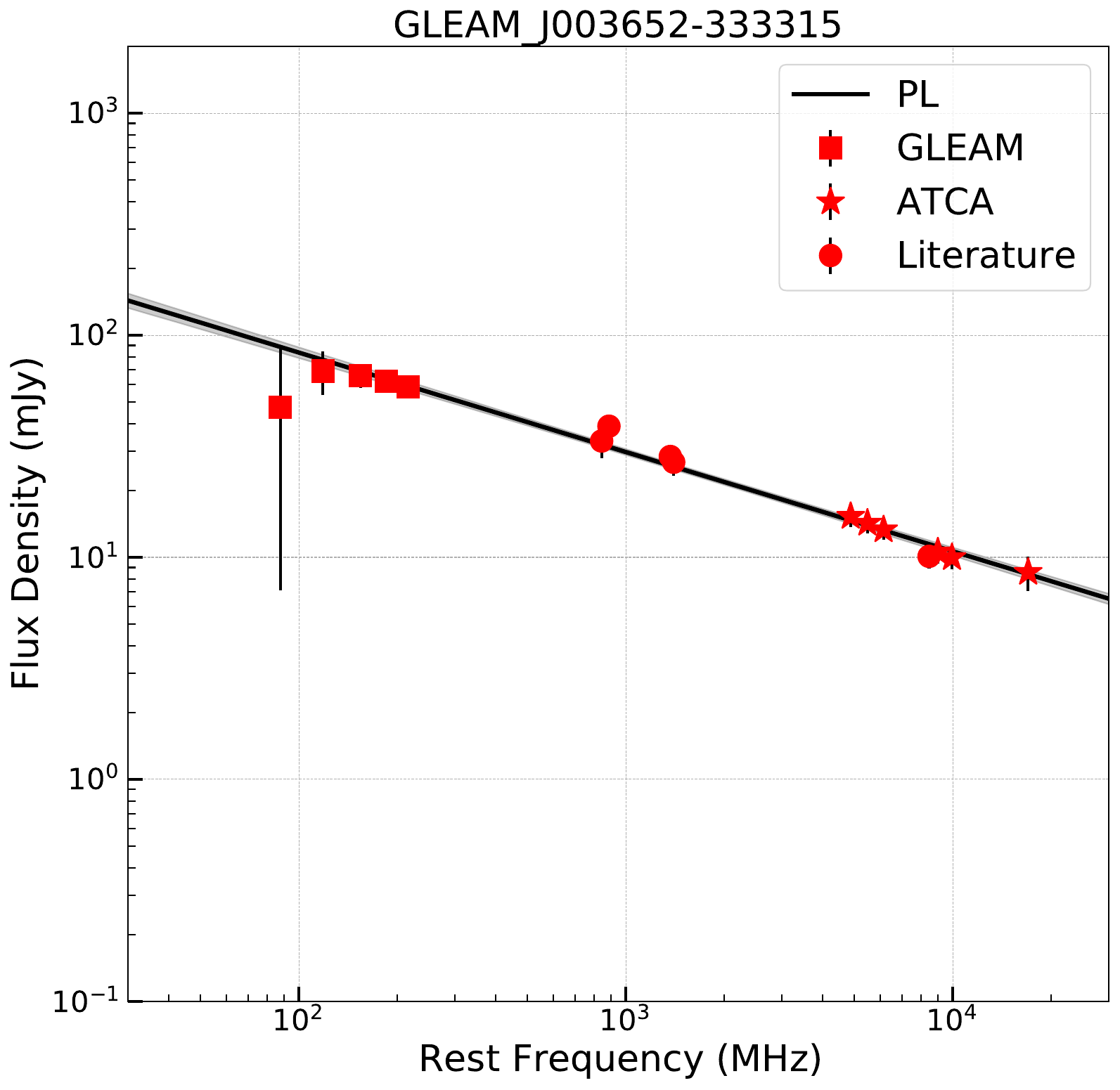}
    \end{subfigure}
    \begin{subfigure}[b]{0.47\textwidth}
        \centering
        \includegraphics[width=\textwidth]{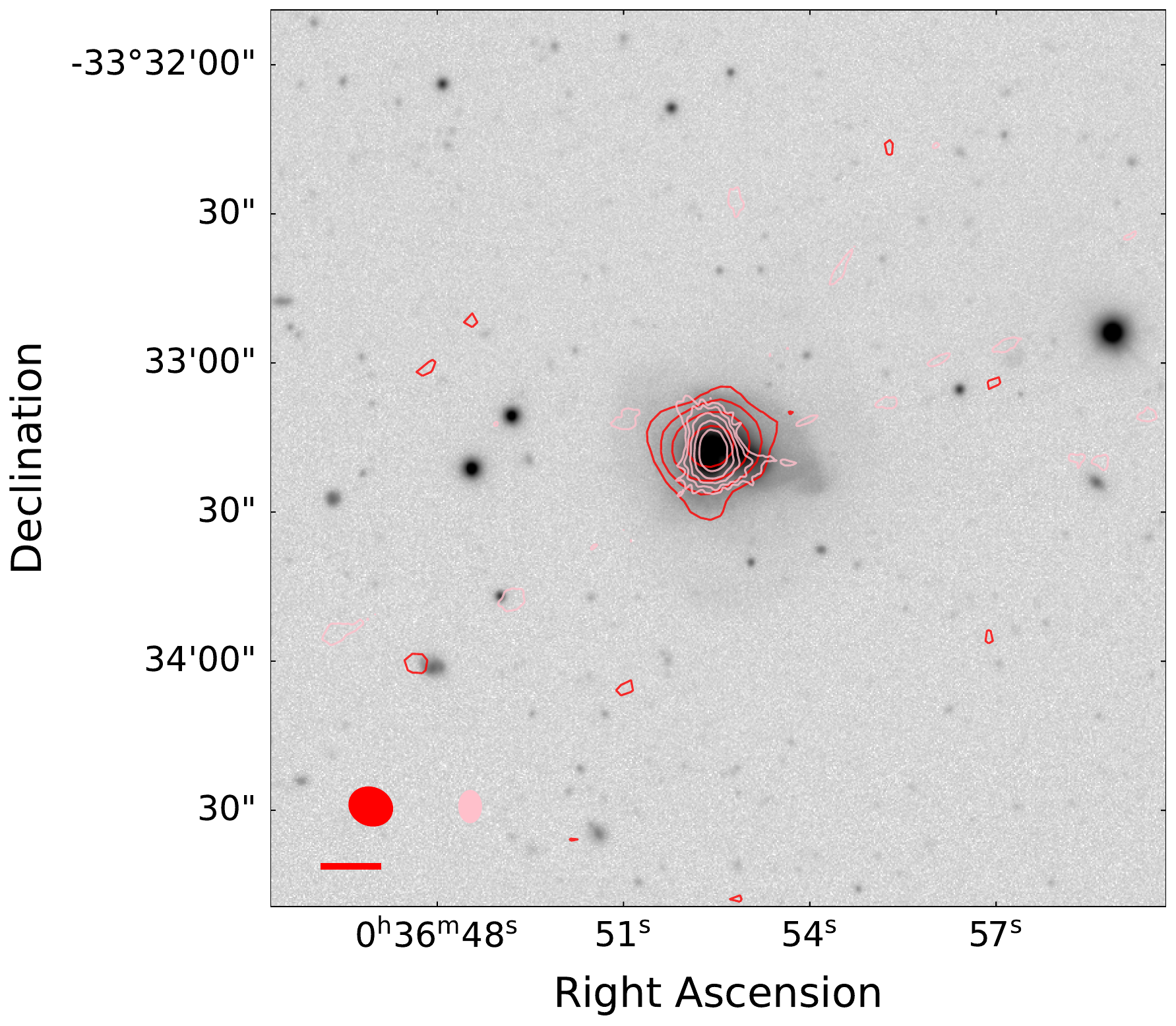}
    \end{subfigure}
    \vspace{0.8em}
    \begin{subfigure}[b]{0.43\textwidth}
        \centering
        \includegraphics[width=\textwidth]{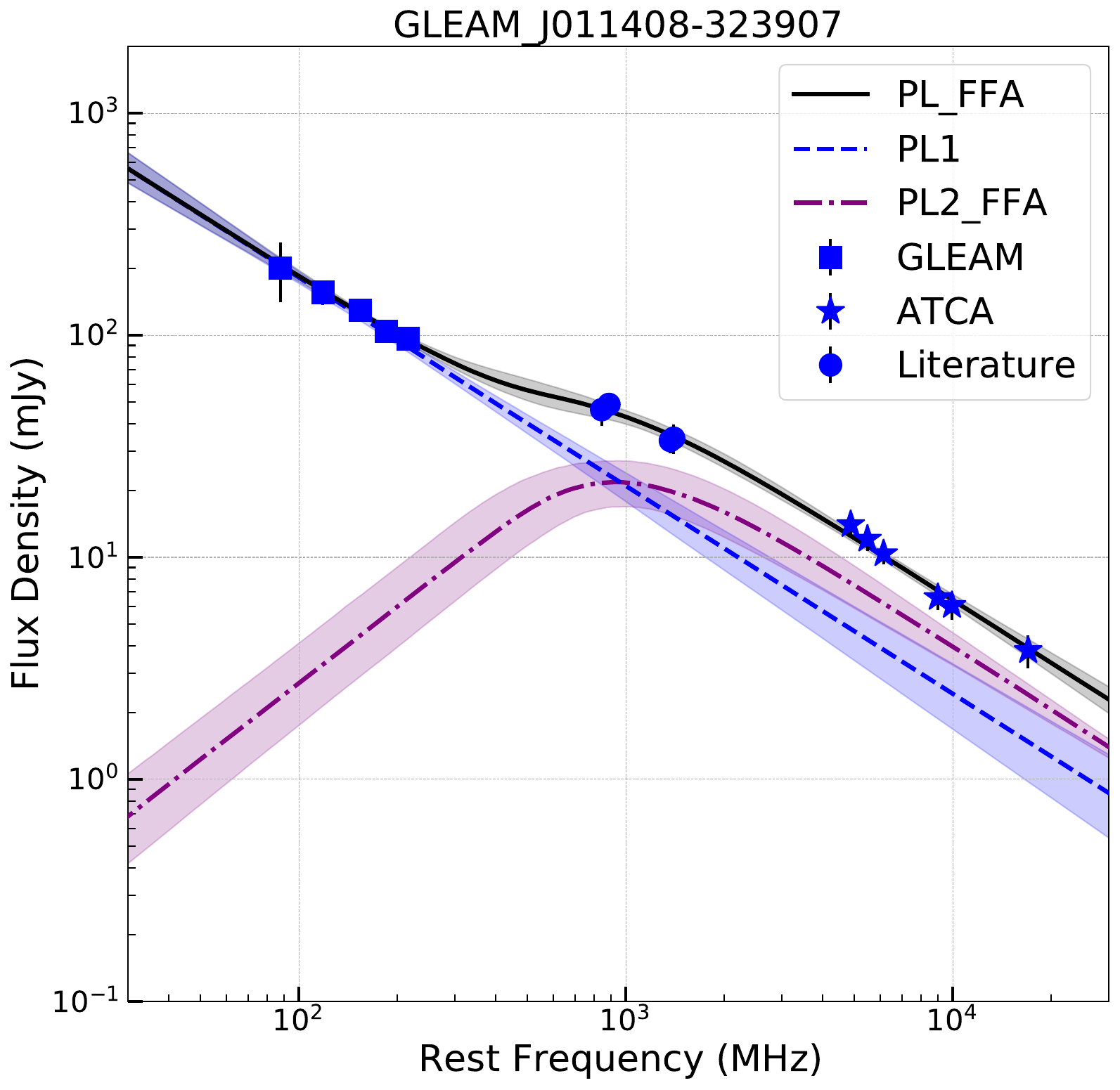}
    \end{subfigure}
    \begin{subfigure}[b]{0.47\textwidth}
        \centering
        \includegraphics[width=\textwidth]{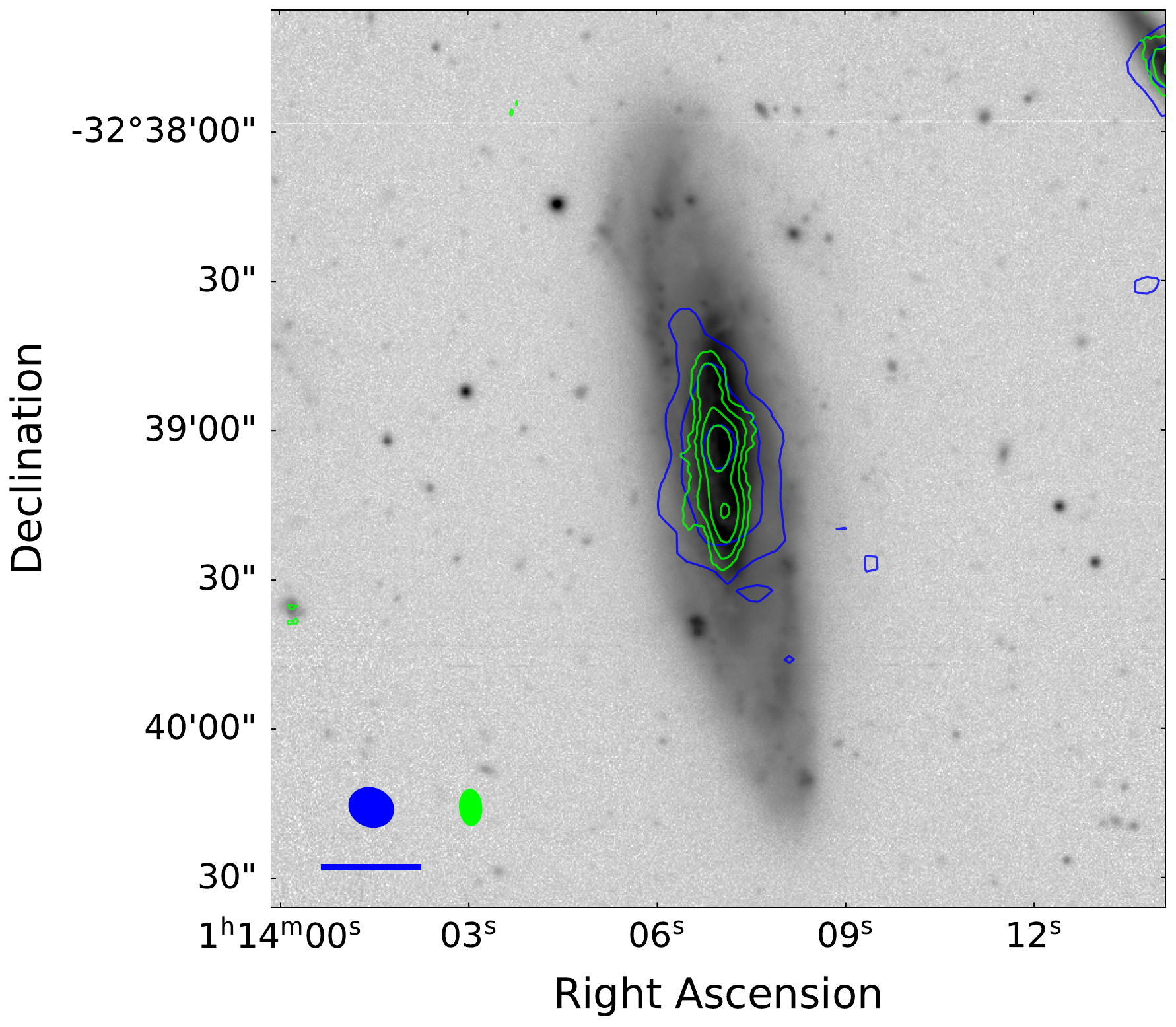}
    \end{subfigure}
    \caption{Left: The preferred radio SED model for each galaxy in the SFG sample with observed data points in blue or red for members of the control or LFTO samples respectively. The overlaid black line indicates the full model with the highlighted regions representing the 1-$\sigma$ uncertainties sampled by {\tt EMCEE}. Right: The optical image of each galaxy overlaid with contours from RACS-mid at 1.37\,GHz in blue/red and ATCA 9.5\,GHz in green/pink for members of the control/LFTO samples respectively. Radio contours for both frequencies start at the 4$\sigma$ level and increase by factors of $\sqrt{3}$. The FWHM beams for RACS-mid and ATCA are given by the blue/red and green/pink ellipses respectively. The scale bar at the bottom left denotes 5\,kpc.}\label{fig:seds}
\end{figure*}

\begin{figure*}[hbt!]
    \centering
    \begin{subfigure}[b]{0.43\textwidth}
        \centering
        \includegraphics[width=\textwidth]{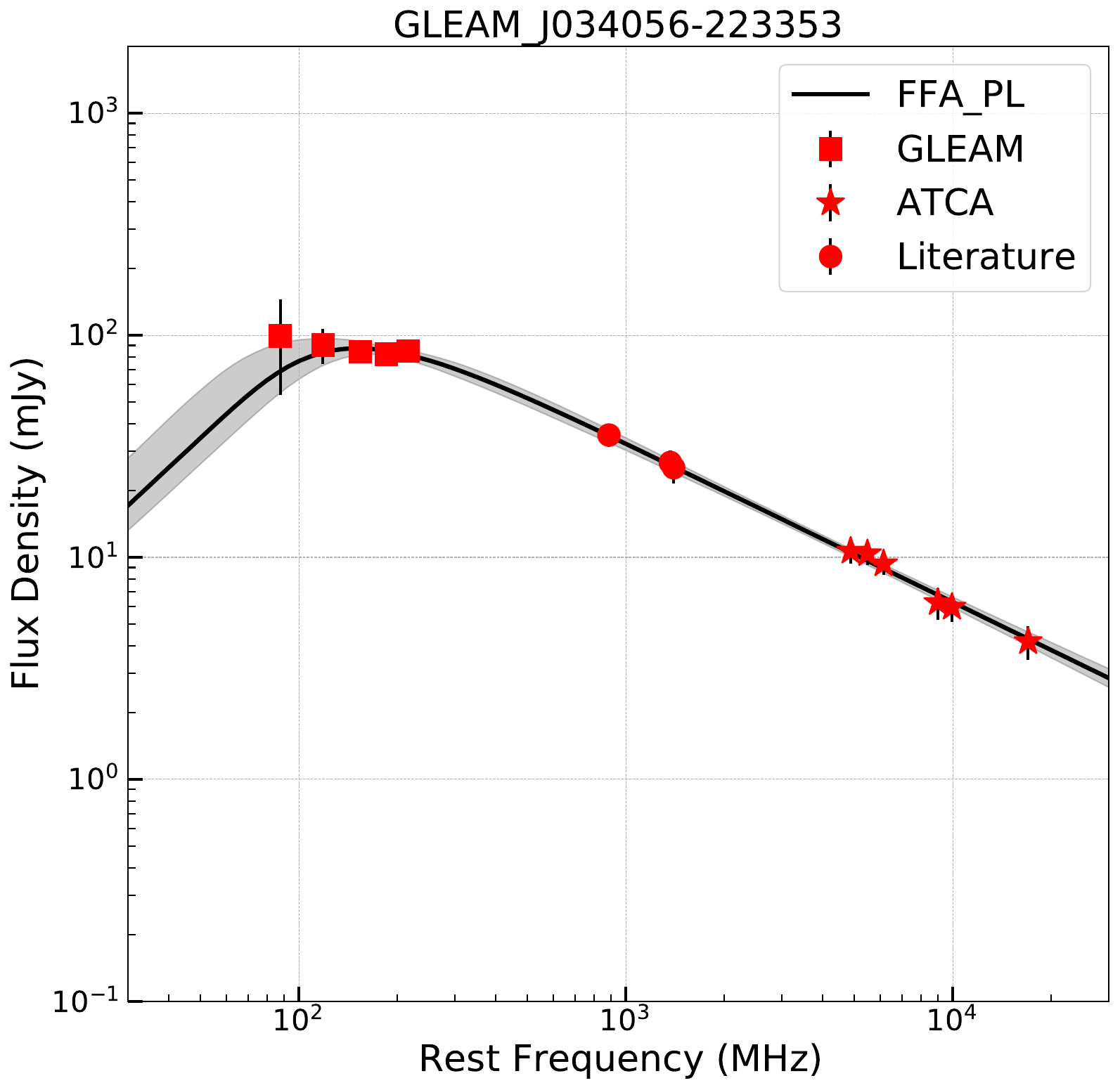}
    \end{subfigure}
    \begin{subfigure}[b]{0.47\textwidth}
        \centering
        \includegraphics[width=\textwidth]{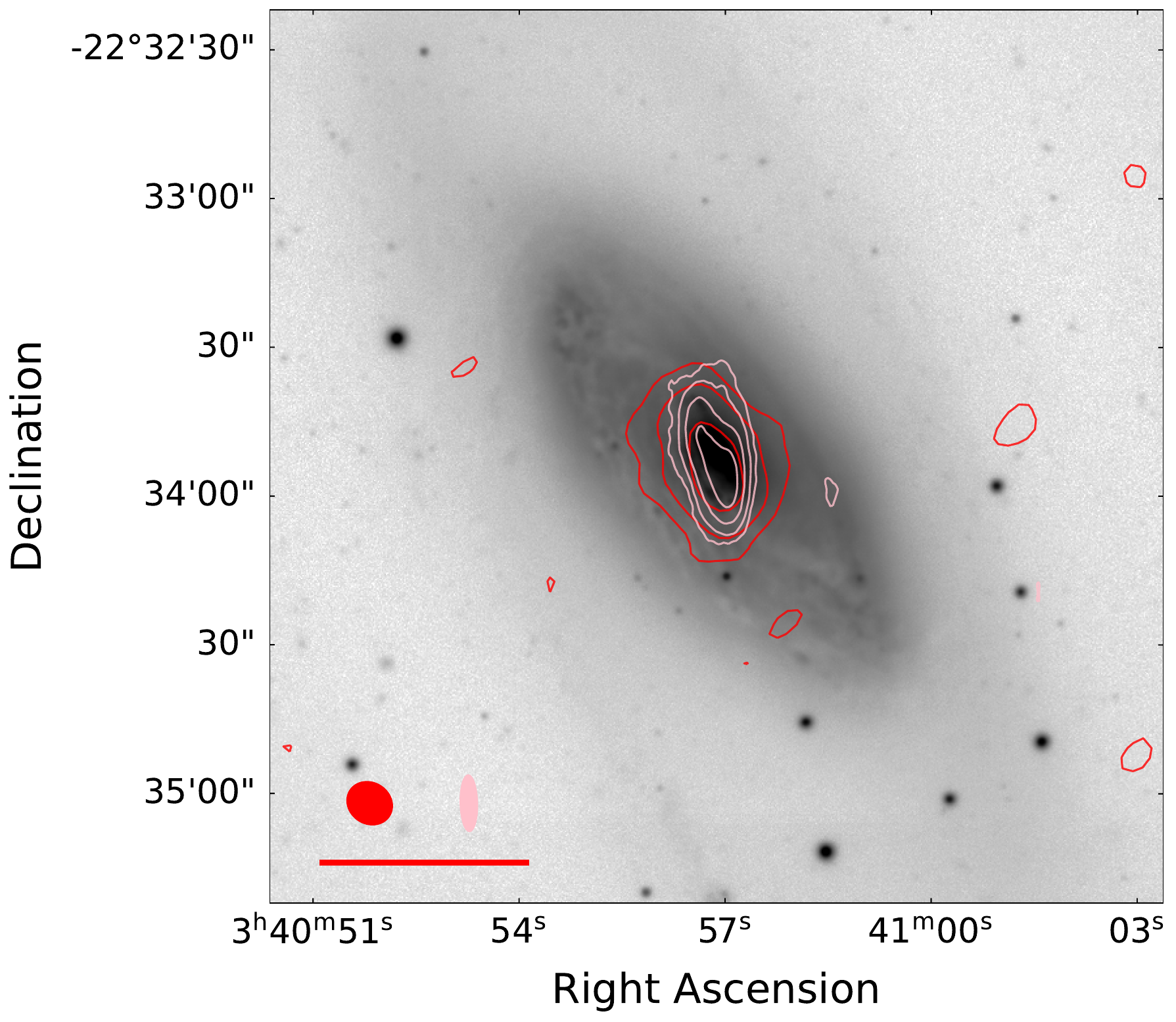}
    \end{subfigure}
    \vspace{0.8em}
        \begin{subfigure}[b]{0.43\textwidth}
        \centering
        \includegraphics[width=\textwidth]{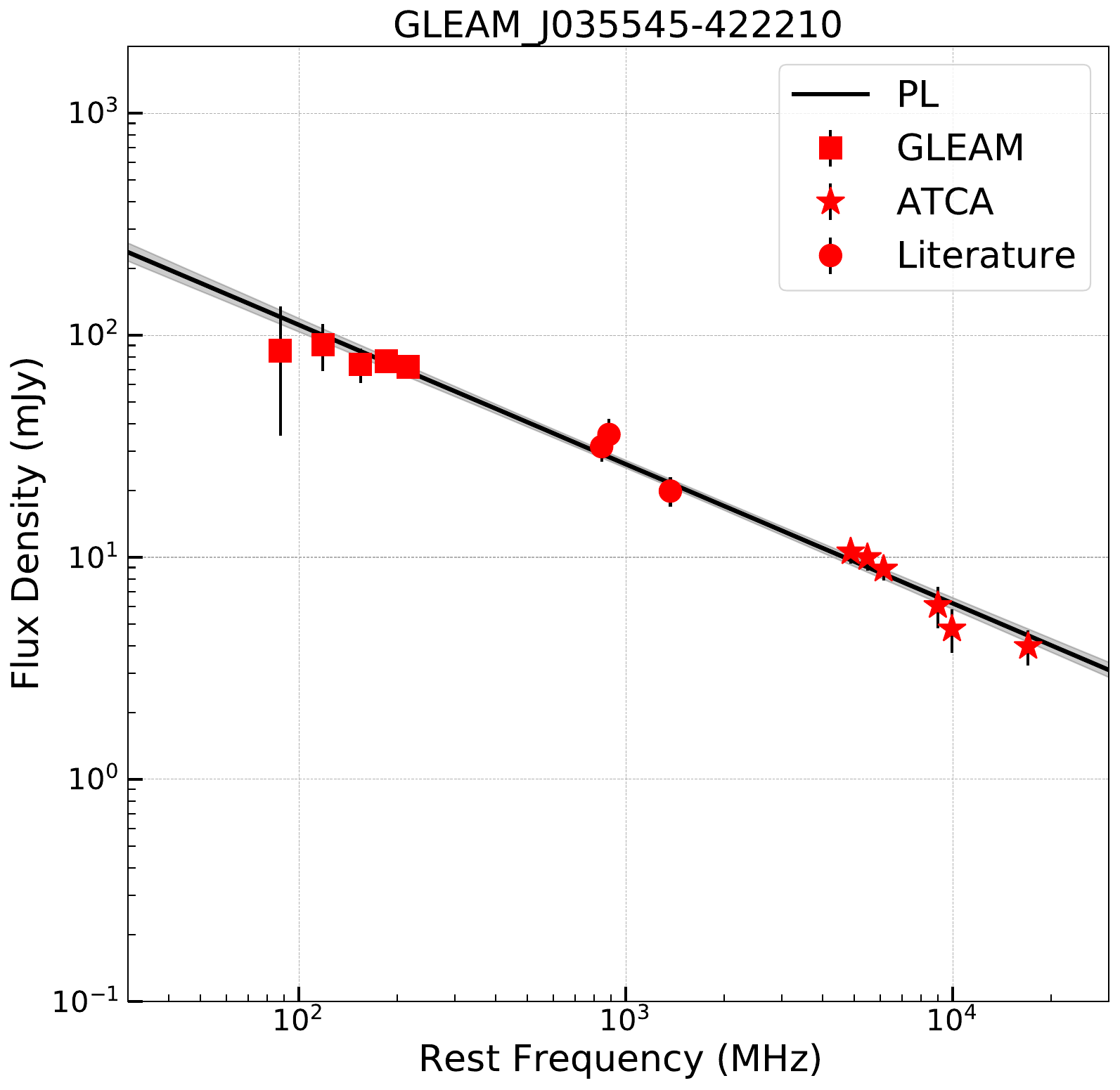}
    \end{subfigure}
    \begin{subfigure}[b]{0.47\textwidth}
        \centering
        \includegraphics[width=\textwidth]{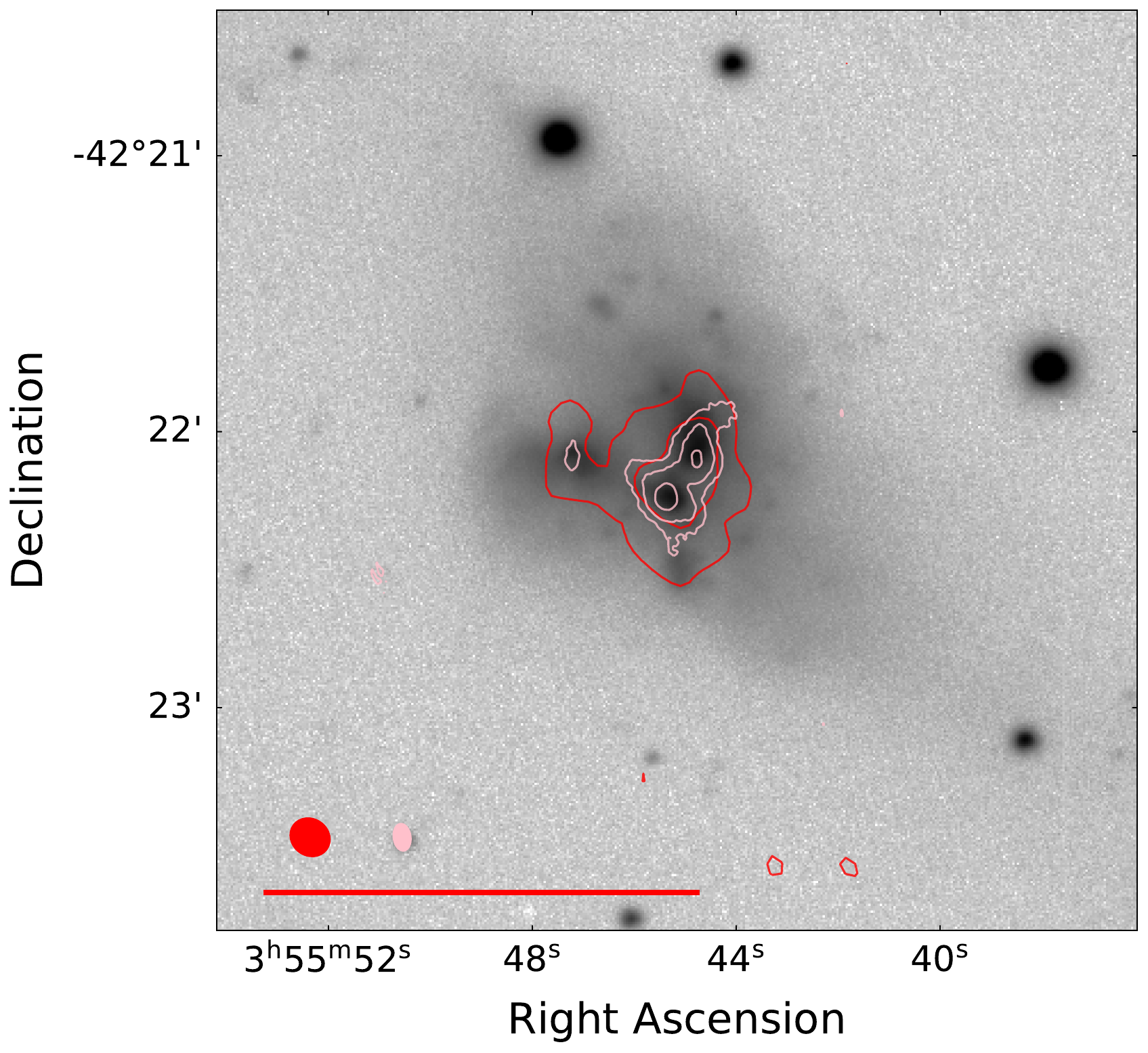}
    \end{subfigure}
        \vspace{0.8em}
        \begin{subfigure}[b]{0.43\textwidth}
        \centering
        \includegraphics[width=\textwidth]{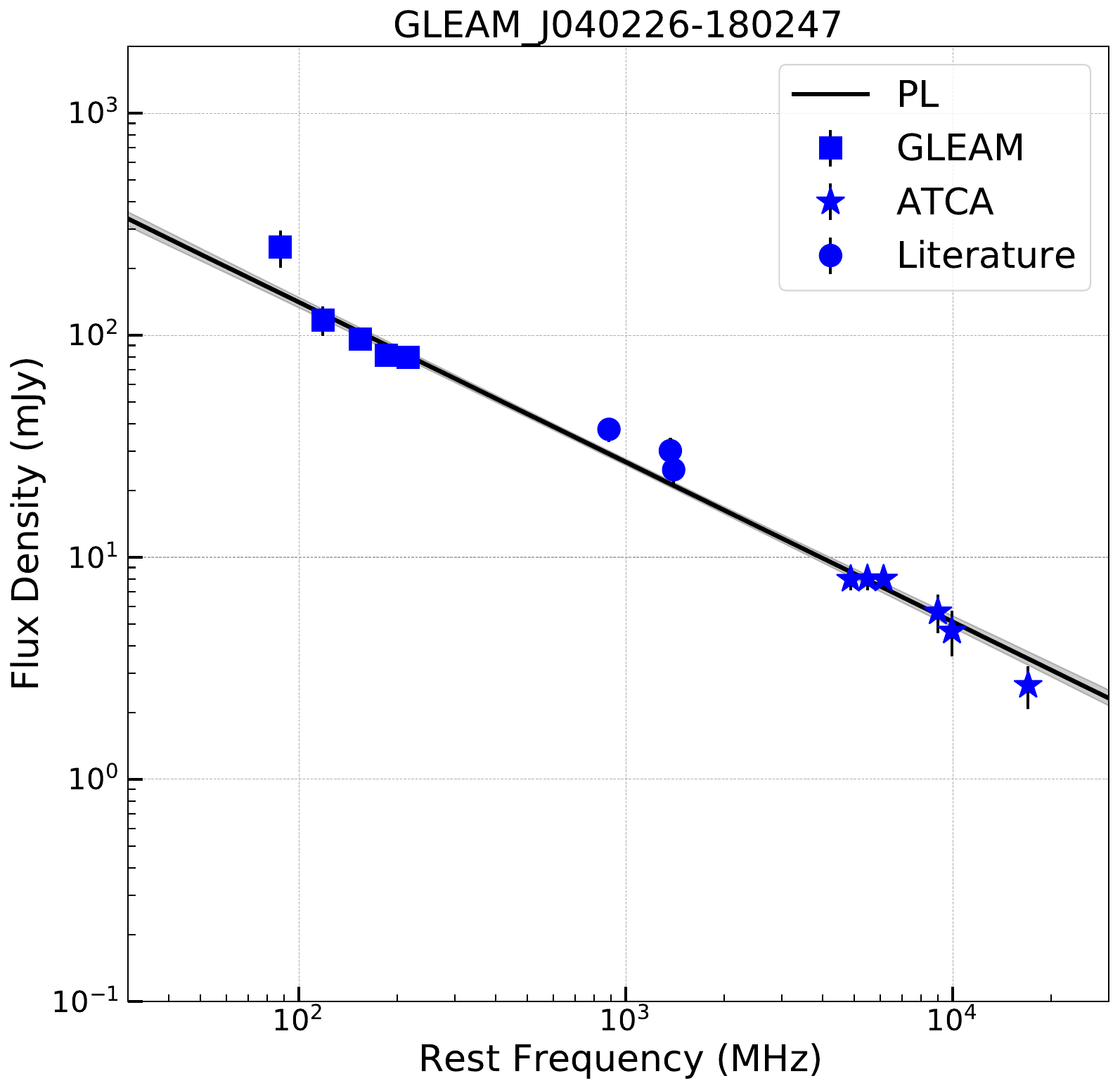}
    \end{subfigure}
    \begin{subfigure}[b]{0.47\textwidth}
        \centering
        \includegraphics[width=\textwidth]{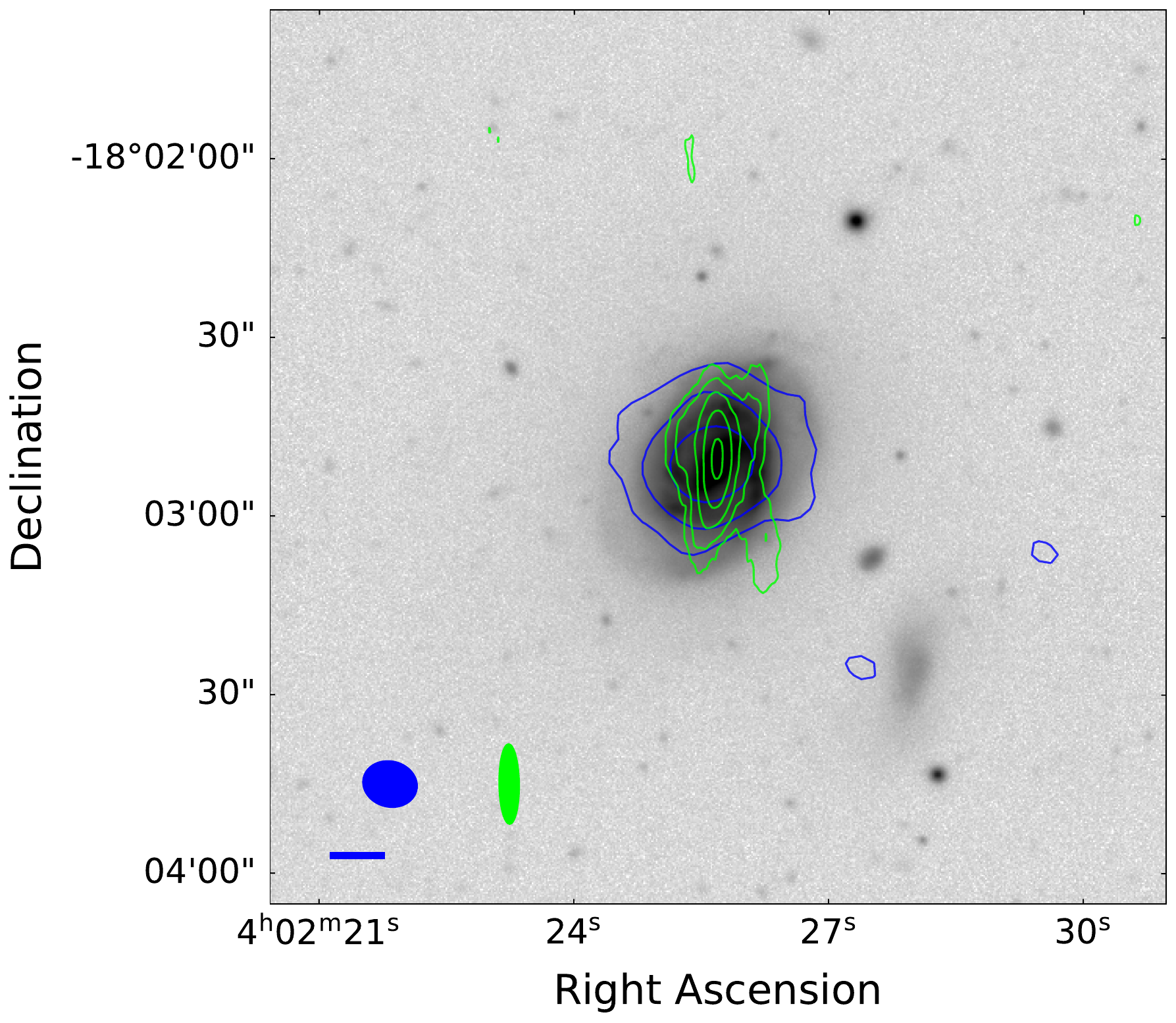}
    \end{subfigure}
    \caption{Figure 13. Continued}
\end{figure*}

\begin{figure*}[hbt!]
    \centering
    \begin{subfigure}[b]{0.43\textwidth}
        \centering
        \includegraphics[width=\textwidth]{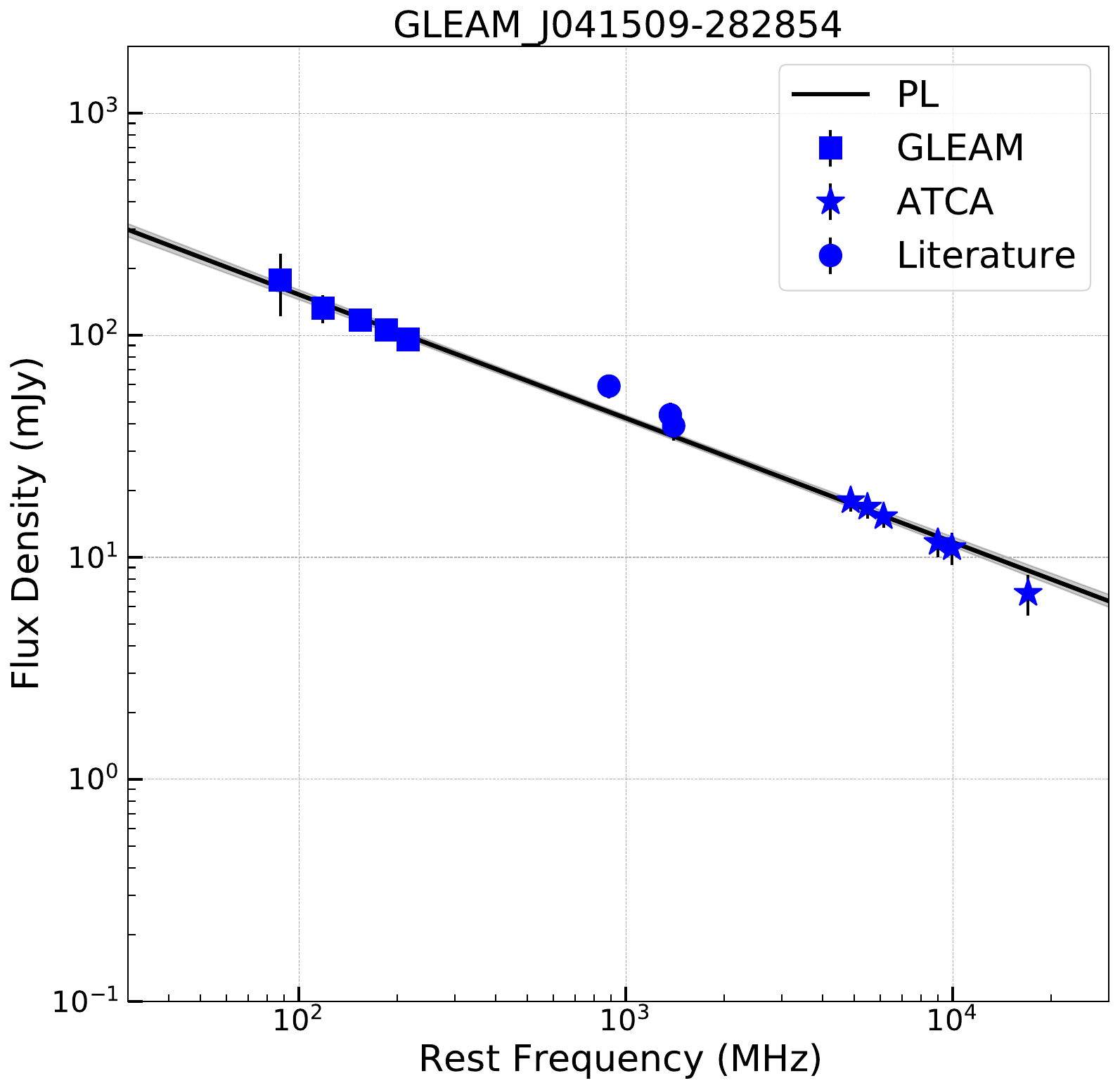}
    \end{subfigure}
    \begin{subfigure}[b]{0.47\textwidth}
        \centering
        \includegraphics[width=\textwidth]{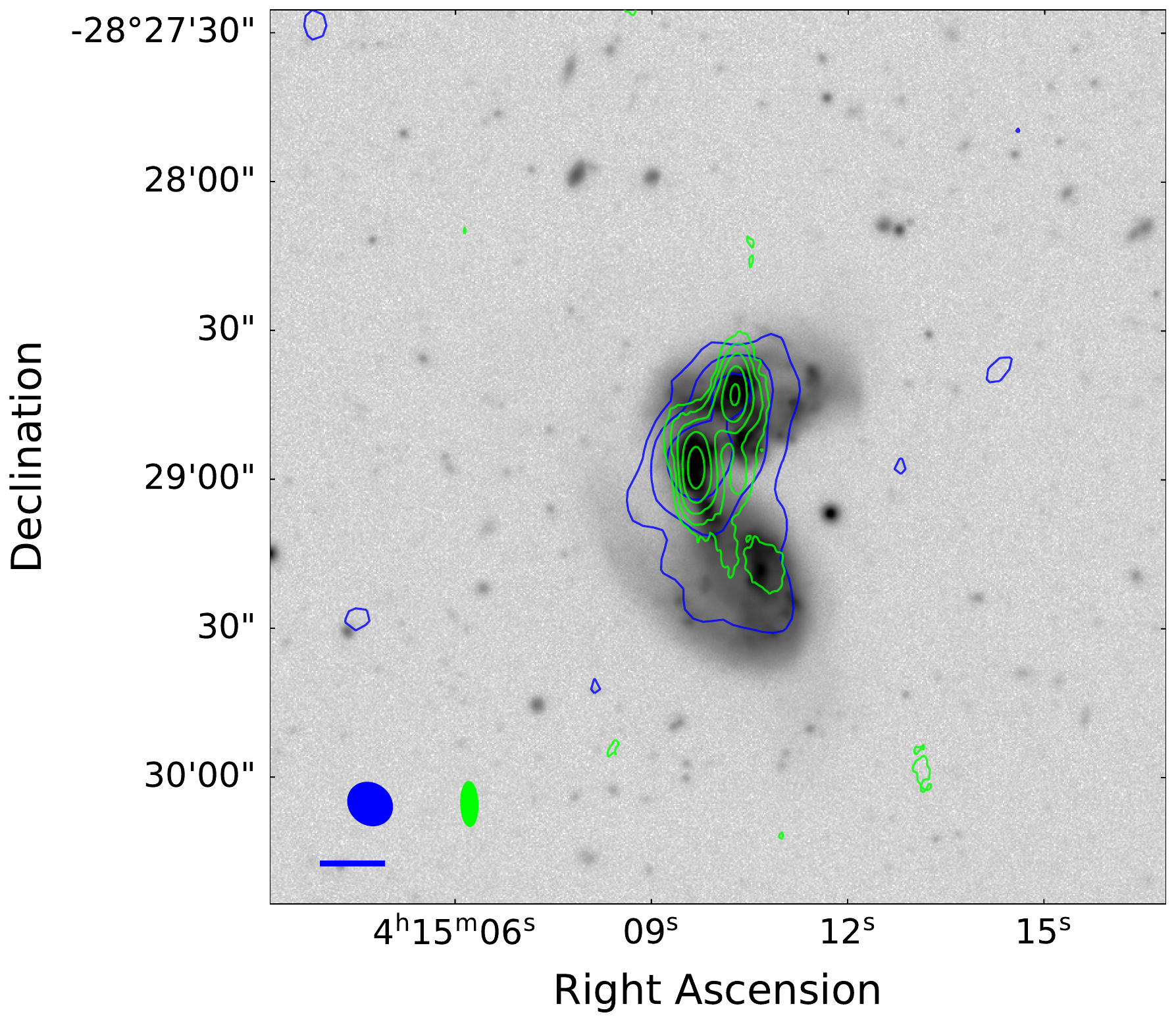}
    \end{subfigure}
    \vspace{0.8em}
        \begin{subfigure}[b]{0.43\textwidth}
        \centering
        \includegraphics[width=\textwidth]{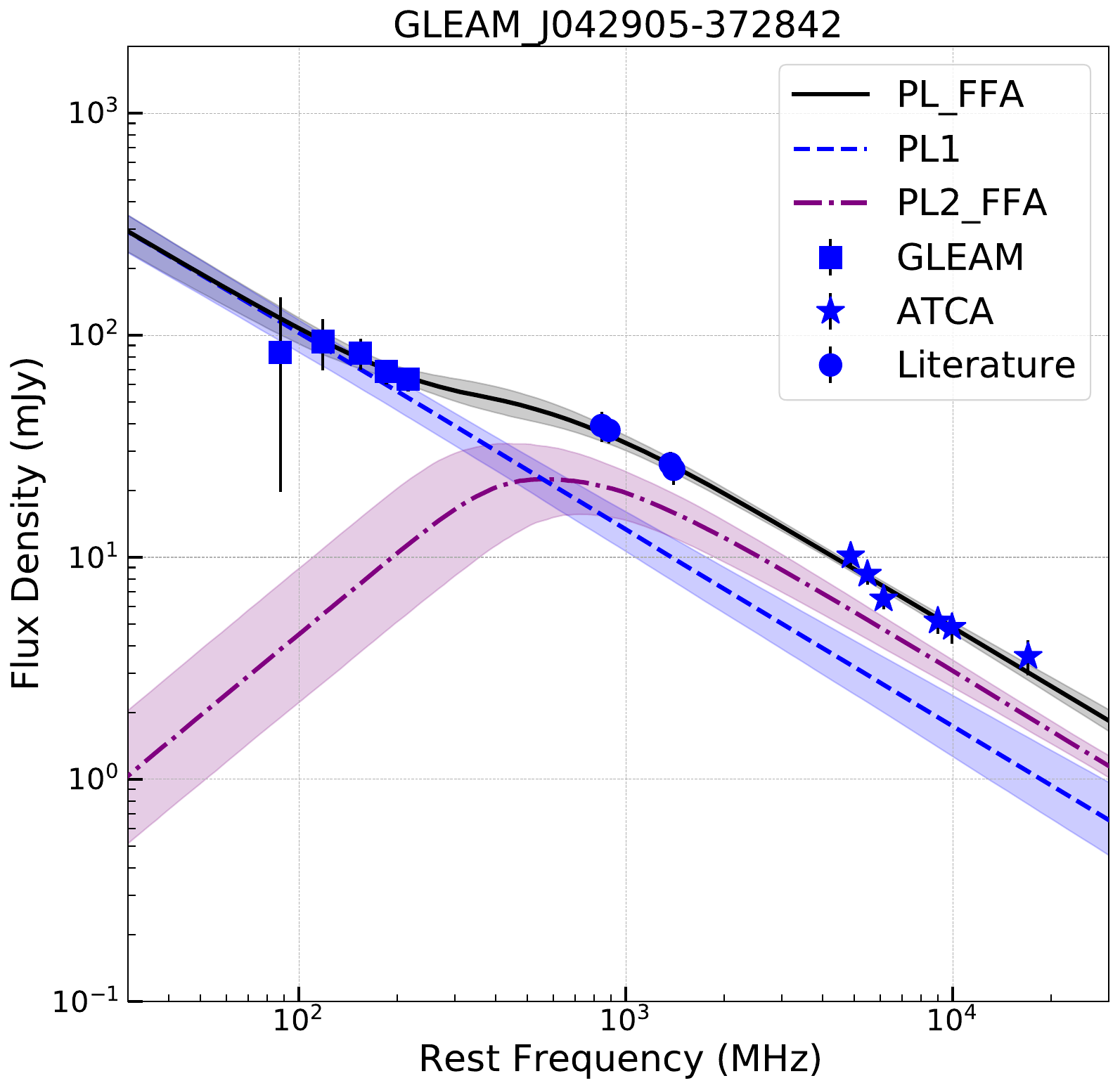}
    \end{subfigure}
    \begin{subfigure}[b]{0.47\textwidth}
        \centering
        \includegraphics[width=\textwidth]{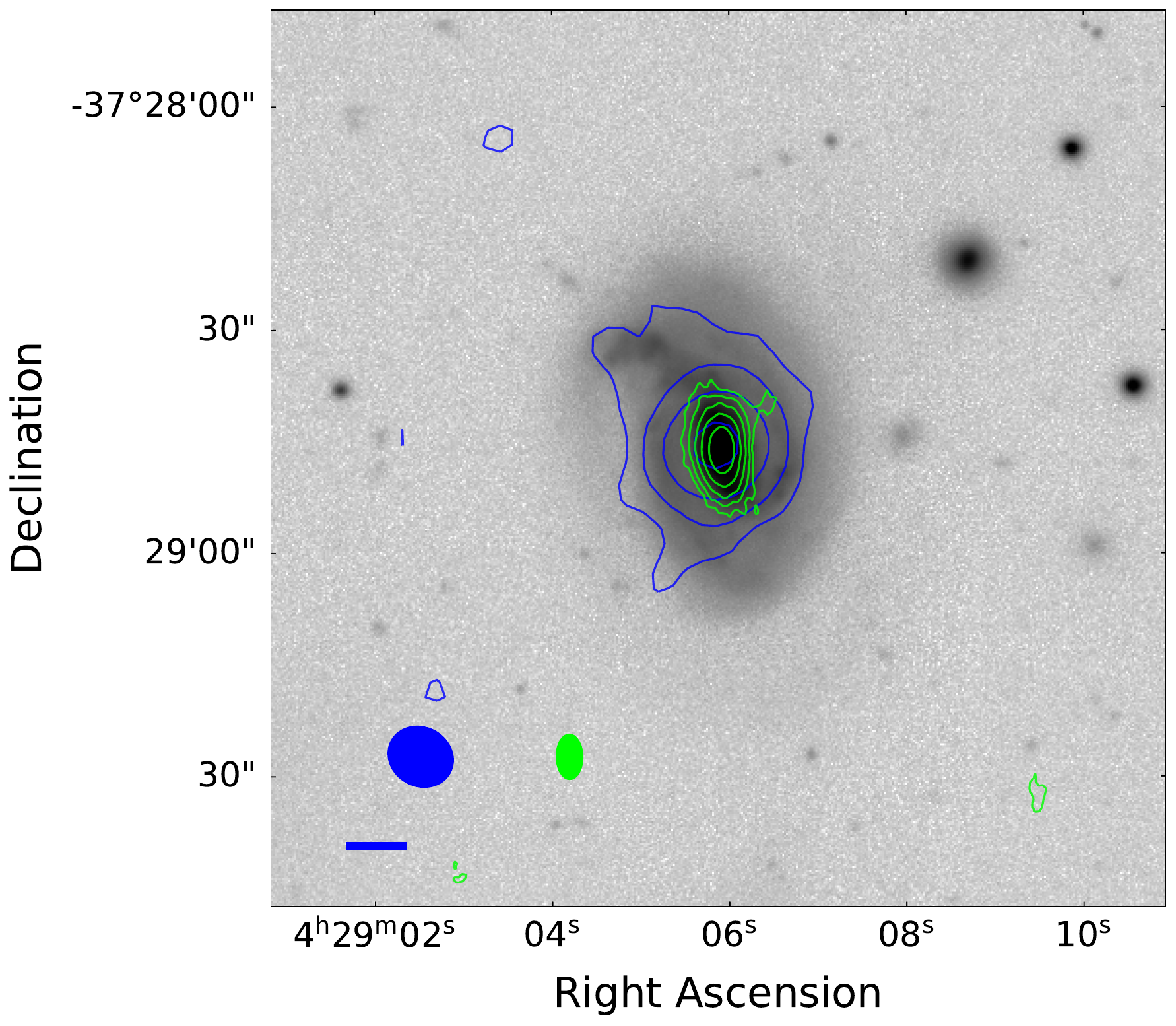}
    \end{subfigure}
        \vspace{0.8em}
        \begin{subfigure}[b]{0.43\textwidth}
        \centering
        \includegraphics[width=\textwidth]{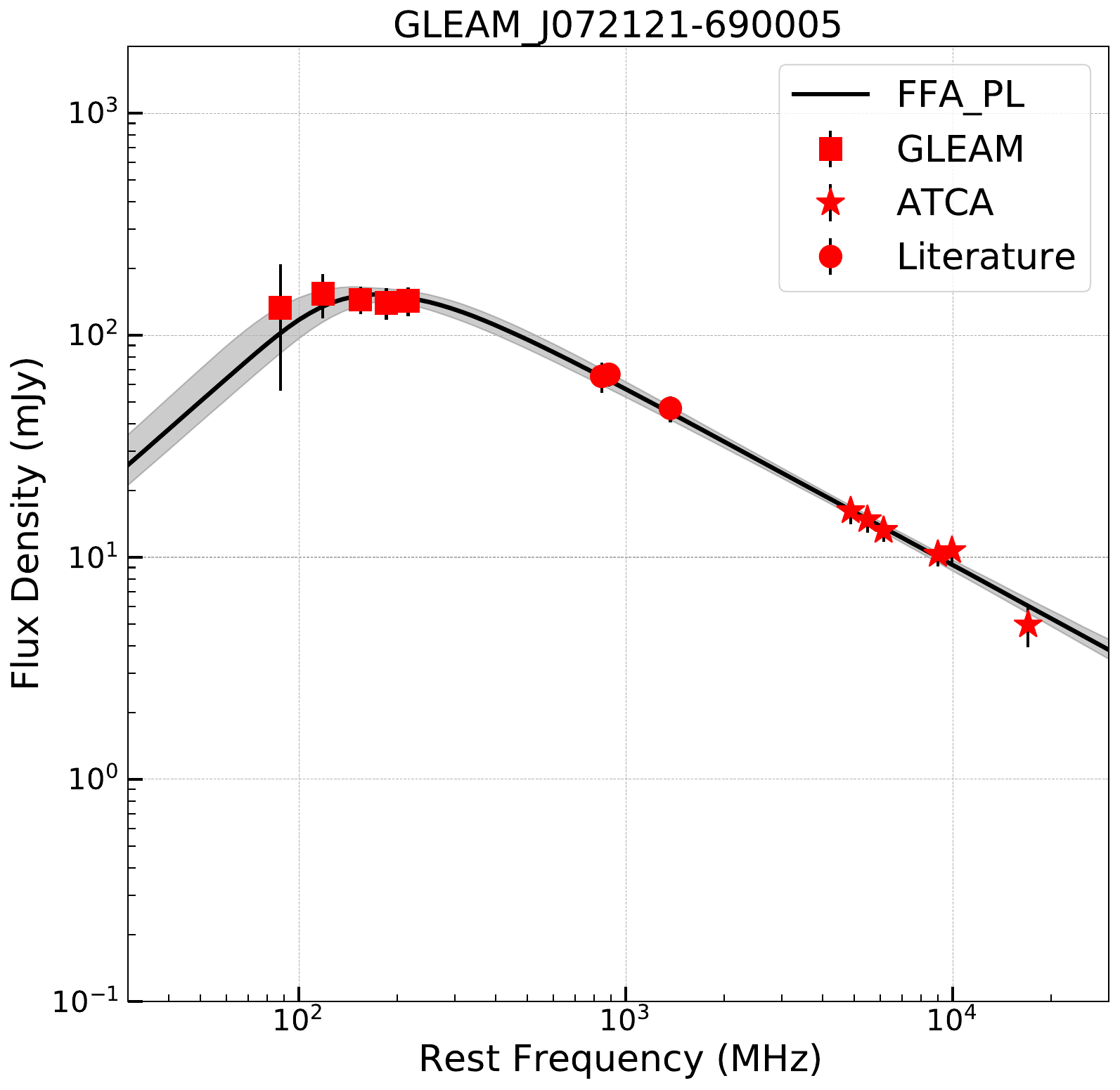}
    \end{subfigure}
    \begin{subfigure}[b]{0.47\textwidth}
        \centering
        \includegraphics[width=\textwidth]{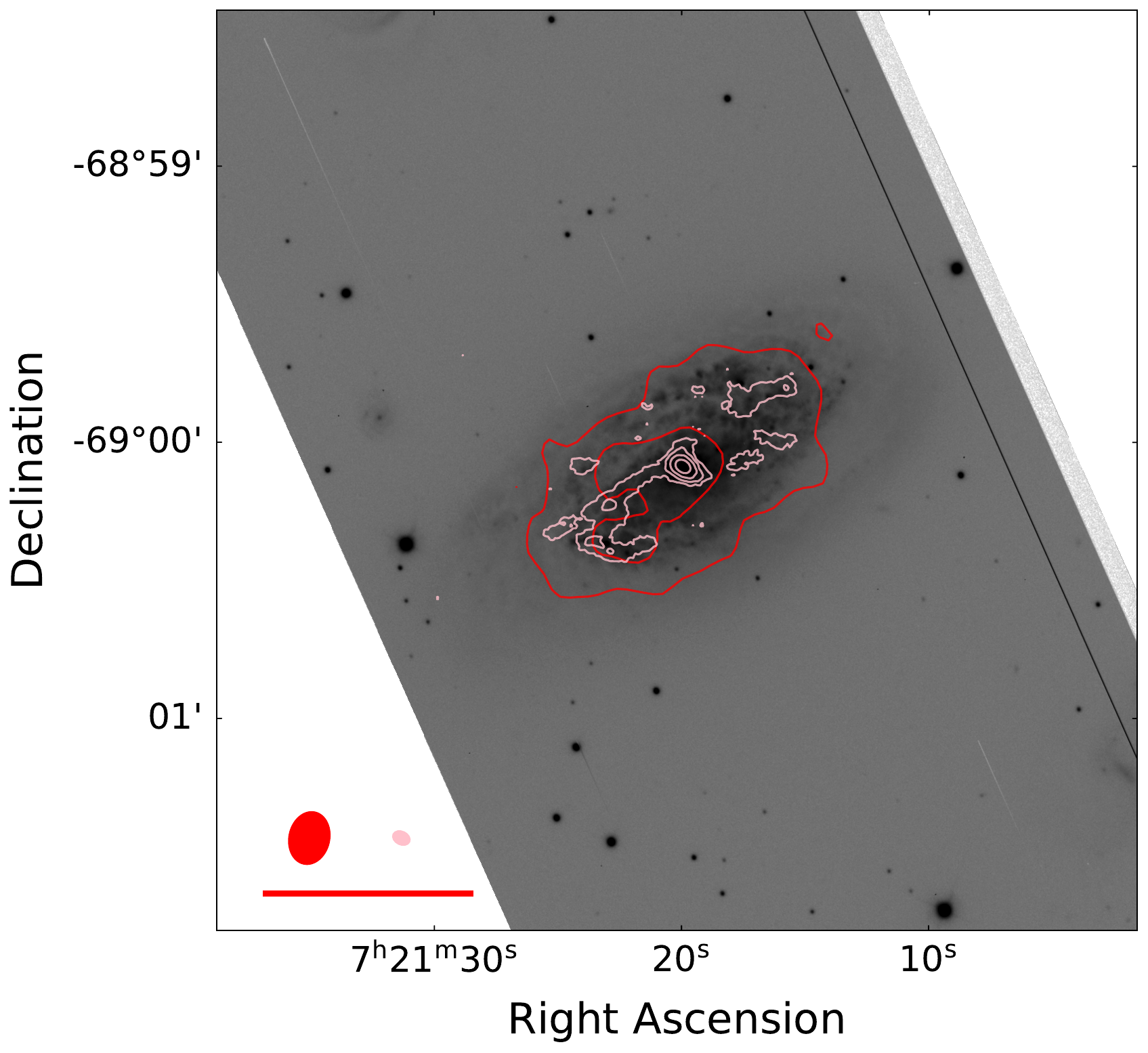}
    \end{subfigure}
    \caption{Figure 13. Continued}
\end{figure*}

\begin{figure*}[hbt!]
    \centering
    \begin{subfigure}[b]{0.43\textwidth}
        \centering
        \includegraphics[width=\textwidth]{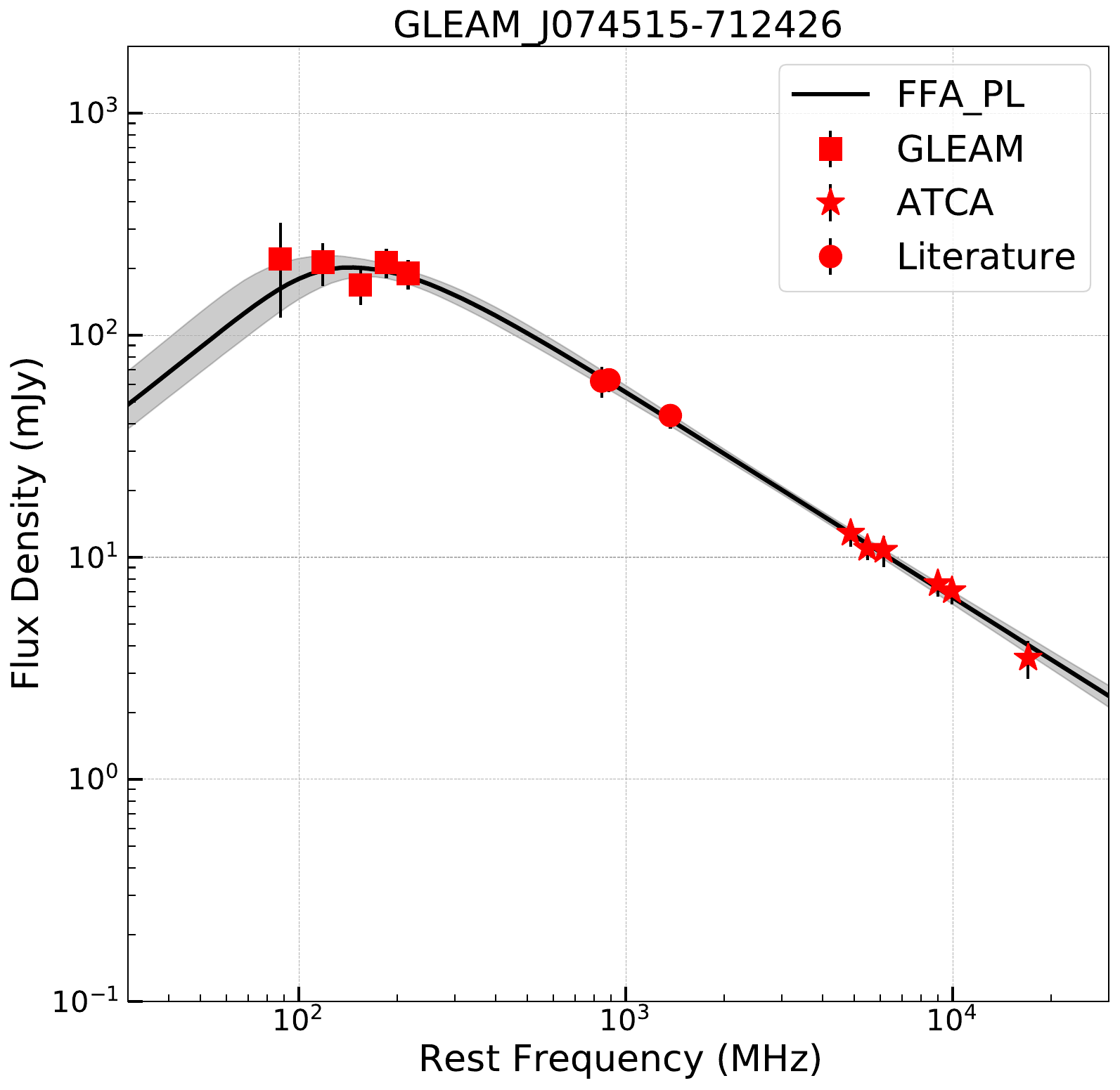}
    \end{subfigure}
    \begin{subfigure}[b]{0.47\textwidth}
        \centering
        \includegraphics[width=\textwidth]{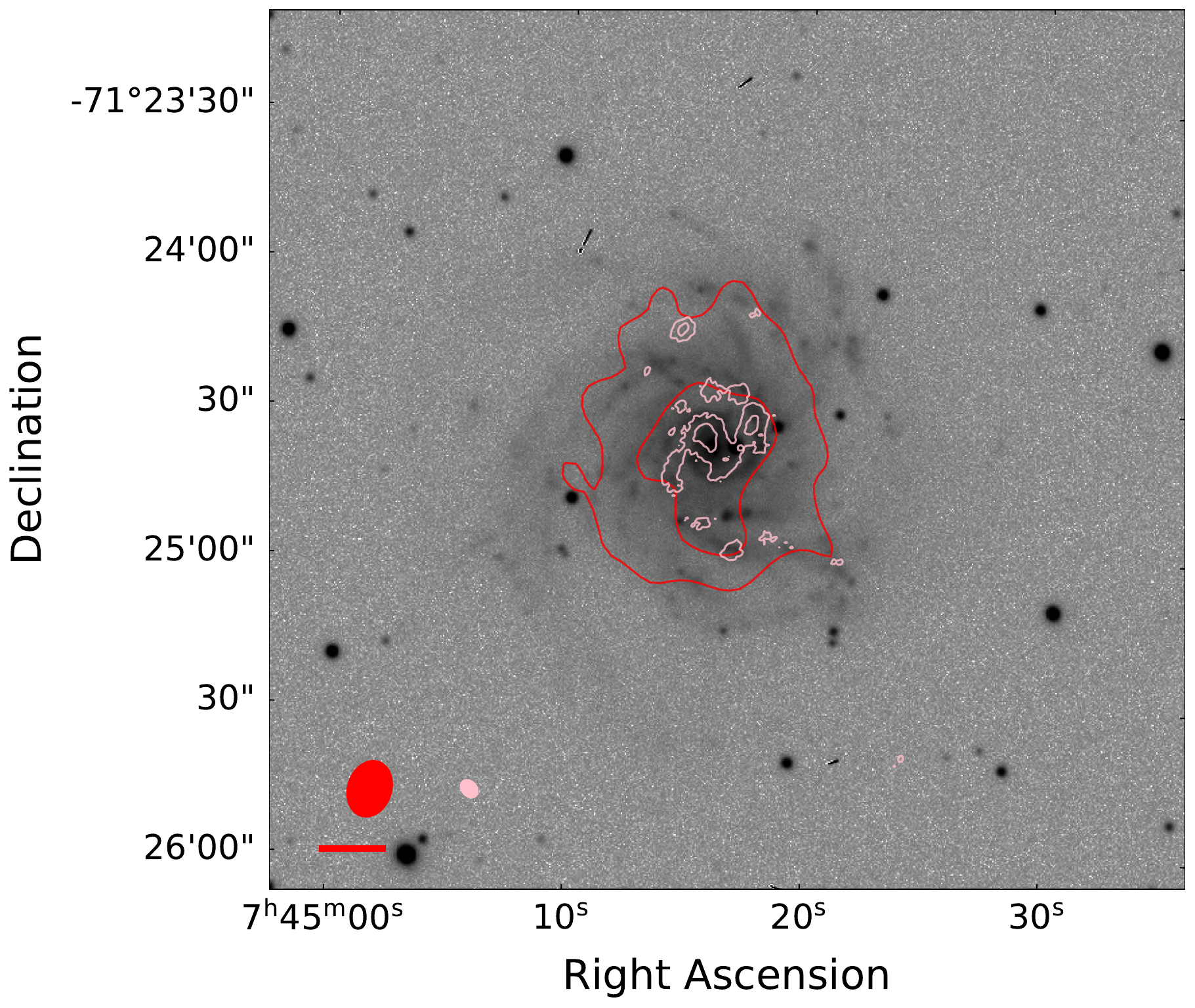}
    \end{subfigure}
    \vspace{0.8em}
        \begin{subfigure}[b]{0.43\textwidth}
        \centering
        \includegraphics[width=\textwidth]{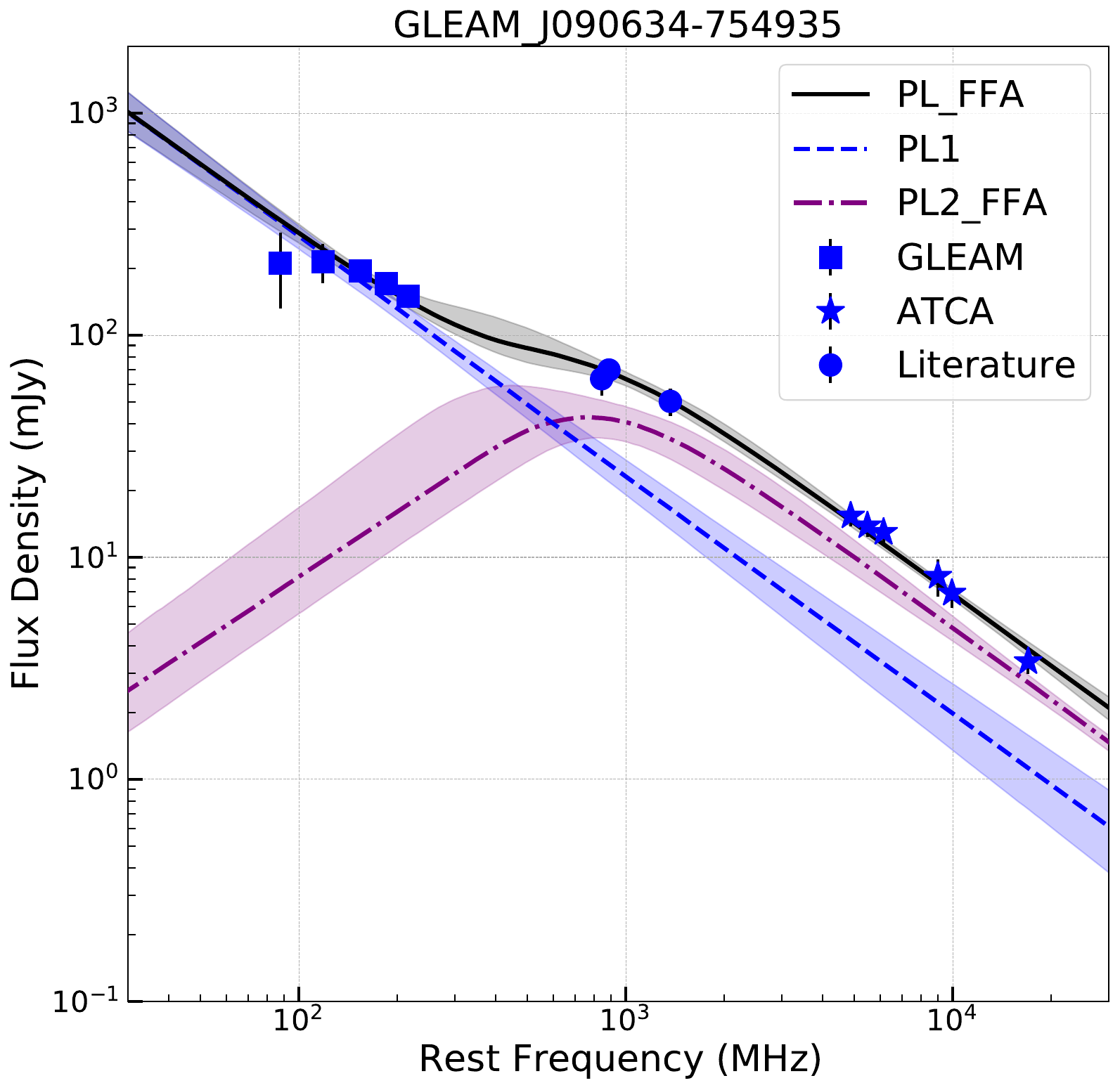}
    \end{subfigure}
    \begin{subfigure}[b]{0.47\textwidth}
        \centering
        \includegraphics[width=\textwidth]{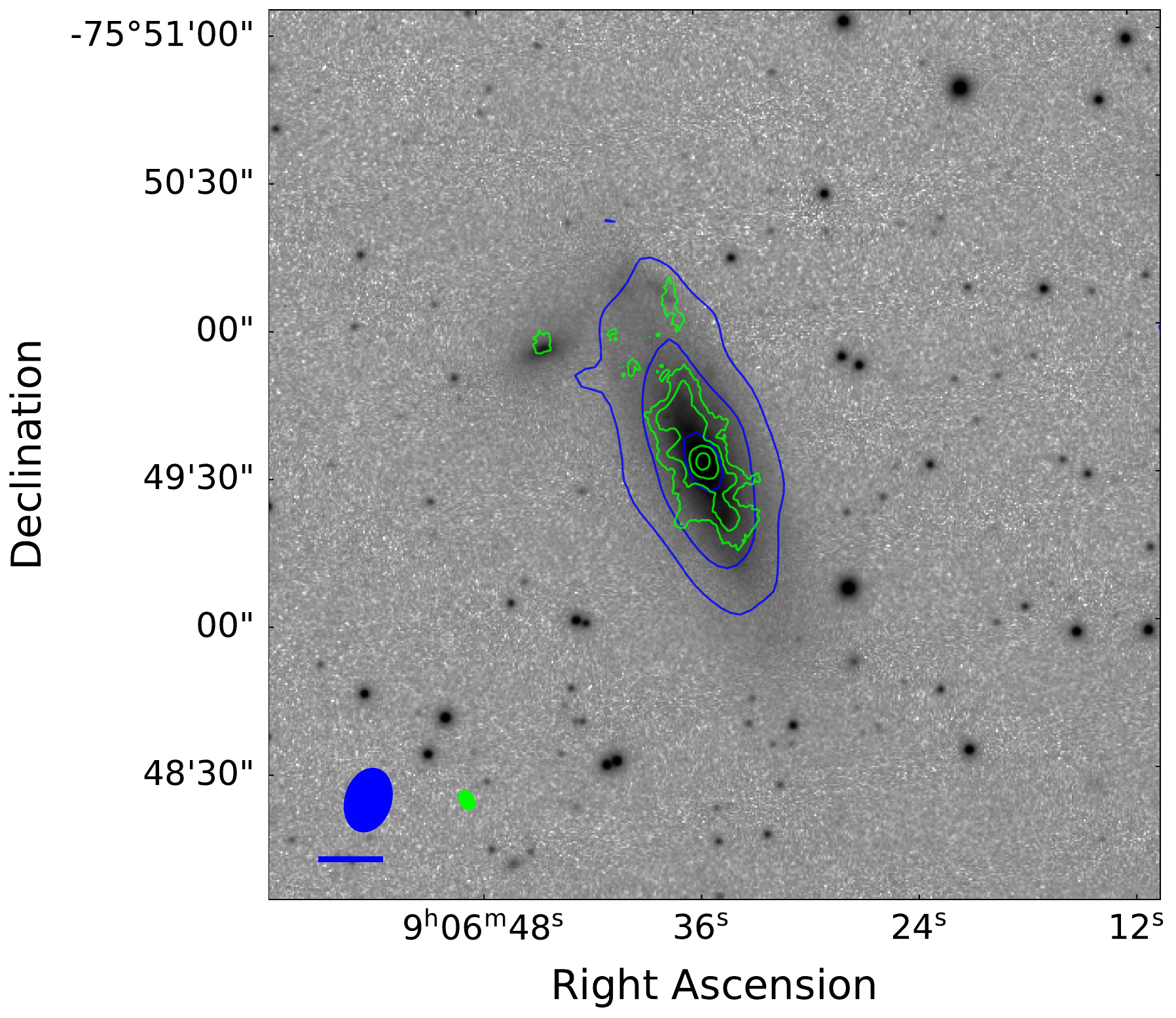}
    \end{subfigure}
        \vspace{0.8em}
        \begin{subfigure}[b]{0.43\textwidth}
        \centering
        \includegraphics[width=\textwidth]{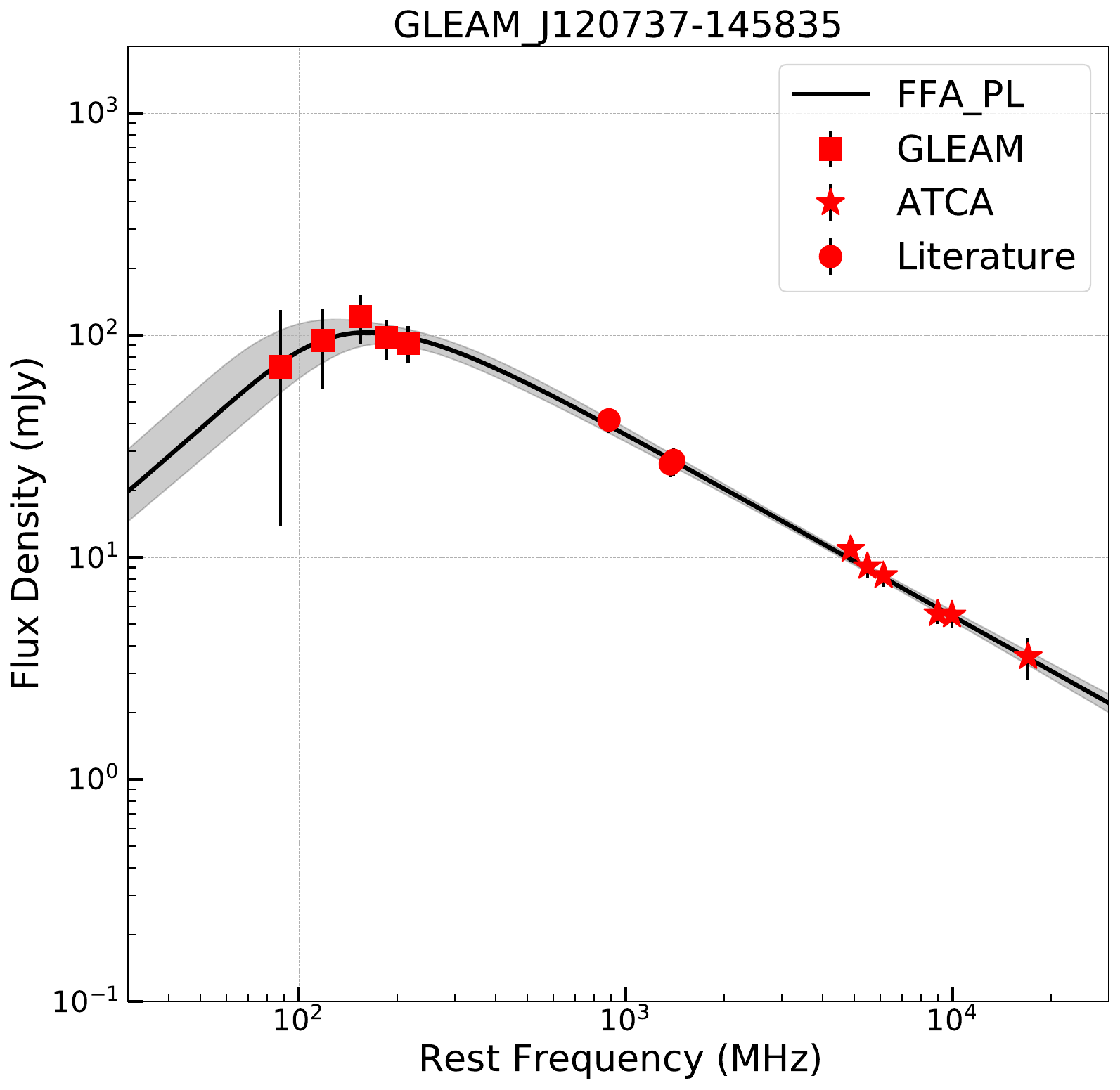}
    \end{subfigure}
    \begin{subfigure}[b]{0.47\textwidth}
        \centering
        \includegraphics[width=\textwidth]{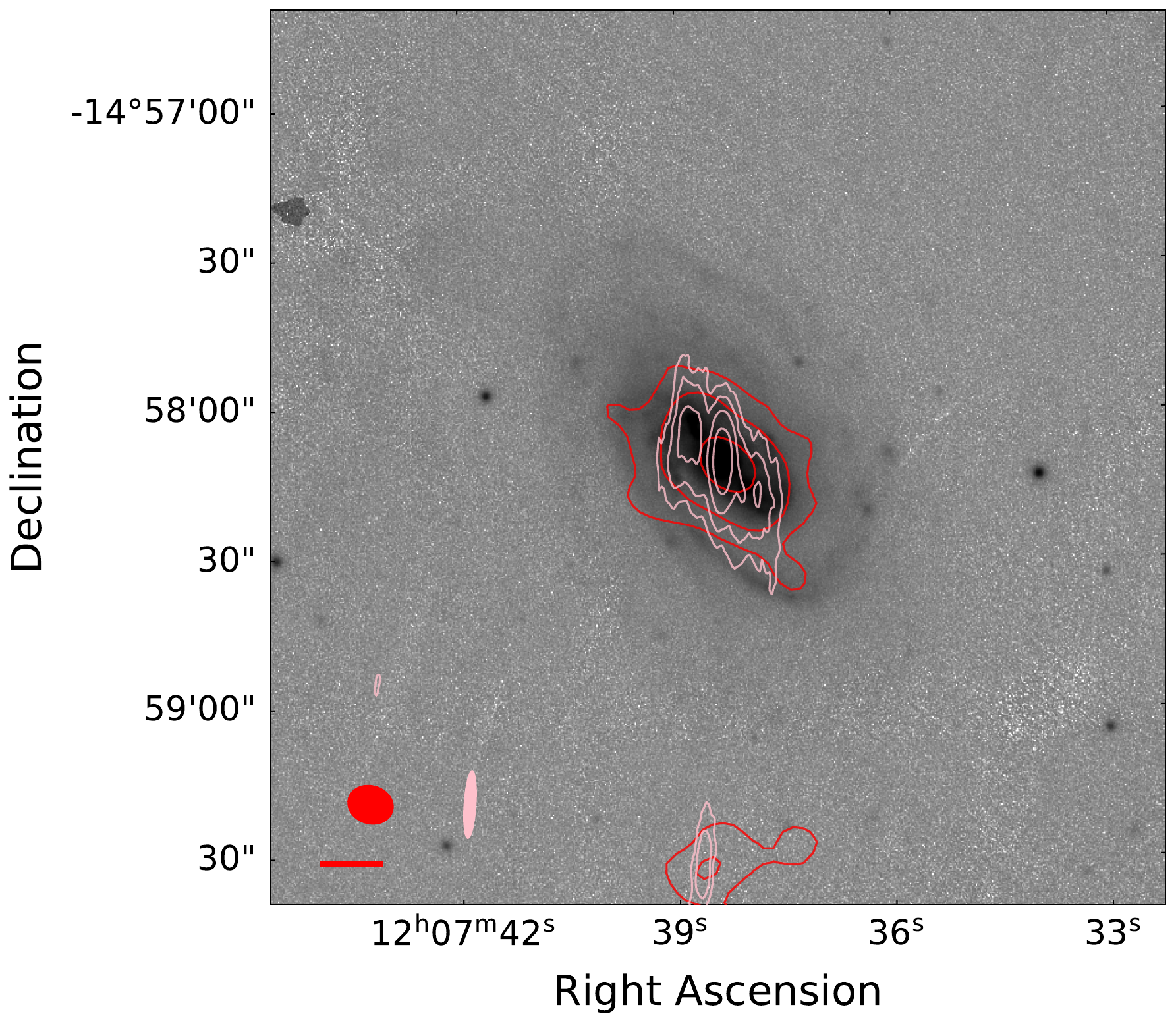}
    \end{subfigure}
    \caption{Figure 13. Continued}
\end{figure*}

\begin{figure*}[hbt!]
    \centering
    \begin{subfigure}[b]{0.43\textwidth}
        \centering
        \includegraphics[width=\textwidth]{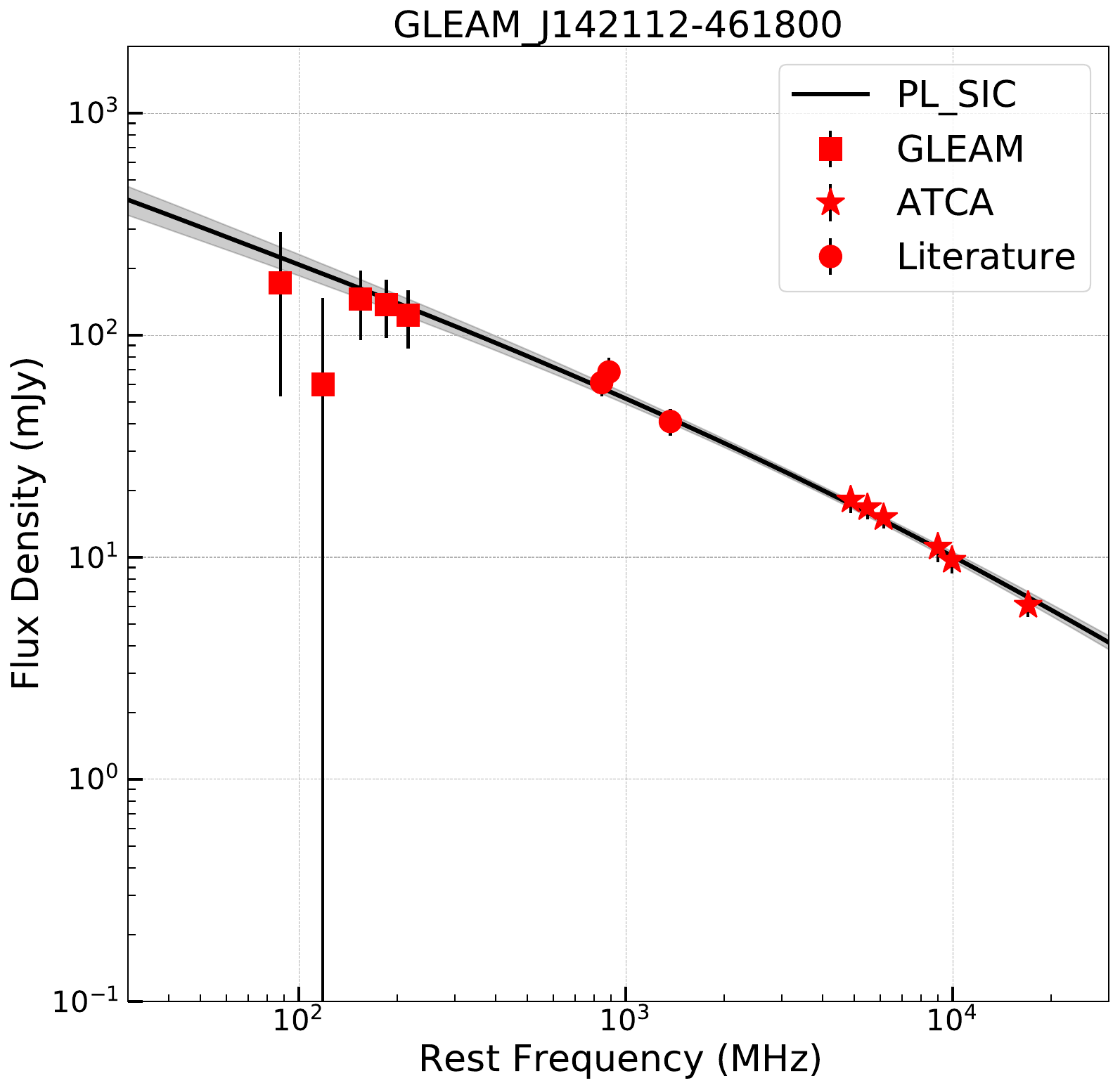}
    \end{subfigure}
    \begin{subfigure}[b]{0.47\textwidth}
        \centering
        \includegraphics[width=\textwidth]{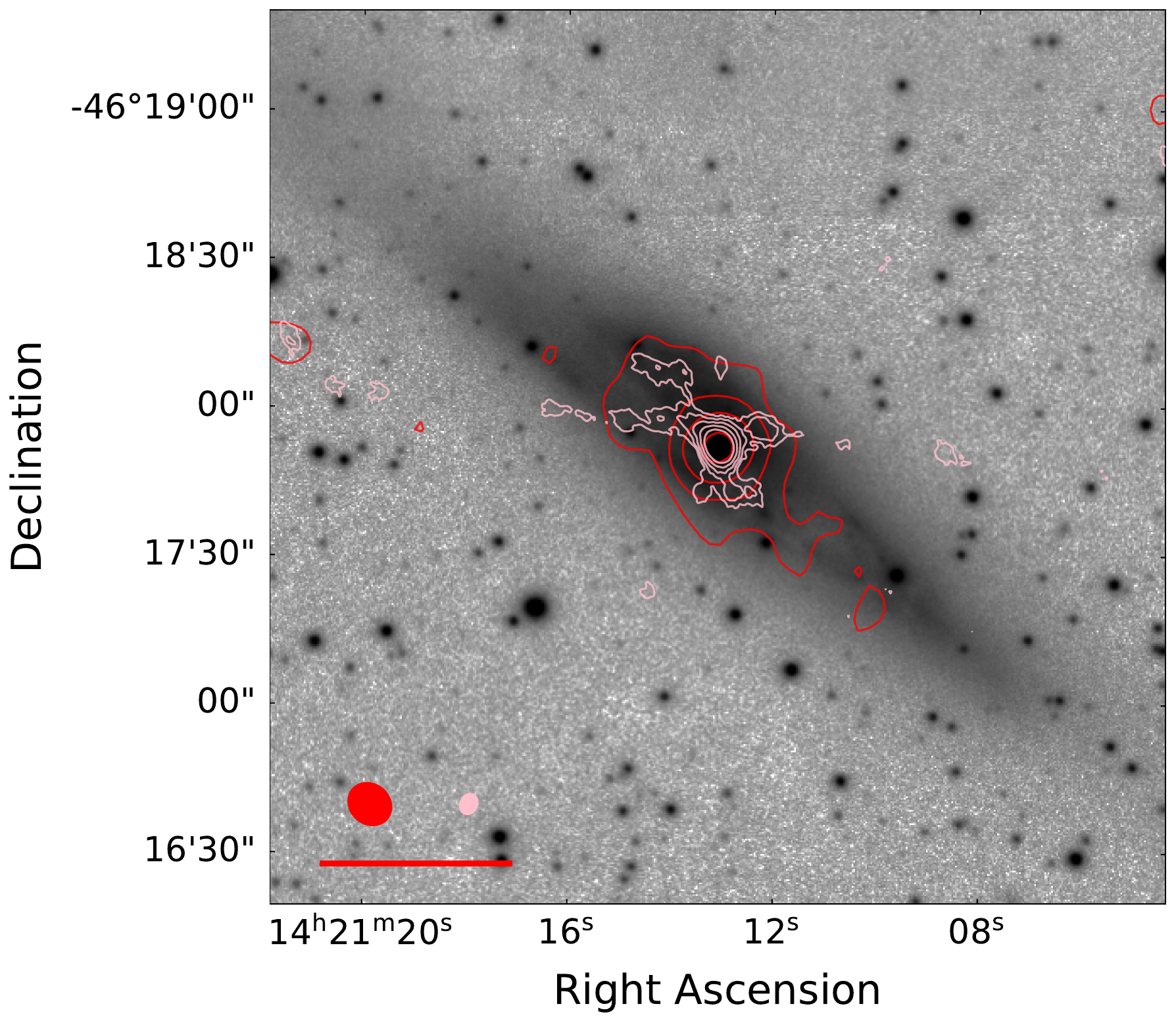}
    \end{subfigure}
    \vspace{0.8em}
        \begin{subfigure}[b]{0.43\textwidth}
        \centering
        \includegraphics[width=\textwidth]{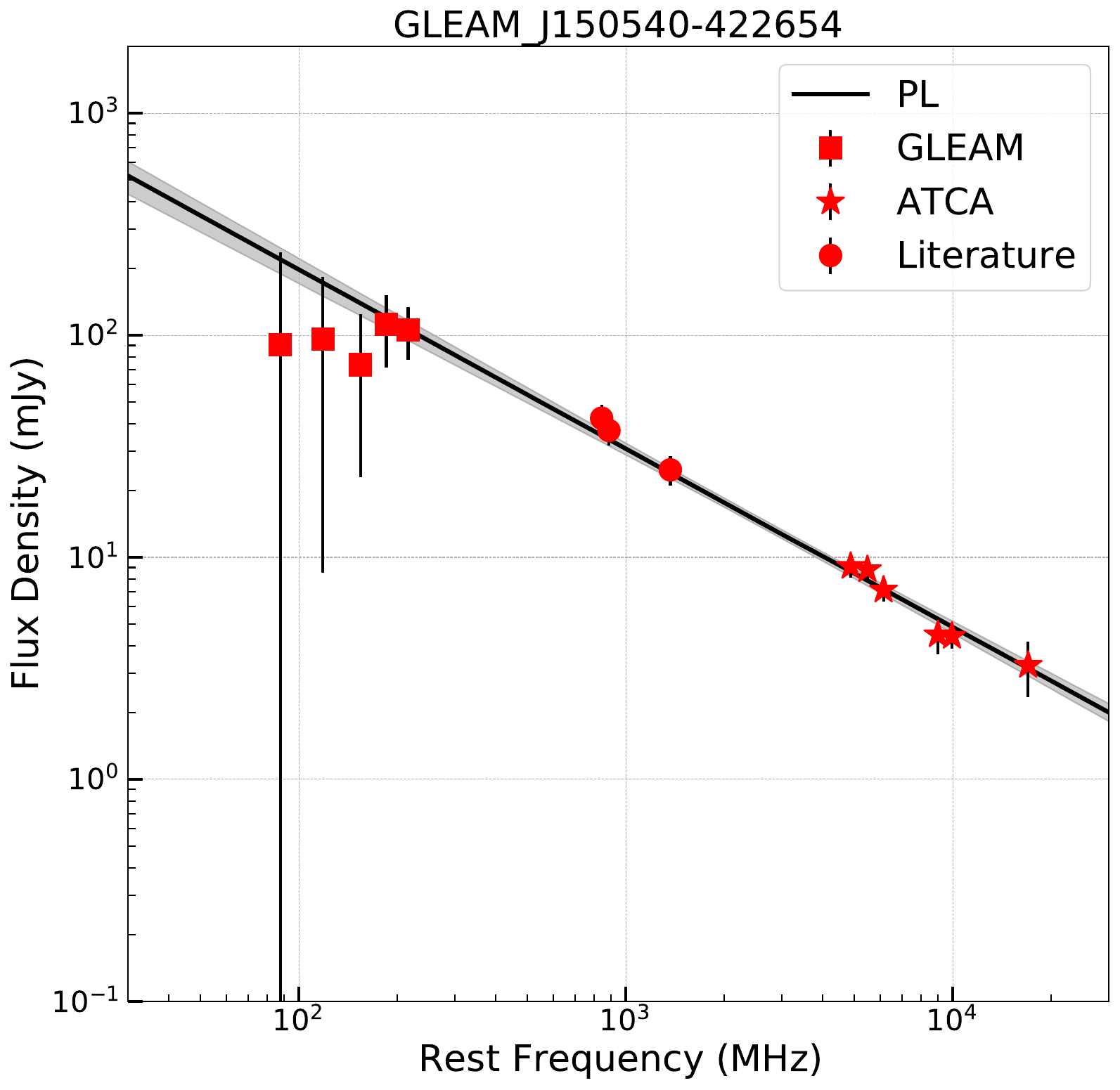}
    \end{subfigure}
    \begin{subfigure}[b]{0.47\textwidth}
        \centering
        \includegraphics[width=\textwidth]{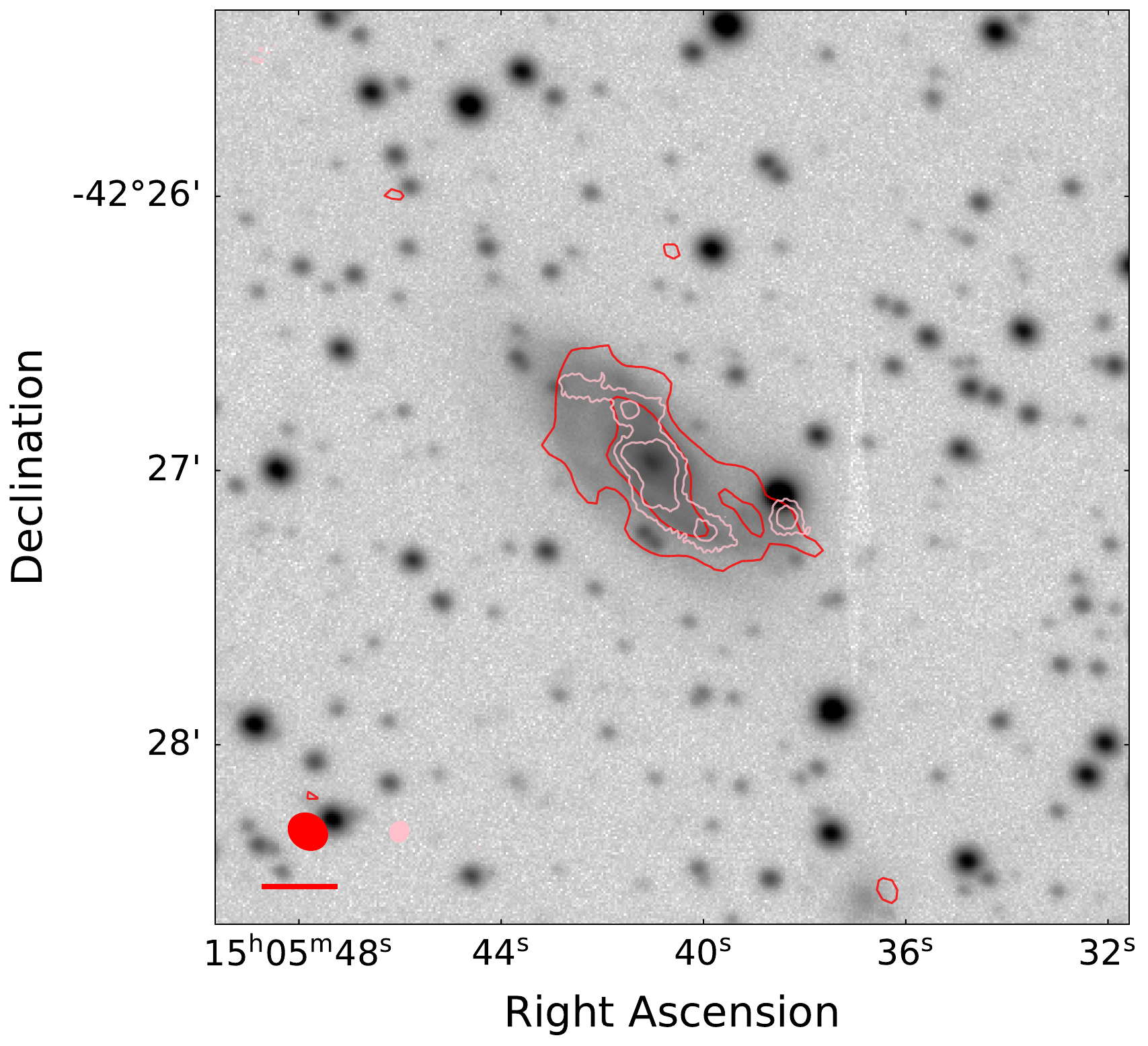}
    \end{subfigure}
        \vspace{0.8em}
        \begin{subfigure}[b]{0.43\textwidth}
        \centering
        \includegraphics[width=\textwidth]{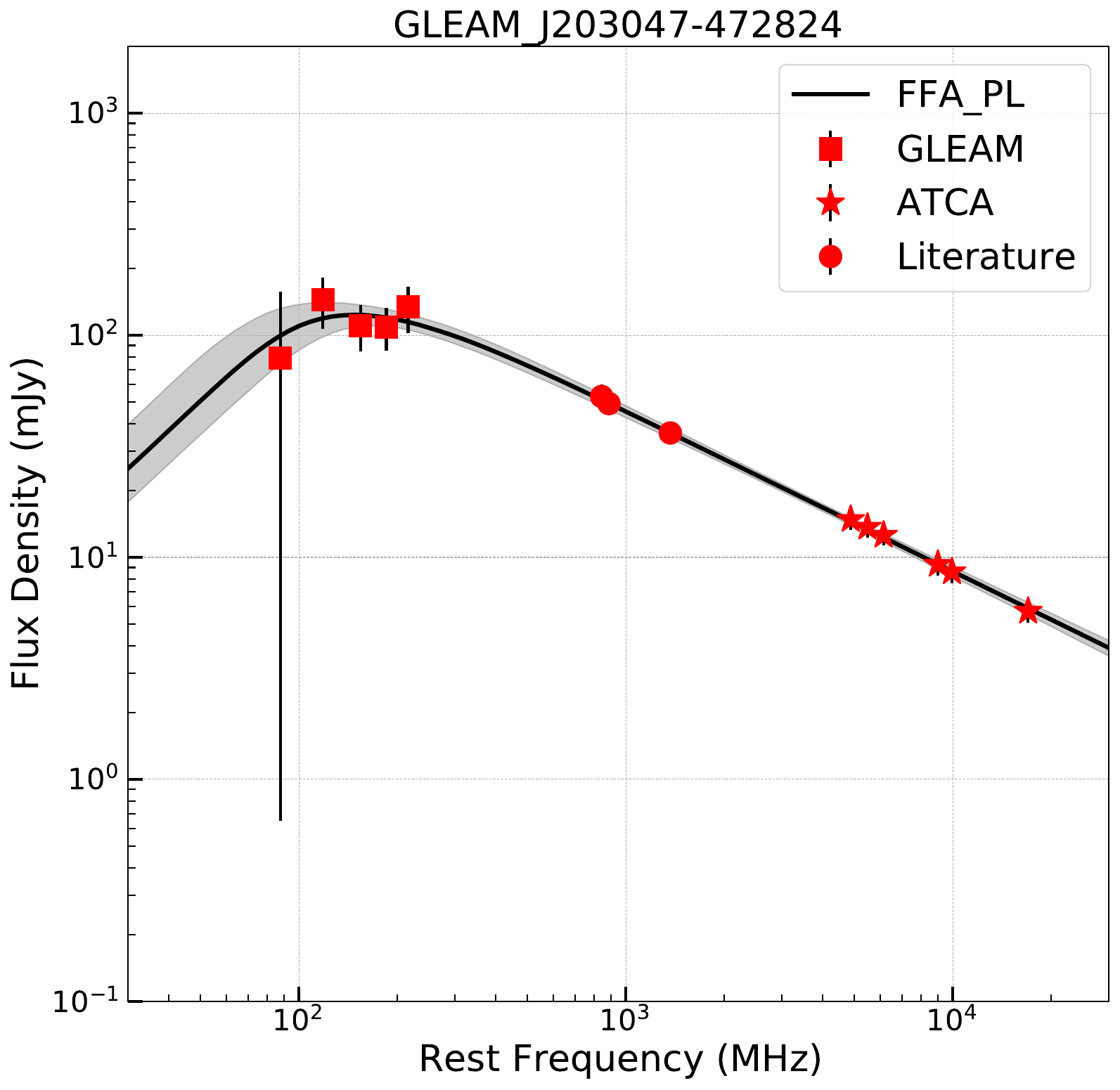}
    \end{subfigure}
    \begin{subfigure}[b]{0.47\textwidth}
        \centering
        \includegraphics[width=\textwidth]{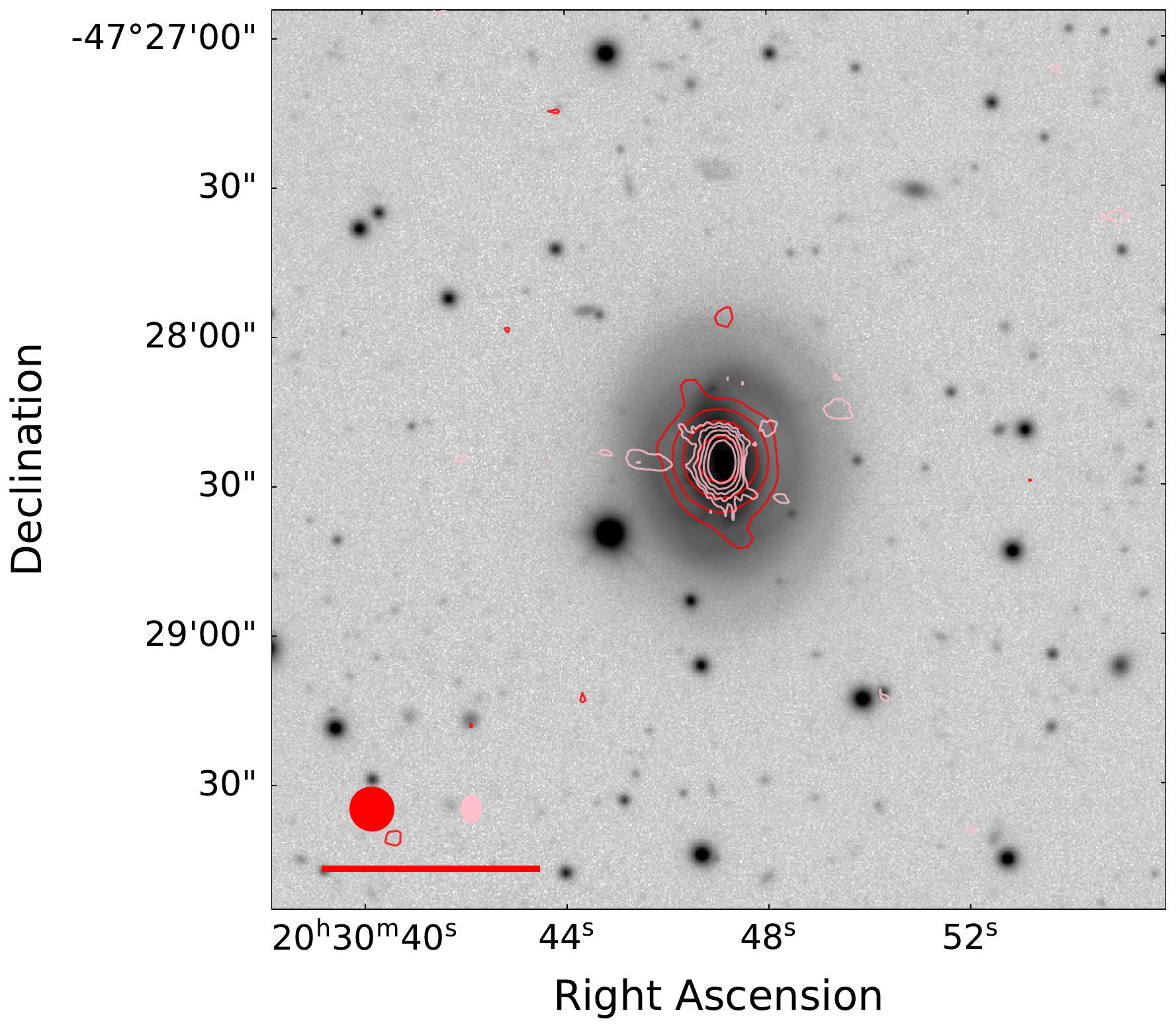}
    \end{subfigure}
    \caption{Figure 13. Continued}
\end{figure*}

\begin{figure*}[hbt!]
    \centering
    \begin{subfigure}[b]{0.43\textwidth}
        \centering
        \includegraphics[width=\textwidth]{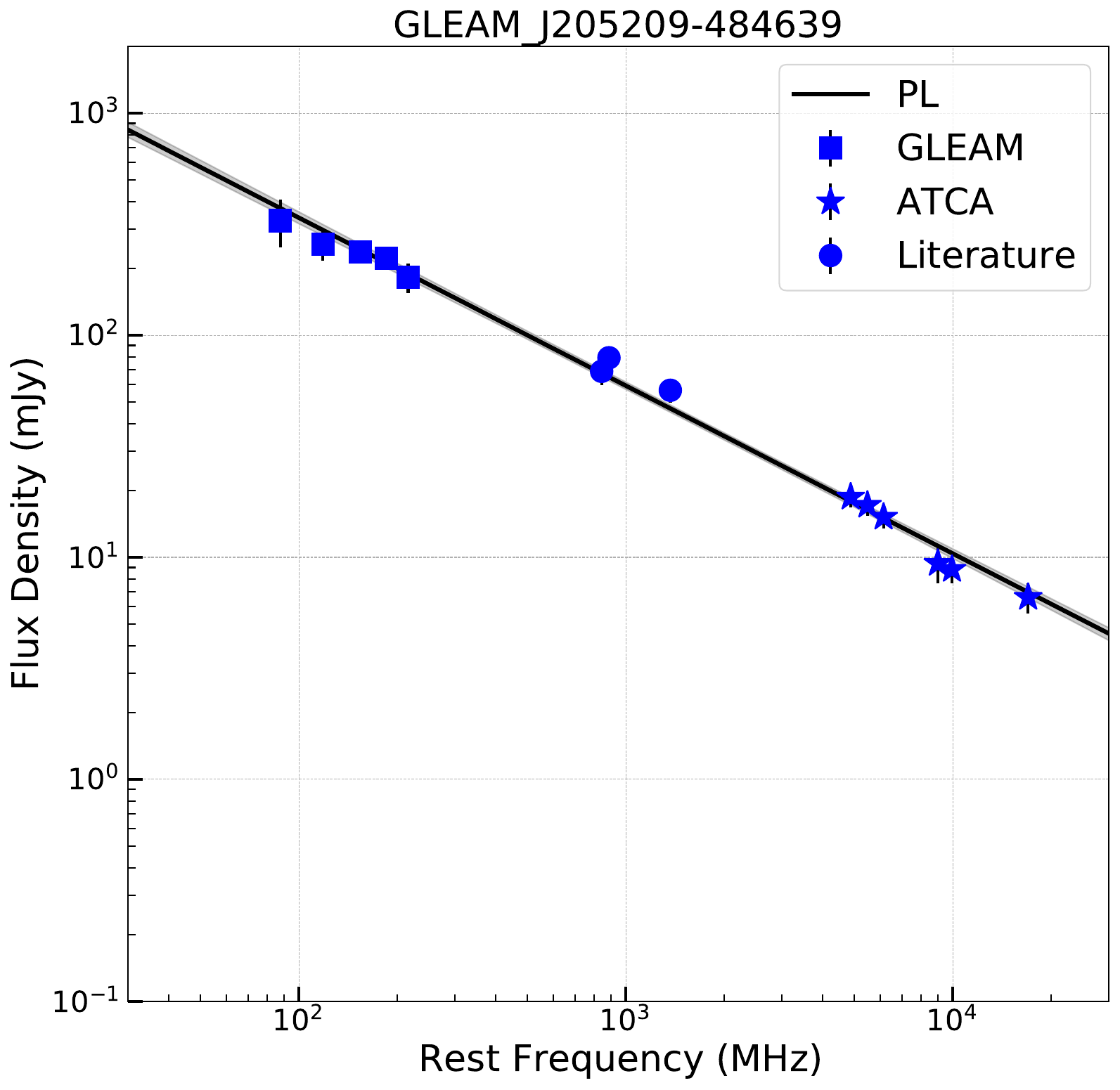}
    \end{subfigure}
    \begin{subfigure}[b]{0.47\textwidth}
        \centering
        \includegraphics[width=\textwidth]{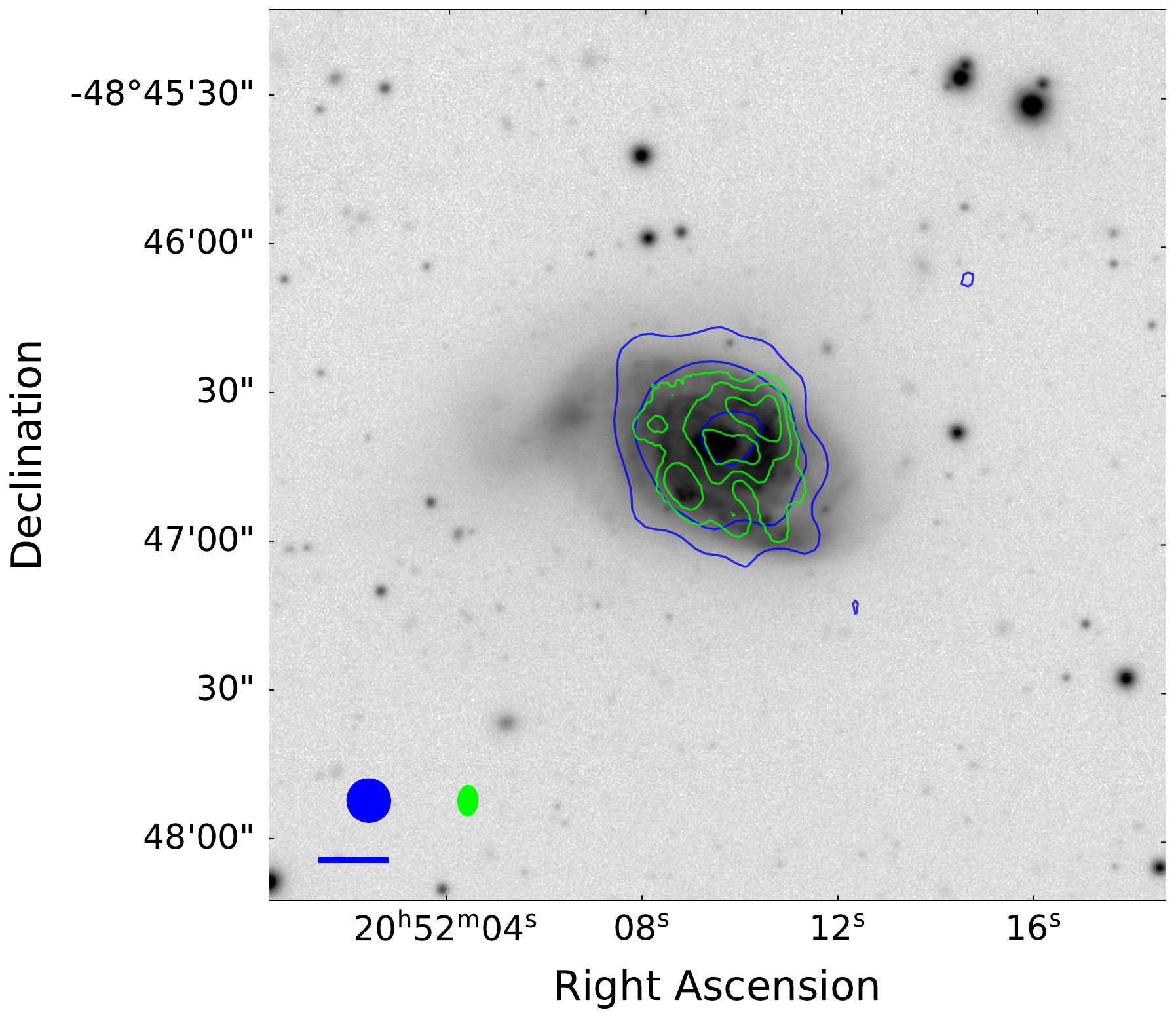}
    \end{subfigure}
    \vspace{0.8em}
        \begin{subfigure}[b]{0.43\textwidth}
        \centering
        \includegraphics[width=\textwidth]{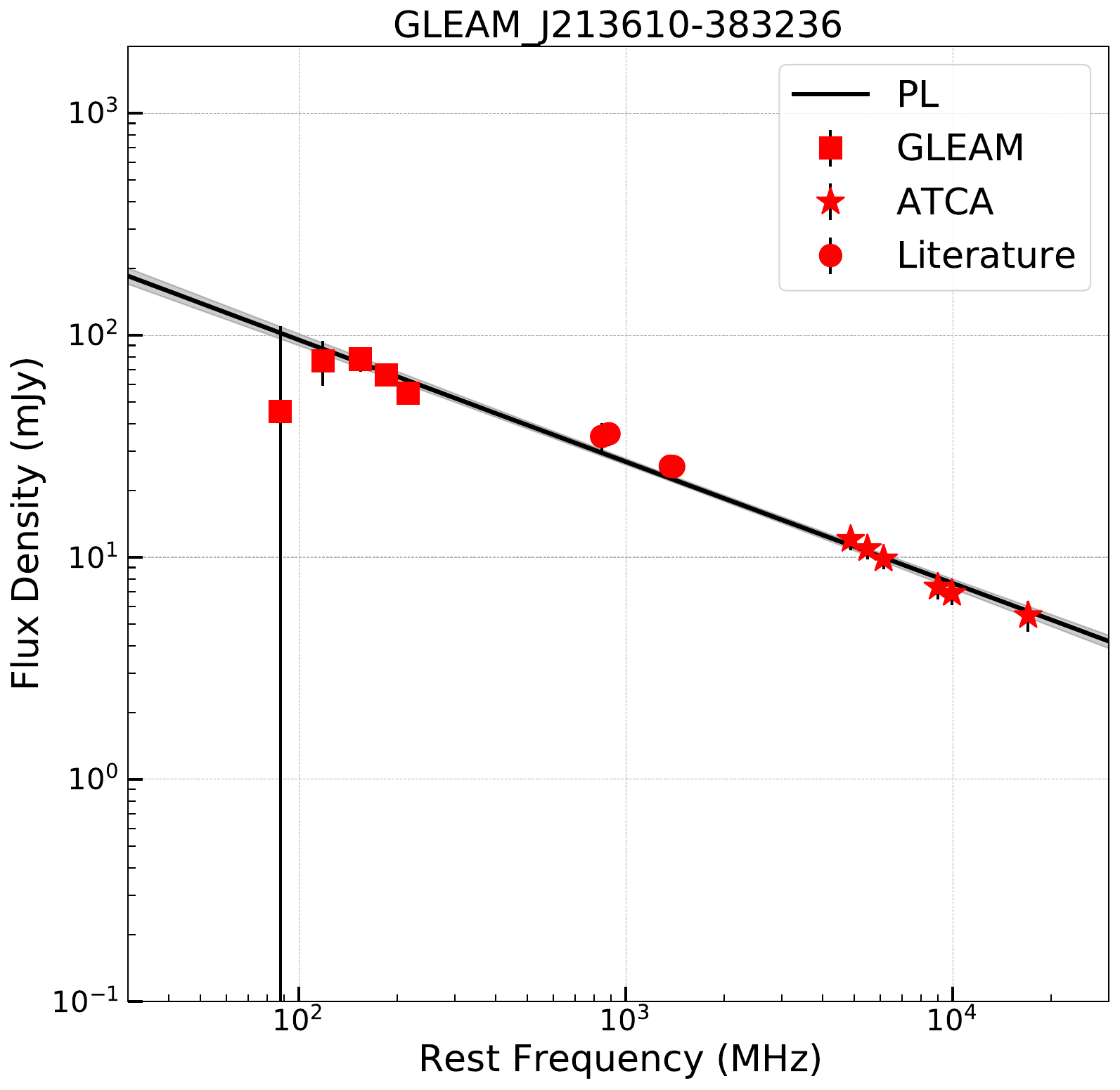}
    \end{subfigure}
    \begin{subfigure}[b]{0.47\textwidth}
        \centering
        \includegraphics[width=\textwidth]{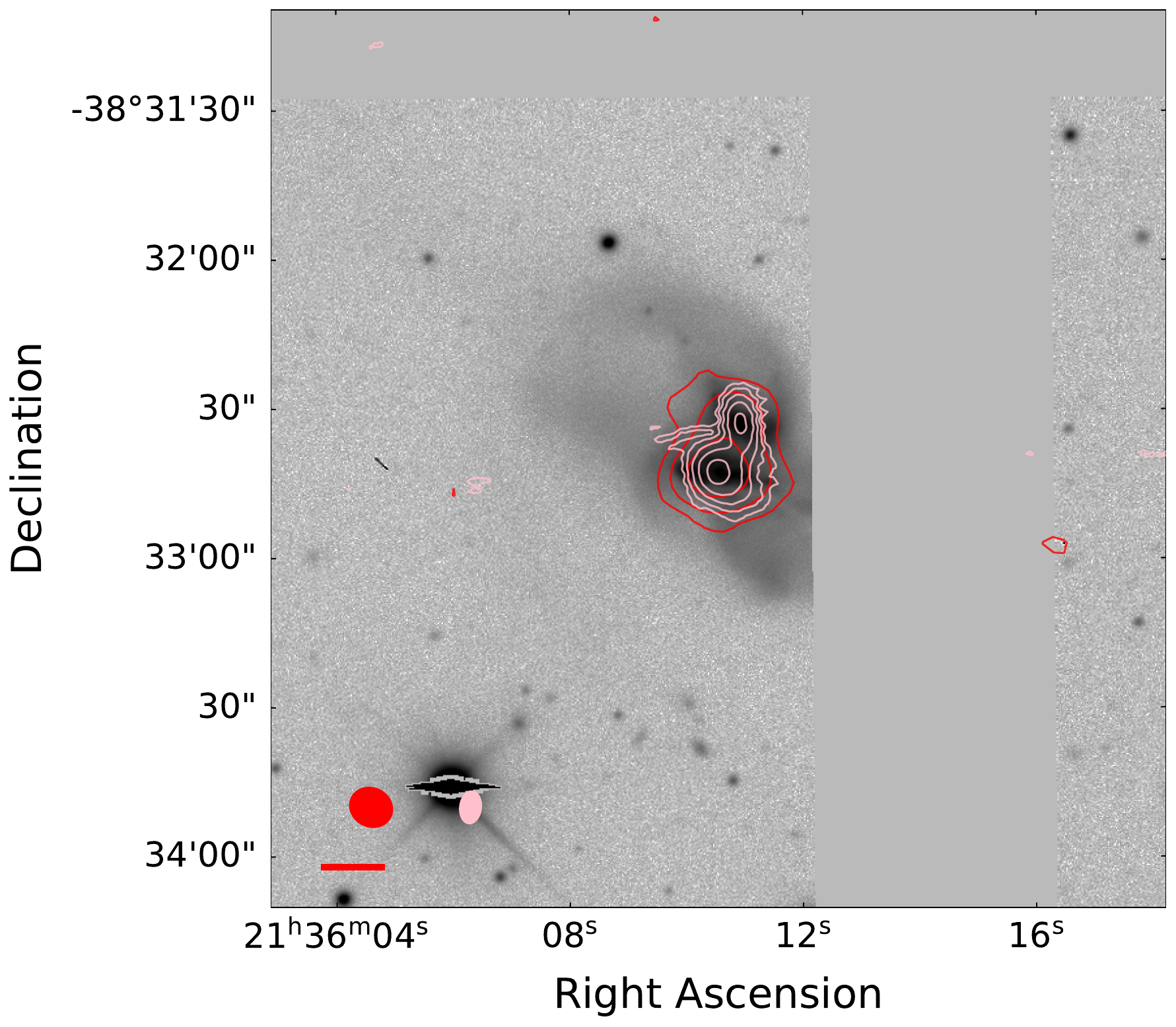}
    \end{subfigure}
    \caption{Figure 13. Continued}
\end{figure*}